# Non-Fermi-Liquid Behavior of Superconducting $SnH_4$

Ivan A. Troyan,[1] Dmitrii V. Semenok,[2,*] Anna G. Ivanova,[1] Andrey V. Sadakov,[3] Di Zhou,[2,*] Alexander G. Kvashnin,[4] Ivan A. Kruglov,[5,6] Oleg A. Sobolevskiy,[3] Marianna V. Lyubutina,[1] Dmitry S. Perekalin,[7] Toni Helm,[8] Stanley W. Tozer,[9] Maxim Bykov,[10] Alexander F. Goncharov,[11] Vladimir M. Pudalov,[3,12] and Igor S. Lyubutin[1,*]

[1] Shubnikov Institute of Crystallography, Federal Scientific Research Center Crystallography and Photonics, Russian Academy of Sciences, 59 Leninsky Prospekt, Moscow 119333, Russia
[2] Center for High Pressure Science and Technology Advanced Research (HPSTAR), Beijing 100094, China
[3] P. N. Lebedev Physical Institute, Russian Academy of Sciences, Moscow 119991, Russia
[4] Skolkovo Institute of Science and Technology, Skolkovo Innovation Center, Bolshoi Blv. 30, Building 1, Moscow 121205, Russia
[5] Dukhov Research Institute of Automatics (VNIIA), Moscow 127055, Russia
[6] Moscow Institute of Physics and Technology, 9 Institutsky Lane, Dolgoprudny 141700, Russia
[7] A.N. Nesmeyanov Institute of Organoelement Compounds, Russian Academy of Sciences, 28 Vavilova str., Moscow 119334, Russia
[8] Hochfeld-Magnetlabor Dresden (HLD-EMFL) and Würzburg-Dresden Cluster of Excellence, Helmholtz-Zentrum Dresden-Rossendorf (HZDR), Dresden 01328, Germany
[9] National High Magnetic Field Laboratory, Florida State University, Tallahassee, Florida 32310, USA
[10] Institute of Inorganic Chemistry, University of Cologne, Cologne 50939, Germany
[11] Earth and Planets Laboratory, Carnegie Institution for Science, 5241 Broad Branch Road NW, Washington, D.C. 20015, United States
[12] National Research University Higher School of Economics, Moscow 101000, Russia

**Corresponding Authors:** Di Zhou (di.zhou@hpstar.ac.cn), Dmitrii Semenok (dmitrii.semenok@hpstar.ac.cn), Igor S. Lyubutin (lyubutinig@mail.ru)

## Abstract

We studied chemical interaction of Sn with $H_2$ by X-ray diffraction methods at pressures of 180-210 GPa. A previously unknown tetrahydride $SnH_4$ with a cubic structure (*fcc*) exhibiting superconducting properties below $T_C = 72$ K was obtained; the formation of a high molecular $C2/m$-$SnH_{14}$ superhydride and several lower hydrides, *fcc* $SnH_2$ and $C2$-$Sn_{12}H_{18}$, was also detected. The temperature dependence of critical current density $J_C$(T) in $SnH_4$ yields the superconducting gap $2\Delta(0) = 23$ meV at 180 GPa. $SnH_4$ has unusual behavior in strong magnetic fields: *B,T*-linear dependences of magnetoresistance and the upper critical magnetic field $B_{C2}(T) \propto (T_C - T)$. The latter contradicts the Wertheimer–Helfand–Hohenberg model developed for conventional superconductors. Along with this, the temperature dependence of electrical resistance of *fcc* $SnH_4$ in non-superconducting state exhibits a deviation from what is expected for phonon-mediated scattering described by the Bloch-Grüneisen model, and is beyond the framework of the Fermi liquid theory. Such anomalies occur for many superhydrides, making them much closer to cuprates than previously believed.

# Introduction

The study of high-temperature superconductivity is one of the most important problems in condensed matter physics. Compressed polyhydrides are promising high-temperature superconductors, which can be obtained at pressures of 1-2 Mbar. Since 2015, many remarkable hydride superconductors have been experimentally discovered. These are $H_3S$ (with $T_C = 200$ K) [1], $LaH_{10}$ ($T_C = 250$ K) [2,3], $ThH_9$ ($T_C = 146$ K) and $ThH_{10}$ ($T_C = 161$ K) [4], $YH_6$ ($T_C = 226$ K) [5,6] and $YH_9$ ($T_C = 243$ K) [6], $CeH_9$ and $CeH_{10}$ ($T_C = 110$-$120$ K) [7], $CaH_6$ ($T_C = 215$ K) [8,9]. Most of superconducting polyhydrides contain elements of groups IIIA, IVB with $d^0$-$d(f)^1$ valent electrons [10], whereas polyhydrides of *p*-elements, and in particular Sn, have not been studied enough.

In the last 10 years, the chemistry of polyhydrides of group IV elements (C, Si, Ge, Sn, Pb) has been intensively studied by means of the density functional theory (DFT). The experimental synthesis of such polyhydrides is carried out at high-pressure in diamond anvil cells (DACs). C, Si, and Ge hydrides are mainly low-symmetry molecular covalent compounds with moderate superconducting properties [11, 12, 13]. The higher molecular silicon polyhydride $SiH_4(H_2)_2$ was observed in experiments of Strobel et al. and Wang et al. [14,15]. Resistive transitions with $T_C$ up to 79 K for silicon polyhydrides were obtained under high pressure by the group of M. Eremets et al. [16] Theoretical calculations for Ge-hydrides at 200-300 GPa, predicted stable phases of $GeH_3$ and $Ge_3H_{11}$ with low symmetry, and various polymorphic modifications of the well-known $GeH_4$. [13,17] A new molecular compound, possibly $GeH_4(H_2)_2$, was obtained experimentally at 7.5 GPa. [18] According to theoretical calculations, metallic $P2_1/c$-$GeH_4(H_2)_2$ is a superconductor with $T_C$ of 76–90 K at 250 GPa. [19]

The increase in metallicity of Sn to Pb results in the appearance of thermodynamically stable polyhydrides in the high-pressure phase diagram. For instance, theory predicts the formation of $C2/m$-$SnH_{12}$, $C2/m$-$SnH_{14}$ [20] and various $PbH_4$, $PbH_6$, $PbH_8$ [21-23] lead polyhydrides under high pressure. However, attempts to synthesize Pb polyhydrides have been unsuccessful to date [24]. The recent observation by Hong et al [25] of a sharp drop in electrical resistance in unidentified Sn polyhydride at 71 K motivated us to investigate in detail the structure and superconducting properties of Sn polyhydrides under high pressure.

Three series of XRD experiments were carried out at different synchrotron facilities such as ESRF in 2017, PETRA in 2020 and APS in 2022. At pressures around 160-210 GPa, we observed X-ray diffraction patterns of the studied samples, which are attributed to different Sn hydrides formed. Most XRD patterns have a characteristic set of cubic ($Fm\bar{3}m$) reflections. In the experiment performed at APS, a single-crystal XRD at a pressure of about 190 GPa was carried out. An analysis of the diffraction pattern confirmed the cubic structure ($Fm\bar{3}m$) of the Sn sublattice in $SnH_4$. As we have found, tin tetrahydride is unusual superconductor, which demonstrates anomalous behavior of electrical resistance, T-linear upper critical magnetic field, and H-linear magnetoresistance in a wide range of temperatures and magnetic fields.

# 2. Results

## 2.1. *High-pressure synthesis of $SnH_4$ with a cubic (fcc) structure*

In this work, tin polyhydrides were synthesized and studied by X-ray diffraction (XRD). The previously unknown tetrahydride $SnH_4$ with a cubic structure (*fcc*) was synthesized at high pressures in diamond anvil cells (DACs) in two different ways. In the first approach, we used the reaction of solutions of $SnCl_4$ and $LiAlH_4$. Gaseous stannane $SnH_4$ was emanated, then condensed upon cooling with liquid nitrogen [26]. Using a cryogenic filling at a pressure of about 2.0 bar, liquefied $SnH_4$ was supplied through a thin capillary into the working volume of the DAC, and then compressed to high pressure at room temperature (Supporting Figure S7). Monitoring the successful loading of gaseous $SnH_4$ and its compression provided additional confirmation of the hydrogen content in the resulting high-pressure phase *fcc* $SnH_4$. In the second approach, $SnH_4$ was synthesized during the interaction of pre-compressed piece of metallic Sn and hydrogen produced by laser heating of $NH_3BH_3$ under high pressure. A table summarizing all the performed experiments can be found in the Supporting Information, Tables S1 and S5.

Let us first consider the results of the initial experiment we performed at the ID27 beamline of ESRF in 2017 (Fig. 1). Our preliminary study of the compression of pure Sn (Fig. 1a, DAC S1) showed that at 190 GPa it exists in two modifications: hexagonal (*hcp*) and cubic (*bcc*), the latter one is dominant. In fact, the phase transition *bcc* → *hcp* starts already at 160 GPa [27]. However, the enthalpies of formation of these modifications differ very little (according to DFT calculations it is $\Delta H$ = 9.5 meV/atom), which explains the very low rate of this transformation. As follows from the XRD pattern in Fig. 1a, at 190 GPa, the Sn unit cell volume is V($Im\bar{3}m$-Sn) = 26.9 $Å^3$, which agrees with results of a previous publication [28]. Interesting, but the unit cell volumes of *hcp* Sn and *bcc* Sn are very similar.

During the experiment with stannane, frozen $SnH_4$ was placed in DAC S2 and compressed to 180 GPa. It was found that with an increase in pressure, rapid metallization of $SnH_4$ occurs at around 10 GPa (Supporting Figure S9), accompanied by the disappearance of the Raman signal. At about 160 GPa, a new cubic modification of $SnH_4$ is formed. Figure 1b shows that the Bragg peaks of the experimental XRD patterns of $SnH_4$ at pressures 160-180 GPa are best indexed by a face-centered cubic lattice (*fcc)* with unit cell volume 19.6 $Å^3$/Sn at 180 GPa. The obtained XRD patterns also contain an impurity (broadened peaks at 2θ = 9.5° and 10.2°) which can be attributed to *hcp* Sn formed during $SnH_4$ dissociation when loaded and compressed in the DAC S2. Stannane is thermodynamically unstable and under ambient conditions it gradually decomposes to Sn and $H_2$.

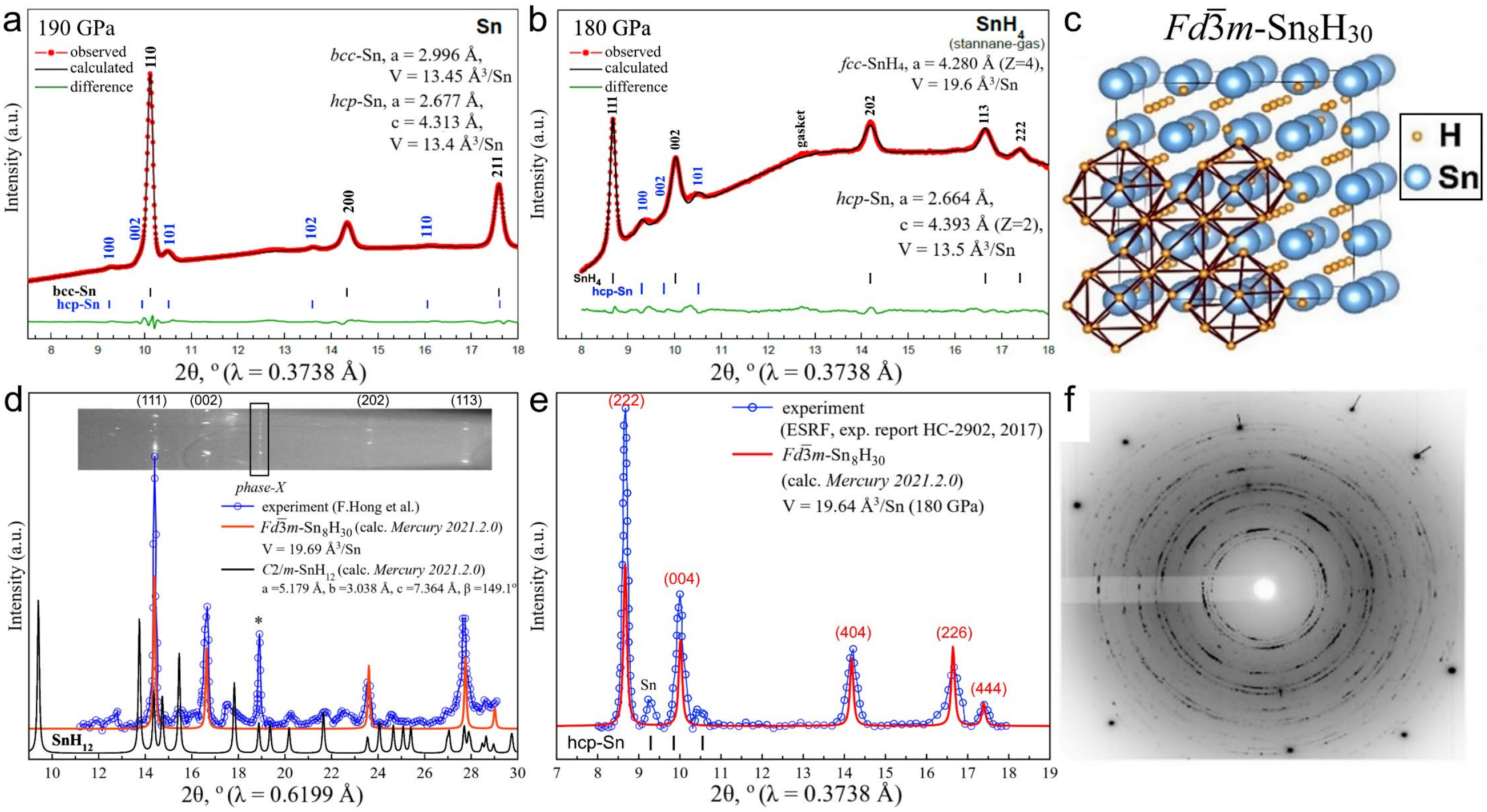

**Figure 1.** X-ray diffraction patterns and the Le Bail refinements of the unit cell parameters of (a) Sn at 190 GPa in DAC S1, and (b) $SnH_4$ and *hcp* Sn obtained after compression of gaseous stannane to 180 GPa in DAC S2 (ESRF-2017). (c) The crystal structure of theoretical model, $Fd\bar{3}m$-$Sn_8H_{30}$, which best fits the experimental pattern. (d) XRD pattern of Sn hydride obtained by Hong et al. [25] (blue curve, 206 GPa) and its fit by $Fd\bar{3}m$-$Sn_8H_{30}$ model (red curve). Asterisk indicate uninterpreted reflection (phase X). Obviously, the diffraction pattern does not correspond to the $C2/m$-$SnH_{12}$ structure (black curve) proposed by Hong et al. (e) Fitting of the experimental XRD pattern after background subtraction (ESRF-2017) using theoretical model $Fd\bar{3}m$-$Sn_8H_{30}$ (red curve) and *hcp* Sn phase (black dashes). It gives much better result than $C2/m$-$SnH_{12}$. (f) Typical diffraction pattern of *fcc* $SnH_4$ at 170 GPa (PETRA-2020, DAC D2).

Compressed Sn hydrides were also experimentally studied by Hong et al. [25] The authors obtained a similar X-ray diffraction pattern (Fig. 1d), indexed as *C*2/*m*-$SnH_{12}$ ($a$ = 5.179 Å, $b$ = 3.038 Å, $c$ = 7.364 Å, $\beta$ = 149.11°). This structure was previously predicted by DFT methods [20]. It can be seen that the experimental XRD clearly does not match the theoretically predicted XRD of $SnH_{12}$ (black curve in Fig. 1d). At the same time, the experimental XRD pattern (blue curve in Fig. 1d) fits well with the theoretical model of *fcc* $SnH_4$, namely $Fd\bar{3}m$-$Sn_8H_{30}$ (= $SnH_{3.75}$, red curve in Fig. 1d), which is discussed in details in the Supporting Information ("Crystal structure search" section). The unit cell volume of the $Sn_8H_{30}$ is $V$ = 19.64 $Å^3$/Sn (or 628.68 $Å^3$ for Z = 32) at 180 GPa. Using the data obtained in the ESRF-2017 experiment, we refined unit cell parameters of $Fd\bar{3}m$-$Sn_8H_{30}$, which were found to be in a good agreement with the results of calculations (Supporting Tables S5, S8).

### *2.2. Synthesis of molecular $SnH_{14}$*

The next experiments were performed at the P02.2 PETRA III beamline in 2020. As has already been shown, stannane compression leads to formation of mixtures of *hcp* Sn and cubic $SnH_4$. This indicates an insufficient amount of hydrogen for the complete conversion of Sn to polyhydrides. Therefore, to obtain tin hydrides with a higher hydrogen content (similar to $SrH_{22}$ [33] and $BaH_{12}$ [34]), we decided to use ammonia borane $NH_3BH_3$ (AB) as source of $H_2$ in DACs D2 and M2.

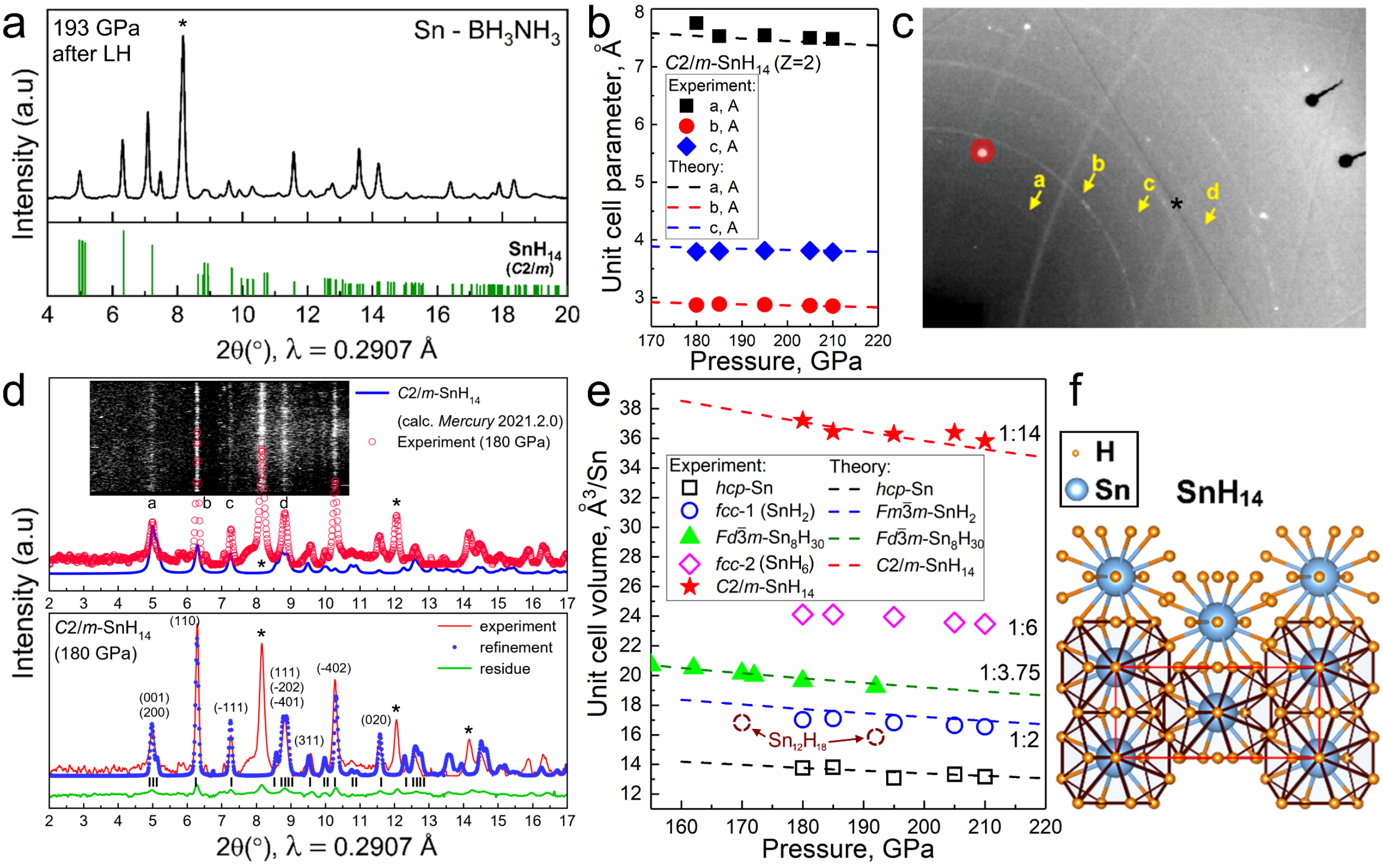

**Figure 2.** X-ray diffraction patterns and the Le Bail refinements of the unit cell parameters of Sn hydrides in DACs D2 and M2. (a) XRD pattern of Sn in $NH_3BH_3$ media after the laser heating (LH) at 193 GPa. Predicted XRD of *C*2/*m*-$SnH_{14}$ are shown on the bottom panel. (b) Experimental and theoretical dependences of the unit cell parameters on the pressure for *C*2/*m*-$SnH_{14}$. (c) Typical diffraction pattern at 180 GPa (PETRA-2020, λ = 0.2904 Å). Asterisks denote uninterpreted peaks, "a-d" denote main reflections from $SnH_{14}$. (d) Comparison of experimental and calculated XRD reflection intensities for $SnH_{14}$ ("a-d"). Inset: diffraction image ("cake"). Bottom panel: the Le Bail refinement of *C*2/*m*-$SnH_{14}$. Unidentified reflections are marked by asterisks. The experimental data, fitted line, and residues are shown in red, blue, and green, respectively. (e) Pressure–unit cell parameters diagram for all tin hydrides synthesized in our experiments: stars, rhombuses, triangles, circles and squares show the experimental data, lines depict the theoretical calculations. (f) Crystal structure of $SnH_{14}$ polyhydride.

The use of AB has yielded interesting results. We found that in addition to $SnH_4$, several new compounds are formed, whose XRD patterns mostly correspond to the higher molecular polyhydride $C2/m$-$SnH_{14}$ predicted in 2016 [29] (Fig. 2). The refinements of the unit cell parameters of synthesized $SnH_{14}$ (Figs. 2a, b, d) are in a good agreement with the theoretical DFT calculations (Supporting Table S9).

This polyhydride has an orthorhombic *Immm*-Sn sublattice and a hydrogen sublattice, consisting of $H_2$ ($d_{HH}$ = 0.833 Å) and $H_3$ ($d_{HH}$ = 0.92 Å) fragments. According to molecular dynamics modeling, $C2/m$-$SnH_{14}$ is dynamically stable in the anharmonic approximation (Supporting Figure S13). Moreover, $SnH_{14}$ exhibits the properties of a typical metal with a high density of electronic states at the Fermi level (DOS). The DOS projection on hydrogen atoms, $N_F$(H) = 0.32 states/eV/f.u., is larger than that on Sn atoms ($N_F$(Sn) = 0.23 states/eV/f.u.). The hydrogen sublattice exhibits anisotropic properties, and is in a solid (not glassy, like in $P1$-$SrH_{22}$ [30]) state. Due to the presence of large amount of molecular hydrogen in the structure of $SnH_{14}$, its Debye temperature is rather high ($\theta_D \sim 1500$ K, $\omega_{\log}$ up to 1252 K). Also, the superconducting properties are well pronounced, the electron-phonon coupling (EPC) coefficient is $\lambda = 1.25$, and the critical temperature is $T_C$ = 107–133 K at 200 GPa ($\mu^*$ = 0.15-0.1) in the harmonic approximation (Supporting Figures S12-S13). This significantly distinguishes molecular tin polyhydrides from similar compounds of barium ($BaH_{12}$, which is a low-$T_C$ superconductor, $T_C \sim 20$ K) and strontium ($SrH_{22}$, which is a semiconductor).

In our last experiment performed in 2022 at APS, Sn and AB were again used. The studies were carried in DAC S3 with a wide opening angle (~70°) and diamond anvils of the Böhler-Almax type to obtain a single-crystal XRD pattern (Supporting Figure S2), where one phase was indexed by the *fcc* $SnH_4$ (*R*-factor is 4.75 %). A co-product of the synthesis is a lower hydride, whose X-ray diffraction pattern can be indexed by $C2$-$Sn_{12}H_{18}$ (= $Sn_2H_3$, $R_1$ = 5.83%, Supporting Figure S4) with an experimental unit cell volume of 15.85 Å$^3$/Sn at 192 GPa. This phase was found during the USPEX crystal structure search. Obtained $Sn_{12}H_{18}$ consists of [$SnH_4$] structural blocks with $d_{Sn\text{-}H}$ = 1.87-1.97 Å (at 200 GPa) and multiple Sn-H-Sn hydrogen bonds.

Summarizing the structural studies of tin hydrides at high pressures, we can conclude that the Sn-H system is rich in various compounds. Structural interpretation of most of these compounds is difficult because of the low enthalpy of formation of Sn hydrides, which does not allow reliable use of computational methods to predict thermodynamically stable phases. Obviously, the main product of the reaction of tin with hydrogen is *fcc* $SnH_4$ ($V$ = 19.6 Å$^3$/Sn at 180 GPa). Among the higher polyhydrides $C2/m$-$SnH_{14}$ ($V$ = 35.8 Å$^3$/Sn at 193 GPa) and, possibly, *fcc* $SnH_{6\pm x}$ ($V$ = 24.1 Å$^3$/Sn at 180 GPa, x = 0.5) were found. Among the lower hydrides *fcc* $SnH_{2+x}$ ($V$ = 17.0 Å$^3$/Sn at 185 GPa, x = 0.5) and $C2$-$Sn_{12}H_{18}$ ($V$ = 15.85 Å$^3$/Sn at 192 GPa) were synthesized.

### 2.3. *Transport properties of fcc* $SnH_4$

Knowing the crystal structure of new hydride compounds, it is very important to experimentally study their physical properties. Therefore, we studied the transport properties of the obtained *fcc* $SnH_4$ using a four-electrode Van der Pauw circuit sputtered on diamond anvils. At 180 GPa, a sharp drop in the sample resistance (by a factor of $10^3$) at $T_C$ = 72 K with the width of $\Delta T_C$ = 2 K was revealed (Figure 3a). It has been established that in external magnetic fields up to 16 T the value of the critical temperature $T_C$ decreases almost linearly, which is typical for superconductors (Figure 3b, c). In the external magnetic fields ($H_{ext}$) we observed significant broadening of superconducting transitions in the range of 2 - 10 K (Fig. 3d), which correlates with data of Ref. [25]. According to previously known data[31-33], the superconducting transitions of some polyhydrides practically do not broaden in relatively weak magnetic fields ($\mu_0 H_{ext} < 0.5 B_{C2}(0)$). Here, due to much lower critical field for $SnH_4$ we were able to trace the $B_{C2}$(T) dependence over almost the entire temperature range. As Figures 3b,d show, $SnH_4$ manifests pronounced broadening of the superconducting transition in external fields.

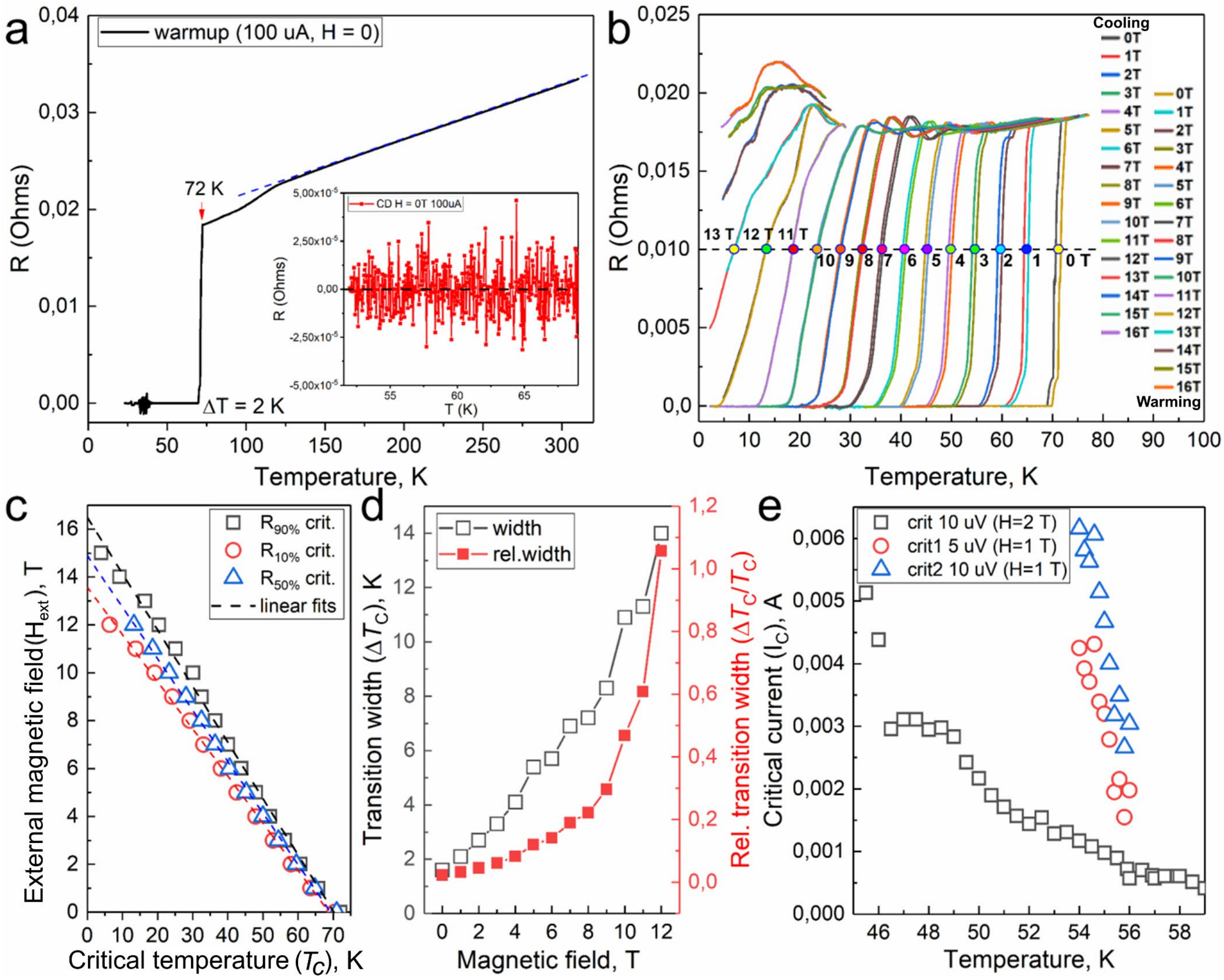

**Figure 3.** Electric transport properties of *fcc* $SnH_4$ at 180 GPa studied in magnetic fields. (a) Temperature dependence of the electrical resistance of the sample. Fully linear *R(T)* is observed from 300 K to 120 K. At about 30 K, the electrode system becomes unstable. Inset: residual resistance at $T < T_C$. (b) Displacement of the superconducting transition in external magnetic fields up to 16 T. WA - heating cycle, CD - cooling cycle. (c) Experimental temperature dependence of $T_C$ ($H_{ext}$). Using linear interpolation by different criteria (90, 50 and 10% of the normal state resistance) the values of upper critical magnetic field ($B_{C2}$) were obtained. (d) Relative and absolute broadening of superconducting transitions in external magnetic fields. (e) Temperature dependence of the critical current of the Sn hydride sample in external magnetic fields of 1 and 2 T.

Linear extrapolation of $B_{C2}(T)$ to $T_C$ = 0 K allows the upper critical magnetic field of $SnH_4$ to be estimated as $B_{C2}(0)$ = 14-16 T. This is a very low value compared to other hydride superconductors. For example, $YH_6$ [5] has the $B_{C2}(0)$ value of at least 110-160 T, and for $LaH_{10}$ [2] it is at least 140-160 T.

From measurements of the temperature dependence of resistance, we focus firstly on a pronounced feature (kink) around 120 K (Fig. 3a, Supporting Figures S17b, S18b) which we observed in several experiments. Hong et al. [25] also saw a similar kink in the *R(T)* dependence for $SnH_4$. Such anomalous behavior of *R(T)* was previously revealed in many hydrides, for instance, in $BaH_{12}$ [34], $H_3S$ [1] and $LaH_{10}$ [2,35]. Additional examples are given in Supporting Figures S20-S21. Secondly, resistance is completely linearly dependent on temperature ($\beta = R_0^{-1} dR/dT = 1.77 \times 10^{-3}$ $K^{-1}$) in the range from 120 to 320 K (see Fig.3a). Because of these two anomalies, the resistance of $SnH_4$ in the normal (non-superconductive) state can hardly be fitted with the Bloch-Grüneisen model [36] of phonon-mediated scattering. Indeed, an attempt to use the Bloch-Grüneisen fit [36] leads to a very low Debye temperature ($\theta_D$) of about 100 K, which is not typical for polyhydrides under high pressure.

The linear dependence of the upper critical field on temperature requires even more detailed discussion. For superconductors described by the Bardeen-Cooper-Schrieffer (BCS) theory [37], the generally accepted model for the $B_{C2}$(T) dependence is the Werthamer–Helfand–Hohenberg (WHH) model, which predicts flattening of the $B_{C2}$(T) dependence at low temperatures [38]. However, for many compressed polyhydrides (e.g., $YH_4$, $LaH_{10}$ at low pressure [39]) the $B_{C2}$(T) function is almost linear down to temperatures of 1-2 K. This behavior may be explained by the presence of two superconducting gaps [40-43] (Supporting Figure S19). Indeed, for a number of polyhydrides ($Fm\bar{3}m$-$LaH_{10}$, $Fm\bar{3}m$-$YH_{10}$, $P6_3/mmc$-$YH_9$) the solution of the anisotropic Migdal-Eliashberg equations indicates the presence of two superconducting gaps [44-46]. However, this explanation is unsatisfactory, since two-gaps superconductivity is not universal in hydrides (e.g., $Im\bar{3}m$-$CaH_6$ [47]). The $B_{C2}$(T) dependences of hydrides have no ascending segments, and polyhydrides that behave in full compliance with the WHH model are currently unknown. Another explanation is related to the mesoscopic inhomogeneity of the sample, the presence of regions with different composition and hydrogen content and, consequently, with different $T_C$ and $B_{C2}$ [32,48,49]. As long as "islands" of superconductivity are still bound via the Josephson effect, one can still detect the superconductivity of the sample. The proposed explanation agrees with the appearance of an ascending feature in *R(T)*, indicating a significant increase in resistance at low temperatures (Fig. 3b), possibly, due to the complex and shunted trajectory of the electric current.

One of the distinguishing features of superconductors is the existence of a self-field critical current density $J_C(T)$, when superconductivity gets destroyed and the electrical resistance of the material becomes nonzero. The low upper critical magnetic field, as expected, leads to low values of the critical current of the sample: $I_C$ = 6 mA at 55 K in a magnetic field of 1 T, that corresponds to $J_C(0) \sim 200$ A/mm$^2$ for the sample with 1 μm thickness and 30 μm diameter (Fig. 3e, Supporting Figure S17a). The critical current measurements can be used to estimate the superconducting gap in $SnH_4$. Talantsev et al. [50] proposed the following model for s-wave superconductors

$$J_c(T) = \frac{B_{c1}(T)}{\mu_0 \lambda(T)}, \qquad (1)$$

where λ(T) is the penetration depth, and $B_{C1}$(T) – is the lower critical magnetic field. Detailed equations for this model are provided in Supporting Information, eq. S5. Surprisingly, superconductivity in tin hydride is quite easily suppressed by electric current. Fit of experimental $J_C(T)$ data (Supporting Figure S17) yields $2\Delta(0) \approx 23$ meV and $2\Delta(0)/k_B T_C = 3.83$, in a reasonable agreement with the BCS value of 3.52. At the same time, according to the Talantsev-Tallon model, the self-field critical current density in $SnH_4$ at 0 K can reach $J_C(0) = 3.1$ kA/mm$^2$.

The results of studying the magnetoresistance (MR) of $SnH_4$ at 180 GPa in pulsed magnetic fields up to 65 T revealed a curious result. The dependence of the electrical resistance $\delta R \propto \mu^2 B^2$ on magnetic field $B_{ext}$ (where $\mu$ is the mobility of carriers) [51] is quadratic only in relatively weak magnetic fields. Then, rather quickly (when $B_{ext} > B_{crit}$) this dependence becomes linear $\delta R \propto (B - B_0)$ , see Figs. 4a, c. Taken separately, each of the two effects – linear $\delta R$(T) and linear $\delta R$(B) dependences may find analogies in other fields of physics. For example, linear $\delta R$(T) occurs in strongly interacting two-dimensional systems of electrons [52] where it originates from interaction assisted impurity scattering. Linear $R$(B) is known for polycrystalline samples (Kapitza linear magnetoresistance) and originates from scattering by the grain boundaries in quantizing magnetic fields [53-55], or is found in materials with Dirac cone spectrum. However, the combination of the two effects in one and the same material leaves the most probable only one analogy - with high-$T_C$ cuprates which exhibit non-Fermi-liquid behavior in the normal state[56,57]. These so called strange metals are known to be materials between insulators and metals[58]. Following the interpretation widely used for cuprate superconductors, we can say that the combination of the anomalous T and B-linear behavior of electrical resistance and magnetoresistance, as well as the upper critical field of *fcc* $SnH_4$ allows to characterize the tin tetrahydride in the normal state as a non-Fermi-liquid strange metal.

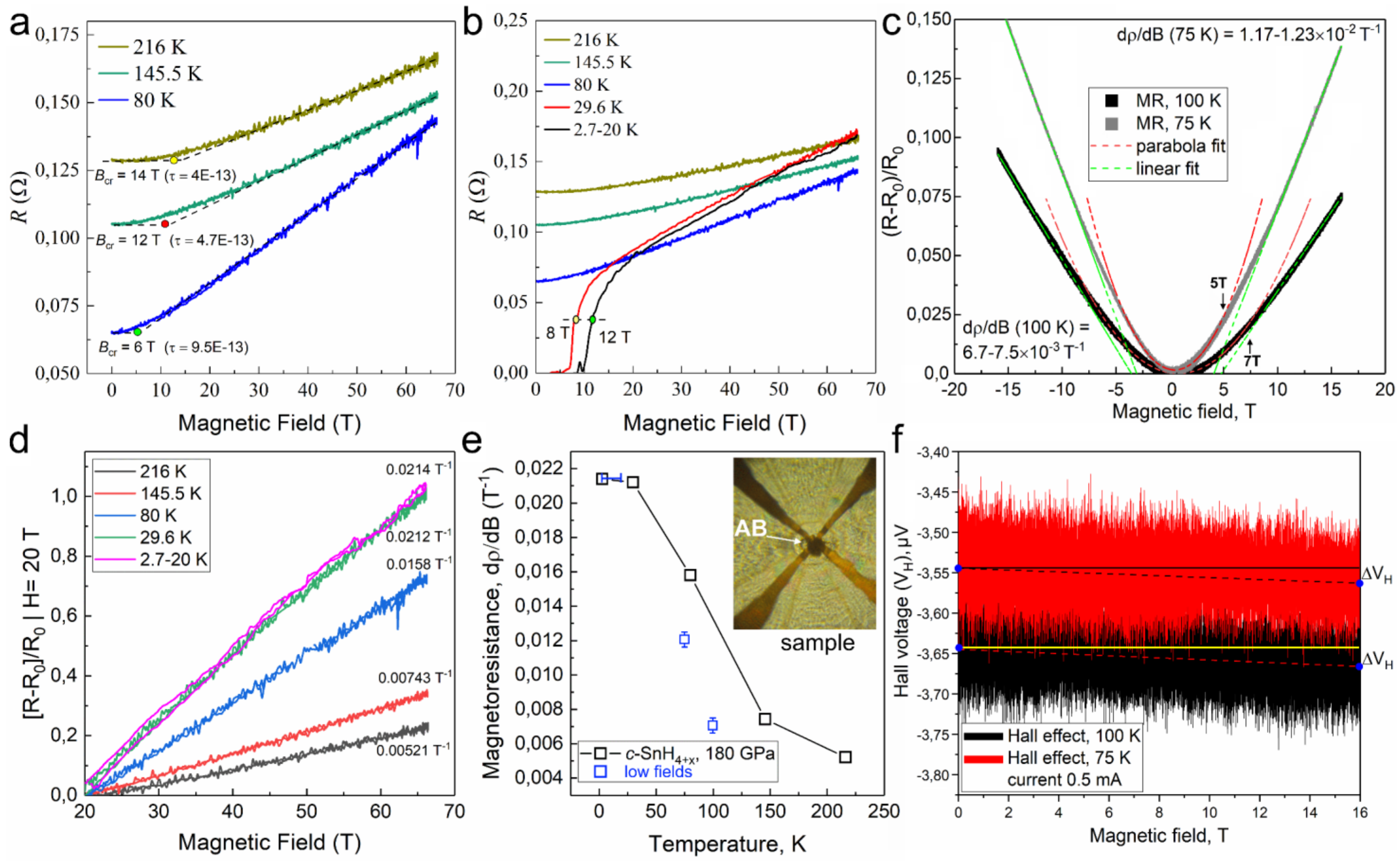

**Figure 4.** Magnetoresistance (MR) of the *fcc* $SnH_4$ at 180 GPa measured in pulse magnetic fields. (a) Dependence of the electrical resistance on the applied magnetic field at temperatures $T > T_C$. A pronounced transition from a quadratic to a linear $R(H_{ext})$ dependence is observed. (b) Dependence of the sample resistance on the applied magnetic field over the entire temperature range. Due to the heating by eddy currents during the pulse, the sample temperature of 2.7 K cannot be considered reliably fixed. (c) MR in low magnetic fields up to 16 T at temperatures 75 and 100 K. (d) Linear part of the relative MR calculated for all temperature points starting from a field $B_{ext}$ of 20 T. (e) Dependence of linear MR on temperature. (f) Attempt to measure the Hall effect in Sn hydrides at temperatures 75 and 100 K.

In general, the results of experiments in pulse magnetic fields agree with measurements in steady fields below 16 T: the MR of $SnH_4$ only in the initial part (up to 5-7 T) is a quadratic function of the field. With a further increase in the field, MR becomes completely T-linear. In addition, measurements in weak magnetic fields (0-16 T, at 75-100 K, Figs. 4c,f) allowed us only to estimate the Hall coefficient $R_H \sim 5\times10^{-12}$ m$^3$/C. The Hall voltage was $\Delta V_H \sim 1\text{-}2\times10^{-8}$ V at a current of 0.5 mA. It is impossible to give a more accurate estimate, since we do not know the exact thickness (t) of the sample (t ~ 1-2 μm). Nevertheless, this value of $R_H$ is rather small which allows us to consider the concentration of charge carriers in $SnH_4$ under pressure to be $10^{29}$-$10^{30}$ m$^{-3}$ (within $R_H = 1/n_e e$, the sign of the carrier charge is unknown). Alternatively, the low Hall coefficient may be due to high hole mobility, which is quite unusual for superhydrides under pressure.

Given the almost identical value of the Fermi velocity $V_F = 2.5\times10^5$ m/s in various hydride superconductors [59], we can estimate the mean free path of an electron in $SnH_4$ in terms of the Drude theory [60,61]:

$$l_e = V_F \tau = \frac{m_e R_H}{eRt}, \qquad (2)$$

where R is electrical resistance of the sample. It can be seen that in the $SnH_4$ the mean free path of electrons $l_e \cong 10^{-15} V_F = 3.5$ Å, which is comparable to the size of unit cell. Of course, this is a very small value corresponded to extremely "dirty" and disordered compounds. In a previous study of (La, Nd)$H_{10}$ [32] we found that the electron mean free path is also very short (~1.3 nm) and compatible with the unit cell parameter. This may lead to quantum effects of weak localization [62] even in three-dimensional samples of $SnH_4$, since the corresponding correction (δσ) to the electrical conductivity (σ) may be significant (eq. 3).

$$\frac{\delta\sigma}{\sigma} = -\frac{1}{k_F^2 l_e}\left(\frac{1}{l_e} - \frac{1}{L_\varphi}\right) \cong -1. \qquad (3)$$

Weak localization effects are well known for semiconductors and graphene [63], in particular, they are responsible for the appearance of negative magnetoresistance in thin films of Si, Ge and Te [64]. A similar phenomenon was recently observed for (La,Nd)$H_{10}$, which also exhibits a negative magnetoresistance in the range of 200–250 K [32]. The weak localization can also lead to an unusual $R$(T) dependence and sign reversal of $dR/dT$ [65], as it was observed for lanthanum-cerium (La,Ce)$H_9$ and sulfur $H_3S$ hydrides (Supporting Figures S21c-22c).

## 3. Discussion

Let us first discuss selection of the best theoretical structural model for *fcc* $SnH_4$ in terms of the experimental superconducting properties. The two predicted structures, namely $R\bar{3}m$-$Sn_{12}H_{45}$ and $Fd\bar{3}m$-$Sn_8H_{30}$, are dynamically stable, well describe the experimental powder XRD patterns, and lie near the convex hull of Sn-H system (Supporting Figure S1). They both have electronic band structures (Supporting Figure S11) typical of metals. The low density of electronic states at the Fermi level (≈ 0.5 states/eV/Sn) also has a relatively weak contribution from the hydrogen atoms (half the contribution of Sn), which is usually attributed to low critical temperature of superconductivity [66,67]. The calculated critical temperature of superconductivity for $Fd\bar{3}m$ modification ($T_C$ = 73-91 K, Supporting Figure S22) is close to the experimental one (72-74 K) observed in *fcc* $SnH_4$ at 180 GPa. The results of single-crystal XRD, which do not reveal any deviations from the ideal *fcc* structure, as well as high critical temperature of superconductivity, indicate that only $Fd\bar{3}m$-$Sn_8H_{30}$ theoretical model agrees well with the experimental data for *fcc* $SnH_4$ at 180 GPa.

It is now believed [1] that hydrides belong to the conventional BCS superconductors and their behavior in the normal state can be described in terms of the Fermi liquid model, widely used for metals. However, gradually accumulating experimental data cast doubt on this point of view. First of all, it is necessary to note the change in the sign of the temperature coefficient of electrical resistance (*dR/dT*) from positive to negative along with decreasing pressure in DACs. This was observed in superconducting $H_3S$ [1] and lanthanum-cerium polyhydrides [68] (Supporting Figure S20). The decrease in electrical resistance with increasing temperature is unusual for metals, but is a distinctive feature of insulators. Moreover, a similar behavior was recently observed in a cuprate superconductor Bi-2212 ($Bi_2Sr_2CaCu_2O_{8+}$), studied at high pressures [69]. This experiment demonstrated that pressure ($P$) may play the same role as the degree of doping (x) [70,71], separating the superconducting phase from the pseudo-gap phase with an insulating-like negative *dR/dT*.

As many cuprates have a dome-shaped dependence of the critical $T_C$(x) on the degree of doping (x), so do most superhydrides have a similar $T_C(P)$ dependence with a sharp decrease in $T_C$ with pressure decreasing[1,2,6,7]. Such decompression is usually accompanied by a sharp increase in the EPC parameter λ to the maximum within the Migdal-Eliashberg theory [72,73] values of 3 – 3.7 [74] followed by decomposition of a compound. There is such a strong electron-phonon interaction that even a small shift of atoms causes a large change in the electronic structure, turning the metal into an insulator with local separation of charges (as in ferroelectrics) and formation of polarons [75]. What can happen to the atomic sublattice of hydrogen in superhydrides under these conditions? It would be logical to assume that hydrogen atoms will form more stable $H_2$ molecules, and the compound will begin to exhibit the properties of a semiconductor, which is typical for molecular hydrides. That is exactly the boundary between the superconducting and pseudo-gap phases in polyhydrides.

The behavioral peculiarities of the electrical resistance of superhydrides have not received much attention since discovery of $H_3S$, but many of them (e.g., Supporting Figures S20-S21) show *R(T)* typical of strange metals in one pressure range, and the usual behavior of a Fermi liquid in another one. Our observation that the studied $SnH_4$ changes from metal to insulator as pressure or magnetic field changes clearly points that this material is in the proximity to a quantum critical point. Amazingly, but polyhydrides may turn out to be much closer to cuprate superconductors than previously thought, as Talantsev et al. [76-79] predicted based on an analysis of the $T_C/T_F$ ratio ($T_F$ is the Fermi temperature).

## 4. Conclusions

We have synthesized a series of novel tin hydrides at pressure of 180-210 GPa: $C2/m$-$SnH_{14}$, *fcc* $SnH_4$, *fcc* $SnH_{6+x}$, *fcc* $SnH_2$ and $C2$-$Sn_{12}H_{18}$. The main product of the reaction of Sn with hydrogen at these pressures is the cubic (*fcc*) superconducting tetrahydride $SnH_4$, with the composition being confirmed by XRD analysis performed after compression of cryogenically loaded stannane. The critical superconducting temperature of this compound is $T_C$ = 72 K at 180 GPa, the upper critical field is $B_{C2}(0)$ = 14-16 T, the self-field critical current density is $J_C(0) \sim 3.1$ kA/mm$^2$, and the superconducting gap $2\Delta(0)$ is 23 meV. $SnH_4$ contains covalent Sn-H bonds, has soft modes in phonon spectra and the large electron-phonon coupling parameter $\lambda \approx 2.5$ due to important contribution of anharmonicity to the crystal lattice dynamics.

Another discovered polyhydride $C2/m$-$SnH_{14}$ with ultra-high hydrogen content, belongs to the same family of compounds as $BaH_{12}$ [34], $LaH_{11-12}$ [80], $SrH_{22}$ [30], and confirms the existence and prevalence of such a type of high molecular polyhydrides. Unlike other members of this group, $SnH_{14}$ is expected to be a high-temperature superconductor with $T_C$ = 93 – 112 K.

Tin tetrahydride ($SnH_4$) exhibits a reproducible anomalous linear dependence of electrical resistance on temperature, T-linear dependence of the upper critical field, and a H-linear dependence of magnetoresistance over a wide range of magnetic fields and temperatures. Such a non-Fermi-liquid behavior is reminiscent to the unconventional normal state of cuprate high-temperature superconductors.

## Supporting Information

Supporting Information is available from the authors.

## Author contributions

I.A.T., D.V.S., and A.G.I. contributed equally to this work. I.A.T, D.V.S., D.Z., T.H. and S.W.T. performed the experiments. I.A.K. performed the T-USPEX and anharmonic phonon density of states calculations. A.G.K., D.V.S. and D.Z. prepared the theoretical analysis and calculated the equation of states, electron and phonon band structures, and superconducting properties of tin hydrides. D.V.S. and D.Z. analyzed and interpreted the experimental results and wrote the manuscript. I.A.T., A.V.S., and O.A.S. made the electric transport measurements in low magnetic fields. M.V.L. analyzed Raman experiments and edited the manuscript. D.S.P. built a system for the synthesis and cryogenic loading of stannane. T.H. and S.W.T. carried out the measurements in pulsed magnetic fields at HZDR HLD. I.S.L. and V.M.P. directed the research, analyzed the results and edited the manuscript. All the authors provided critical feedback and helped shape the research.

## Acknowledgements

In situ X-ray diffraction experiments at high pressure were performed on ESRF, stations ID 18 and ID 27, Grenoble, France (proposal No. HC-2902, 2017), as well as on PETRA III, beamline P02.2, Hamburg, Germany (2020), and at HPCAT (sector 16) of the APS, Argonne National Laboratory. HPCAT operations are supported by U.S. Department of Energy (DOE) National Nuclear Security Administration's Office of Experimental Sciences. The APS is a DOE Office of Science User Facility operated for the DOE Office of Science by Argonne National Laboratory under Contract No. DE-AC02-06CH11357. Experiments in pulsed magnetic fields were done in Dresden High Magnetic Field Laboratory (HLD HZDR), Dresden, Germany (2021). This work was supported by HLD-HZDR, member of the European Magnetic Field Laboratory (EMFL). The high-pressure experiments were supported by the Ministry of Science and Higher Education of the Russian Federation within the state assignment of the FSRC Crystallography and Photonics of the RAS and by the Russian Science Foundation (Project No. 22-12-00163). I.A.K. thanks the Russian Science Foundation (grant No. 21-73-10261) for the financial support of the anharmonic phonon density of states calculations and molecular dynamics simulations. SEM, XRF, and XRD studies of the initial alloys were performed using the equipment of the Shared Research Center FSRC Crystallography and Photonics of the RAS. I.A.T. and A.G.I. acknowledge the use of the facilities of the Center for Collective Use "Accelerator Center for Neutron Research of the Structure of Substance and Nuclear Medicine" of the INR RAS for high-pressure cell preparation. The research used resources of the LPI Shared Facility Center. V.M.P, A.V.S. and O.A.S. acknowledge the support of the state assignment of the Ministry of Science and Higher Education of the Russian Federation (Project No. 0023-2019-0005) and the Russian Science Foundation, grant 22-22-00570. I.A.K. thanks the Russian Science Foundation (grant No. 19-73-00237) for the financial support of the development of T-USPEX method and anharmonic phonon density of states calculation algorithm. S.W.T was supported by NSF Cooperative Agreement No. DMR-1157490/1644779 and by the State of Florida. We acknowledge the support of the HLD at HZDR, member of the European Magnetic Field Laboratory (EMFL). We also thank Dr. E. Talantsev (IMP RAS) for calculations within two-gap models, and Prof. Artem R. Oganov for help in using the USPEX code.

## Conflict of Interest

The authors declare no conflict of interest.

## Data availability

The raw and processed data required to reproduce these findings are available to download from GitHub: https://github.com/mark6871/SnHx-Tin-hydrides/, and as Supporting Information for the manuscript.

SUPPORTING INFORMATION

# Non-Fermi-Liquid Behavior of Superconducting $SnH_4$

Ivan A. Troyan,[1] Dmitrii V. Semenok,[2,*] Anna G. Ivanova,[1] Andrey V. Sadakov,[3] Di Zhou,[2,*] Alexander G. Kvashnin,[4] Ivan A. Kruglov,[5,6] Oleg A. Sobolevskiy,[3] Marianna V. Lyubutina,[1] Dmitry S. Perekalin,[7] Toni Helm,[8] Stanley W. Tozer,[9] Maxim Bykov,[10] Alexander F. Goncharov,[11] Vladimir M. Pudalov,[3,12] and Igor S. Lyubutin[1,*]

[1] Shubnikov Institute of Crystallography, Federal Scientific Research Center Crystallography and Photonics, Russian Academy of Sciences, 59 Leninsky Prospekt, Moscow 119333, Russia

[2] Center for High Pressure Science and Technology Advanced Research (HPSTAR), Beijing 100094, China

[3] P. N. Lebedev Physical Institute, Russian Academy of Sciences, Moscow 119991, Russia

[4] Skolkovo Institute of Science and Technology, Skolkovo Innovation Center, Bolshoi Blv. 30, Building 1, Moscow 121205, Russia

[5] Dukhov Research Institute of Automatics (VNIIA), Moscow 127055, Russia

[6] Moscow Institute of Physics and Technology, 9 Institutsky Lane, Dolgoprudny 141700, Russia

[7] A.N. Nesmeyanov Institute of Organoelement Compounds, Russian Academy of Sciences, 28 Vavilova str., Moscow 119334, Russia

[8] Hochfeld-Magnetlabor Dresden (HLD-EMFL) and Würzburg-Dresden Cluster of Excellence, Helmholtz-Zentrum Dresden-Rossendorf (HZDR), Dresden 01328, Germany

[9] National High Magnetic Field Laboratory, Florida State University, Tallahassee, Florida 32310, USA

[10] Institute of Inorganic Chemistry, University of Cologne, Cologne 50939, Germany

[11] Earth and Planets Laboratory, Carnegie Institution for Science, 5241 Broad Branch Road NW, Washington, D.C. 20015, United States

[12] National Research University Higher School of Economics, Moscow 101000, Russia

**Corresponding Authors:** Di Zhou (di.zhou@hpstar.ac.cn), Dmitrii Semenok (dmitrii.semenok@hpstar.ac.cn), Igor S. Lyubutin (lyubutinig@mail.ru)

# Contents

# 1. Methods

## *1.1. Theory*

The computational predictions of thermodynamic stability of the Sn–H phases at high pressures of 180 and 200 GPa were carried out using the variable-composition evolutionary algorithm USPEX [1-3]. The first generation, consisting of 80 structures, was produced using the random symmetric [3] and random topology [4] generators, whereas all subsequent generations contained 20% of random structures and 80% of structures created using heredity, softmutation, and transmutation operators.

The evolutionary searches were combined with structure relaxations using the density functional theory (DFT) [5,6] within the Perdew–Burke–Ernzerhof (PBE) generalized gradient approximation (GGA) functional [7] and the projector augmented wave method [8,9] as implemented in the VASP code [10-12]. The kinetic energy cutoff for plane waves was 600 eV. The Brillouin zone was sampled using $\Gamma$-centered $k$-points meshes with a resolution of $2\pi \times 0.05$ Å$^{-1}$. The same parameters were used to calculate the equations of state of the discovered phases. We also calculated the phonon densities of states of the studied materials using the finite displacements method (VASP and PHONOPY [13,14]). This methodology is similar to that used in our previous works [15,16].

The calculations of the critical temperature of superconductivity $T_C$ were carried out using Quantum ESPRESSO (QE) package [17,18]. We used the tetrahedron method with an offset q-point grid in order to avoid the singularity at $q = \Gamma$. The phonon frequencies and electron–phonon coupling (EPC) coefficients were computed using the density functional perturbation theory [19], employing the plane-wave pseudopotential method and the PBE exchange–correlation functional. The critical temperature ($T_C$) was calculated using the Allen–Dynes formula [20].

The dynamic stability and phonon density of states of $Sn_8H_{30}$, $Sn_{12}H_{45}$ and $SnH_{14}$ were studied using classical molecular dynamics and interatomic potentials (MTP) based on machine learning [21]. It was demonstrated that the MTP can be used to calculate the phonon properties of materials [22]. Moreover, within this approach we can explicitly take into account the anharmonicity of hydrogen vibrations.

To train the potential, we first simulated two models of cubic $SnH_4$, $Sn_8H_{30}$ and $Sn_{12}H_{45}$, as well as $SnH_{14}$ in quantum molecular dynamics in an NPT ensemble at a temperature of 2000 K, with a duration of 5 picoseconds using the VASP code [10-12]. We used the PAW PBE pseudopotentials for the H and Sn atoms, $2\pi \times 0.06$ Å$^{-1}$ $k$-mesh with a cutoff energy of 400 eV. For training of the MTP, a set of structures of tin hydrides was chosen using active learning [23]. We checked the dynamic stability of $Sn_8H_{30}$, $Sn_{12}H_{45}$ and $SnH_{14}$ using the obtained MTPs via several runs of molecular dynamics calculations at 300 K. First, the NPT dynamics simulations were performed in a supercell with about 1000 atoms for 40 picoseconds. During the last 20 picoseconds, the cell parameters were averaged. At the second step, the coordinates of the atoms were averaged within the NVT dynamics with a duration of 20 picoseconds and the final structure was symmetrized.

Then, the phonon density of states (DOS) was calculated within the MTP using the velocity autocorrelator (VACF) separately for each type of atoms [24]:

$$g(\vartheta) = 4\int_0^{\infty} \cos(2\pi\vartheta t)\frac{\langle\overline{\vartheta(0)\vartheta(t)}\rangle}{\langle\overline{\vartheta(0)^2}\rangle}dt, \quad \text{(S1)}$$

where $\vartheta$ is the frequency. The velocity autocorrelator was calculated using molecular dynamics, then the phonon DOS was obtained.

The critical temperature of superconductivity ($T_C$) was estimated using the Allen-Dynes formula

$$T_c = \omega_{log}\frac{f_1 f_2}{1.2}\exp\left(\frac{-1.04(1+\lambda)}{\lambda - \mu^* - 0.62\lambda\mu^*}\right), \quad \text{(S2)}$$

where product of Allen-Dynes coefficients $f_1$ and $f_2$ is

$$f_1 f_2 = \sqrt[3]{1 + \left(\frac{\lambda}{2.46(1 + 3.8\mu^*)}\right)^{\frac{3}{2}}} \cdot \left(1 - \frac{\lambda^2(1 - \omega_2/\omega_{log})}{\lambda^2 + 3.312(1 + 6.3\mu^*)^2}\right). \tag{S3}$$

Polyhydrides at high pressure can be considered as binary alloys of metallic hydrogen and other elements, for example, metals. As demonstrated by J. Appel [25] for binary disordered alloys, the electron-phonon interaction coefficient in this case can be expressed as

$$\lambda = -\sum_{\alpha,\beta} \frac{N_\alpha(0) N_\beta(0)}{N(0)} \cdot \frac{\langle V \rangle_{\alpha\beta}}{p_\alpha p_\beta}, \tag{S4}$$

where $N_\alpha(0)$ and $N_\beta(0)$ are the partial contributions of metal (α) and hydrogen (β) atoms to the density of electronic states at the Fermi level (DOS). $p_\alpha$ and $p_\beta$ are the probabilities of occupying certain positions by these atoms, and <V> are the averaged matrix elements of the electron-phonon interaction.

It should be taken into account that the strength of the electron-phonon interaction is the main driver of the critical temperature of superconductivity. Considering that $N_\alpha(0) = x\, N(0)$, $N_\beta(0) = (1- x)\, N(0)$, we can conclude that the highest λ achieved at x = 0.5 as maximum of x(1-x) function. This corresponds to results of many ab initio calculations: the contribution of hydrogen to the density of electronic states of a superhydride should be approximately 50% of the total density of states. If the contribution of hydrogen to $N(0)$ is small (x << 0.5), we come to superconductivity in metals, which even at high pressure does not exceed $T_C$ ~ 30 K. If the contribution of hydrogen is large (x ~ 1) – this is the case of molecular polyhydrides (such as $BaH_{12}$, $SrH_{22}$), in which isolated hydrogen molecules weakly interact with each other.

There are several techniques to probe order parameter in superconductors. Some of them are based on tunnel effects – Josephson effect and Andreev reflection effects, while others are based on the investigating of the temperature dependence of the superfluid density ($\rho_s$). The latter proved to be a reliable tool to determine not only the quantity and values of energy gaps, but also the symmetry of order parameter – whether it has s-wave, d-wave or even more complex structure [26]. Temperature dependence $\rho_s(T)$ is usually deduced by means of either measuring directly the London penetration depth λ(T), or measuring of lower critical field, or measuring self-field critical current density $J_C(T)$. Given all the limits and difficulties of DAC measurements we were only able to measure the self-field critical current. According to Ref. [26,27] the temperature dependence of $J_C(T)$ is related to the penetration depth λ(T) by formulas

$$J_c(T) = \frac{\hbar}{4e\mu_0\lambda^3(T)}(\ln(\kappa) + 0.5) \cdot \left(\frac{\lambda(T)}{a}\tanh\left(\frac{a}{\lambda(T)}\right) + \frac{\lambda(T)}{b}\tanh\left(\frac{b}{\lambda(T)}\right)\right),$$

$$\frac{\lambda(T)}{\lambda(0)} = \sqrt{1 - \frac{1}{2k_BT}\int_0^\infty cosh^{-2}\left(\frac{\sqrt{\varepsilon^2 + \Delta^2(T)}}{2k_BT}\right)}, \tag{S5}$$

$$\Delta(T) = \Delta(0) \cdot \tanh\left(\frac{\pi k_B T}{\Delta(0)}\sqrt{\eta\left(\frac{\Delta C}{C}\right)\left(\frac{T_C}{T} - 1\right)}\right),$$

where $2a$ – is the width of sample, $2b$ – is the thickness of sample, $\mu_0$ is the permeability of free space, e is the electron charge, κ = λ/ξ is the Ginsburg-Landau parameter, Δ(T) – the superconducting gap, η = 2/3 for s-wave superconductivity, and ΔC/C – is the specific heat capacity jump at the superconducting transition. In these equations, parameters $b$, Δ(0), λ(0) and ΔC/C are refined parameters. The best fit is shown in Supporting Figure S17d.

### *1.2. Experiment*

We used the diamond anvils with a 50 µm culet beveled to 300 µm at 8.5°, equipped with four ~200 nm thick Ta electrodes with ~80 nm gold plating that were sputtered onto the piston diamond. Rhenium or steel ring composite gaskets were used for electrical measurements, and a mixture of $CaF_2$/epoxy resin was used to insulate the gaskets from electrical leads. Sn pieces with a thickness of ~1–2 µm were sandwiched between the electrodes and ammonia borane $NH_3BH_3$ (AB) layer. Ammonia borane served as a pressure-transmitting medium and simultaneously was a source of hydrogen at the laser heating of the sample. The laser heating of the samples above 1500 K at pressures of 170–180 GPa was done by several 100 µs pulses and led to the formation of various Sn hydrides, whose structure was analyzed using the single-crystal (SCXRD [28-30]) and powder X-ray diffraction (XRD). It should be noted that DACs used for XRD have a wide opening angle (60-80°), while DACs prepared for the transport studies cannot be used for XRD due to presence of Ta/Au electrodes and small opening angle.

The XRD patterns of tin hydrides were studies at the ID27, P02 and HPCAT beamlines of ESRF (Grenoble, France), PETRA III (Hamburg, Germany) and APS (Argonne, USA) respectively, using monochromatic synchrotron radiation and an imaging plate detector at room temperature. The X-ray beam was focused in a less than 3 × 3 µm spot. The XRD data were analyzed and integrated using Dioptas software package (version 0.5) [31]. The full profile analysis of the diffraction patterns and the calculation of the unit cell parameters were performed using JANA2006 software [32] with the Le Bail method [33]. The pressure in the DACs was determined via the Raman signal from diamond at room temperature [34]. We should note that pressure in all DACs changes during cooling from 300 K to 4 K by about 5-10 GPa.

Single crystal XRD was performed at the beamline 13 IDD (GSECARS, Advanced Photon Source, Argonne, USA). For the single-crystal XRD measurements samples were rotated around a vertical ω-axis in a range ±32°. The diffraction images were collected with an angular step $\Delta\omega = 0.5°$ and an exposure time of 5-10 s/frame. For analysis of the single-crystal diffraction data (indexing, data integration, frame scaling and absorption correction) we used the CrysAlisPro software package, version 1.171.42.54a. To calibrate an instrumental model in the CrysAlisPro software, i.e., the sample-to-detector distance, detector's origin, offsets of goniometer angles, and rotation of both X-ray beam and the detector around the instrument axis, we used a single crystal of orthoenstatite ($(Mg_{1.93}Fe_{0.06})(Si_{1.93}, Al_{0.06})O_6$, *Pbca* space group, $a$ = 8.8117(2), $b$ = 5.18320(10), and $c$ = 18.2391(3) Å).

Magnetoresistance measurements under high magnetic fields were carried out in a 24 mm bore 72 T resistive pulse magnet (rise time of 15 ms) at the Helmholtz-Zentrum Dresden-Rossendorf (HZDR). Strands of the Litz wire glued to the silver paint were moved closer together to minimize open loop pickup. The DAC was encased with strips of 125 µm × 1 cm wide Kapton tape. All twisted pairs were fixed using the GE7031 varnish (Lake Shore, 50:50 toluene: methanol thinner). A flow He cryostat (VTI) was used, which made it possible to better control the DACs temperature. Cernox thermometers (Lake Shore Cernox X95809) were attached to the DAC body for measurements of the temperature. There was no observable heating from the ramping of the magnetic field at rates up to 100 T/sec at temperatures above 20 K. A high-frequency (3.33 and 33.3 kHz) lock-in amplifier technique was employed to measure the sample resistance in a 72 T pulsed magnet. The magnet can be used on very special occasions to 72 T, but is usually used to 65 T to extend its lifetime. For the measurements in high magnetic fields, we used a four-wire AC method with the excitation current of 0.5–1 mA [35]; the voltage drop across the sample was amplified by an instrumentation amplifier and detected by a lock-in amplifier. Comparing the up and down sweep resistance curves at various field sweep rates, no significant sample heating was observed during a ~150 ms long magnet pulse at temperatures above 35 K. In general, we used the same methodology as in the previous study of (La, Nd)$H_{10}$ [36].

**Table S1**. List of high-pressure DACs used in this experiment, samples loaded into them and products obtained.

| DAC | Experiment | Sample | Products |
|---|---|---|---|
| DAC S1 | ESRF-2017 | Sn | *hcp*-Sn, *bcc*-Sn |
| DAC S2 | ESRF-2017 | $SnH_4$, cryogenically loaded stannane | *fcc*-$SnH_4$, *hcp*-Sn |
| DAC D2 | PETRA-2020 | $Sn/NH_3BH_3$ | *bcc*-Sn, *hcp*-Sn, *fcc*-$SnH_4$, $C2/m$-$SnH_{14}$ |
| DAC M2 | PETRA-2020 | $Sn/NH_3BH_3$ | $C2/m$-$SnH_{14}$, *fcc*-$SnH_{6-8}$, *fcc*-$SnH_4$, *fcc*-$SnH_2$ |
| DAC S3 | APS-2022 | $Sn/NH_3BH_3$ | $C2$-$Sn_{12}H_{18}$, *fcc*-$SnH_4$ |

## 2. Crystal structure search

It should be noted that the formation of tin polyhydrides at high pressures is a process with just a little thermodynamic benefit. This is evidenced by the values of the enthalpy of formation of Sn hydrides, which are ΔH ~ 10–60 meV/atom [37], whereas, those of other metal polyhydrides ΔH ($X + mH_2 \rightarrow XH_{2m}$) are of the order of hundreds meV/atom [37]. In addition, the low enthalpy of formation makes the chemical process more dependent on hyperfine effects such as the zero-point energy (ZPE), entropy factor, spin-orbit coupling (SOC), anharmonicity of the hydrogen sublattice, as well as the choice of pseudopotentials and exchange-correlation functional for DFT calculations. As a result, the real convex hull of the Sn-H system is difficult to predict from the first principles. This explains the failure of all previous theoretical studies of tin polyhydride formation at high pressure [37,38].

To verify that common DFT approaches are poorly suited to predict the existence of stable tin hydrides, we performed a search for the most appropriate crystal structure for $SnH_4$ composition using USPEX code [1-3,39] at 180 and 200 GPa (Supporting Figure S1). As a result, the *P*6/*mmm*, *P*6$_3$/*mmc* and *I*4/*mmm* phases of $SnH_4$ were identified as the most thermodynamically favorable. However, no cubic tin tetrahydrides were found. Moreover, none of the theoretically proposed Sn polyhydrides can explain the diffraction pattern of the material obtained by compressing $SnH_4$ and indexed as a cubic *fcc* lattice (experiment ESRF-2017). However, if we assume that the stoichiometric ratio Sn:H is not 1:4, but 1:3.75 ($Th_4H_{15}$-like structure), which may be the result of a decomposition reaction,

$$8\ SnH_4 \rightarrow Sn_8H_{30} + H_2, \qquad (S6)$$

then we can find a good approximation to the structural solution of the problem. The stoichiometry of $Sn_4H_{15}$ allows for dynamically and thermodynamically stable cubic (*fcc*) metal lattices. The results of an evolutionary structural search among such compositions (up to supercells of $Sn_{32}H_{120}$) at 180 GPa are: $R\bar{3}m$-$Sn_{12}H_{45}$ (distorted $Fd\bar{3}m$) and $Fd\bar{3}m$-$Sn_8H_{30}$ (Supporting Table S2, Figure 1c in the main text). Both structures are metals with unit cell volumes of about 19.7 Å$^3$/atom at 180 GPa. They are dynamically stable if one considers anharmonic effects. Both of these phases lie close to the convex hull of Sn-H system and exhibit *fcc*-like XRD patterns. Generally speaking, the most stable tin hydride at pressures of 170-210 GPa is *P*6$_3$/*mmc*-$SnH_4$ (with and without SOC, ZPE and entropy contribution, Supporting Tables S3, S4). But since such a compound was not observed among the experimental diffraction patterns, we excluded this phase from the consideration. Then, the most thermodynamically stable phase (with accounting ZPE) is $Fd\bar{3}m$-$Sn_8H_{30}$ at 180 GPa, but $R\bar{3}m$-$Sn_{12}H_{45}$ at 200 GPa and 0 K. At 1000 K, the situation does not change: the cubic modification of $Fd\bar{3}m$-$Sn_8H_{30}$ is still slightly above the convex hull at 200 GPa (Supporting Table S4). As shown below, the calculated parameters of the superconducting state of $Fd\bar{3}m$-$Sn_8H_{30}$ are close to the experimental data.

**Table S2.** Calculated convex hull parameters of Sn-H system at 180 GPa and 0 K. $E_{total}$ is the total DFT energy, $E_{form}$ is the energy of formation, X = H/(Sn+H).

| Phase | Number of Sn atoms | Number of H atoms | X | $E_{total}$, eV | $E_{form}$, eV/atom |
|---|---|---|---|---|---|
| *P*6$_3$/*mmc*-Sn | 2 | 0 | 0 | 30.982 | 0 |
| $Fd\bar{3}m$- $Sn_8H_{30}$ | 8 | 30 | 0.789 | 116.208 | -0.043 |
| $R\bar{3}m$- $Sn_{12}H_{45}$ | 12 | 45 | 0.789 | 174.352 | -0.042 |
| $I\bar{4}m2$-$SnH_8$ | 2 | 16 | 0.889 | 27.088 | -0.036 |
| *C*2/*m*-$SnH_{12}$ | 2 | 24 | 0.923 | 25.686 | -0.016 |
| *C*2/*m*-$SnH_{14}$ | 2 | 28 | 0.933 | 25.219 | -0.002 |
| *C*2/*m*-H | 0 | 24 | 1 | -4.879 | 0 |

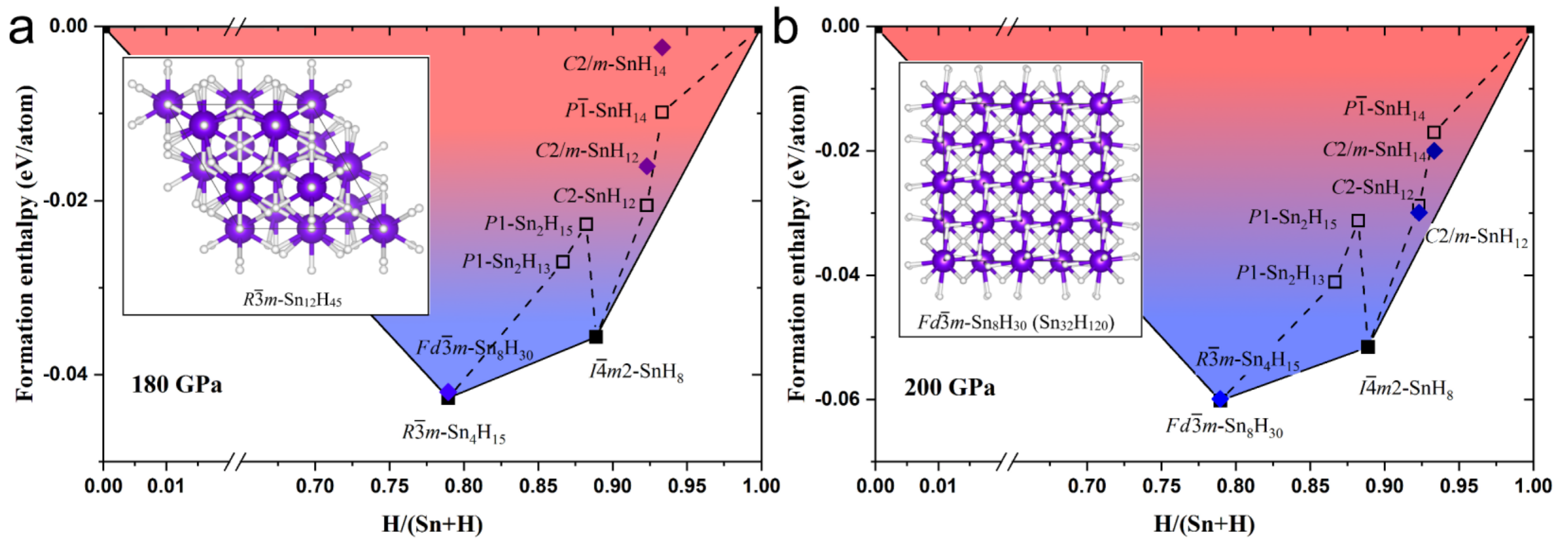

**Figure S1.** Convex hull of the Sn-H system calculated for pressures of 180 and 200 GPa and $T$ = 0 K using the USPEX code without considering the ZPE. The most stable phase $P6_3/mmc$-$SnH_4$ is excluded. The insets show the structures of $R\bar{3}m$-$Sn_{12}H_{45}$ and $Fd\bar{3}m$-$Sn_8H_{30}$ in the pressure range of 180–200 GPa. Consideration of ZPE and temperature effects has a little effect on the convex hull.

**Table S3.** Calculated convex hull parameters of Sn-H system at 200 GPa and 0 K. $H_{form}$ is the DFT enthalpy of formation. ZPE is the zero-point energy in the harmonic approximation.

| Phase | Sn atoms | H atoms | X | E(total), eV | E, eV/atom | $E_{form}$, eV/atom | ZPE, eV | E+ZPE, eV | $H_{form}$, eV/atom |
|---|---|---|---|---|---|---|---|---|---|
| $P6_3/mmc$-Sn | 2 | 0 | 0 | 34.366 | 17.183 | 0 | 0.051 | 34.417 | 0 |
| $Fd\bar{3}m$- $Sn_8H_{30}$ | 8 | 30 | 0.789 | 135.678 | 3.570 | -0.060 | 8.556 | 144.234 | -0.064 |
| $R\bar{3}m$- $Sn_{12}H_{45}$ | 12 | 45 | 0.789 | 203.533 | 3.571 | -0.060 | 12.332 | 215.865 | -0.072 |
| $P6_3/mmc$-$SnH_4$ | 2 | 8 | 0.800 | 33.450 | 3.345 | -0.105 | 2.320 | 35.770 | -0.104 |
| $I\bar{4}m2$-$SnH_8$ | 2 | 16 | 0.889 | 33.705 | 1.872 | -0.052 | 4.034 | 37.738 | -0.082 |
| $C2/m$-$SnH_{12}$ | 2 | 24 | 0.923 | 33.987 | 1.307 | -0.030 | 6.750 | 40.738 | -0.034 |
| $C2/m$-$SnH_{14}$ | 2 | 28 | 0.933 | 34.321 | 1.144 | -0.017 | 7.847 | 42.168 | -0.021 |
| $C2/m$-H | 0 | 24 | 1 | 0.400 | 0.017 | 0 | 6.794 | 7.194 | 0 |

**Table S4.** Extended convex hull parameters of Sn-H system at 200 GPa and 300-1000 K. $F_{vib}$ is the vibrational energy, $G_{form}$ is the Gibbs free energy of formation.

| Phase | $F_{vib}$, eV/atom | | | $G_{form}$, eV/atom | | |
|---|---|---|---|---|---|---|
| Temperature | 300 K | 500 K | 1000 K | 300 K | 500 K | 1000 K |
| $P6_3/mmc$-Sn | -0.164 | -0.379 | -1.076 | 0 | 0 | 0 |
| $Fd\bar{3}m$- $Sn_8H_{30}$ | -0.485 | -1.295 | -4.849 | -0.058 | -0.049 | -0.026 |
| $R\bar{3}m$- $Sn_{12}H_{45}$ | -0.712 | -1.995 | -7.648 | -0.066 | -0.059 | -0.041 |
| $P6_3/mmc$-$SnH_4$ | -0.098 | -0.284 | -1.150 | -0.097 | -0.086 | -0.059 |
| $I\bar{4}m2$-$SnH_8$ | -0.105 | -0.342 | -1.577 | -0.077 | -0.070 | -0.051 |
| $C2/m$-$SnH_{12}$ | -0.134 | -0.440 | -2.118 | -0.031 | -0.025 | -0.013 |
| $C2/m$-$SnH_{14}$ | -0.154 | -0.512 | -2.494 | -0.020 | -0.015 | -0.007 |
| $C2/m$-H | -0.038 | -0.273 | -1.577 | 0 | 0 | 0 |

Thus, in this section, we showed a vivid example of how far the results of theoretical DFT calculations can differ from the results of experiments at high pressure. Although we succeeded to find a satisfactory dynamically stable model ($Sn_8H_{30}$) to describe the properties of experimental cubic $SnH_4$, we have failed to solve the problem of thermodynamic stability of Sn polyhydrides at high pressure (Supporting Figure S1). Strictly speaking, at a given crystal structure $P6_3/mmc$-$SnH_4$, all experimentally found Sn hydrides appear to be displaced from the convex hull at 180-200 GPa and, in terms of the current level of theoretical structural search, can exist only as metastable phases. Obviously, there must be some unusual effect that stabilizes the cubic modification of $SnH_4$, but this problem is beyond the scope of this study.

## 3. X-ray diffraction and unit cell parameters of tin hydrides

Reaction products (APS-2022 experiment, Table S4) contained multiple good-quality single-crystalline domains of novel phases (Supporting Figure S2) and it was possible to find orientation matrices of several single crystals following well established procedures (see Ref. [40] for more details on the procedure).

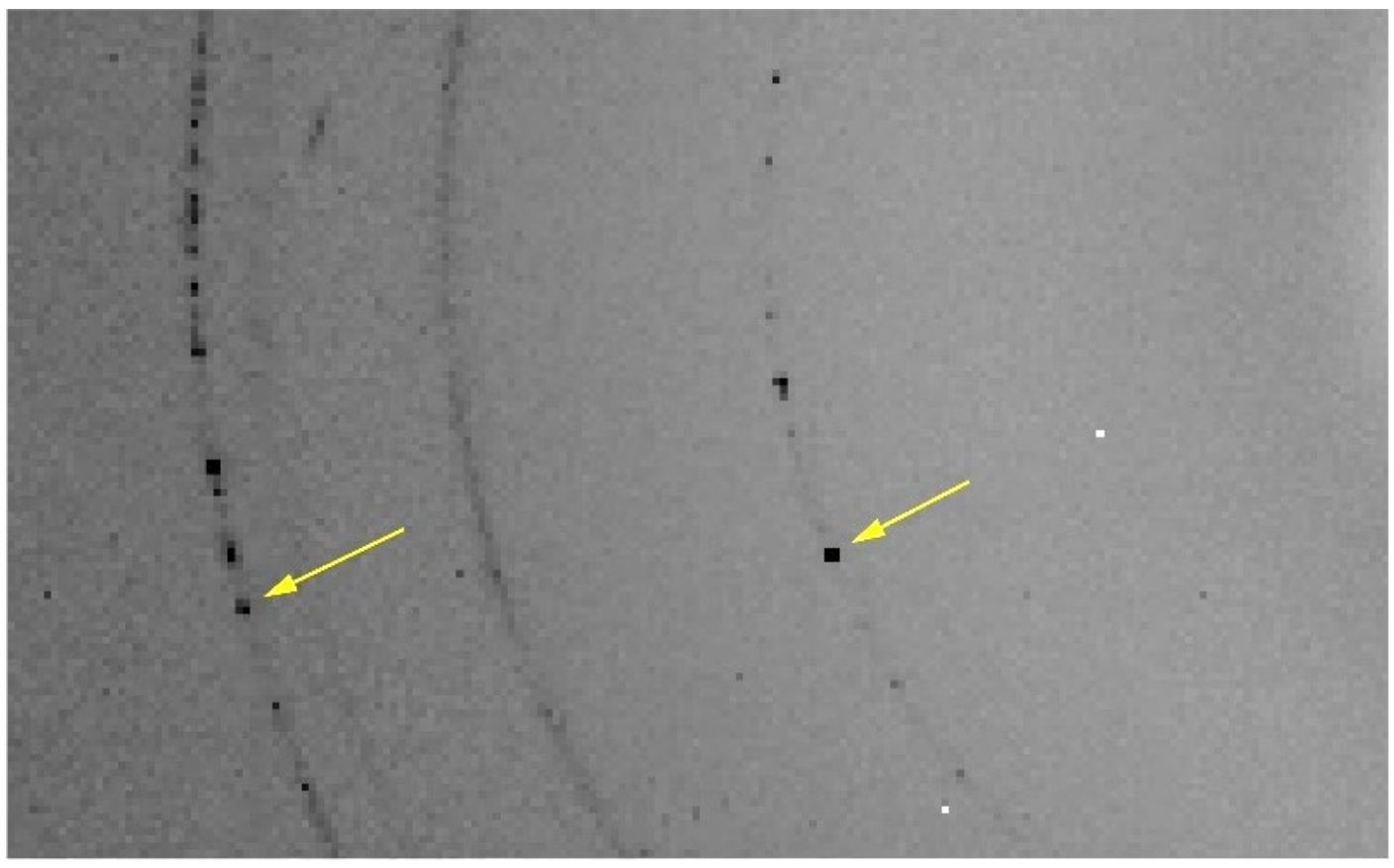

**Figure S2.** Example of the diffraction pattern showing reflections originating from small single crystals, APS-2022. These peaks are distinct and can be separated from the "bad powder" rings.

The dominant phase was initially indexed using a cubic *F*-centered lattice with $a$ = 5.780(5) Å at 171 GPa. This lattice at the first glance does not correspond to any of the theoretically predicted Sn-H phases. Nevertheless, one should take into account that XRD gives information on the positions of heavy atoms only due to the great difference in the atomic scattering factors of Sn and H. Therefore, it is not excluded that detection of weak superlattice reflections is beyond the current experimental capabilities. Unconstrained refinement of the unit cell parameters based on 42 single-crystal reflections revealed slight deviation of the cubic lattice parameters from each other: $a$ = 5.780(5) Å, $b$ = 5.796(10) Å, $c$ = 5.775(9) Å, which might indicate that the lattice is indeed distorted.

A sequence of lattice transformations defined by the following matrices allows to convert the *F*-centered lattice to a monoclinic *C*-centered lattice and refine unit cell with a better final figure of merit:

$$cF \rightarrow mI \begin{pmatrix} 0.5 & 0 & 0.5 \\ -0.5 & 0 & 0.5 \\ 0 & -1 & 0 \end{pmatrix}; \quad mI \rightarrow mC \begin{pmatrix} -1 & 0 & -1 \\ 0 & 1 & 0 \\ 1 & 0 & 0 \end{pmatrix}; \quad cF \rightarrow mC \begin{pmatrix} -0.5 & 1 & -0.5 \\ -0.5 & 0 & 0.5 \\ 0.5 & 0 & 0.5 \end{pmatrix}$$

The *C*-centered lattice obtained by the transformation of *F*-centered cubic lattice agrees well with the predicted $C2$-$Sn_{12}H_{18}$ phase with $a$ = 7.083(15) Å, $b$ = 4.082 Å, $c$ = 4.090 Å, $\beta$ = 125.1°. Structure refinement in a cubic $Fm\bar{3}m$ symmetry ($a$ = 5.780(5) Å) with Sn occupying Wyckoff sites 4*b* and 8*c* converges with an excellent agreement factor $R_1$ = 5.83%, confirming the pseudocubic arrangement of Sn atoms in the calculated $C2$-$Sn_{12}H_{18}$ structure (Table S4, Figure S3).

The second phase (*fcc* $SnH_4$) in the sample at 171 GPa was initially indexed with a cubic *F*-centered lattice with $a$ = 4.244(3) Å. Crystal structure refinement suggests *fcc* arrangement of Sn atoms with a space group $Fm\bar{3}m$ and Sn occupying Wyckoff site 4*a* ($R_1$ = 4.75 %). This lattice parameter is exactly half of that of the predicted $Fd\bar{3}m$-$Sn_8H_{30}$ phase. However, due to the low atomic scattering factor of hydrogen, the supercell could not be confirmed by single-crystal XRD.

Impurity reflections in the XRD patterns (Figure 2 in the main text) can be explained by orthorhombically distorted $Im\bar{3}m$ structure. However, as can be seen from the data of Supporting Tables S5, S7, S8, the resulting volume $V \sim$ 23.5-24.1 Å$^3$ (Z = 2) is too small for *bcc*-Sn. Better agreement with the unit cell volume was obtained for $P6_3/mmc$-Sn structure (for example, $a$ = 2.633 Å, $c$ = 4.581 Å, $V$ = 27.5 Å$^3$, Z = 2), but the refinement quality is lower. This indicates that the strong reflection at $2\theta \approx 8°$ (Figures 2a, d) may have a source other than residual tin.

In our experiment with DAC M2, in addition to $SnH_{14}$, several other tin hydrides were also detected. They can be indexed as two cubic phases: *fcc*-1 (Supporting Figure S5), and *fcc*-2 with a larger unit cell volume. The latter phase can also be indexed as pseudocubic *I*4/*mmm* (Supporting Figure S6). For the *fcc*-1 phase, indexing gives the unit cell volume $V \approx$ 16.5-17.1 Å$^3$/Sn at 180 GPa, which is approximately 3 Å$^3$/Sn less than it is necessary for the $SnH_4$ composition. Such a value of the unit cell volume better matches the composition of $SnH_2$ where Sn has a formal valence of +2. For the second cubic phase *fcc*-2, refinement gives the unit cell volume $V$ = 23.5-24.1 Å$^3$/Sn (Supporting Table S5), which corresponds to the composition $SnH_{6+x}$ (x = 0 - 0.5, see Figure 2e), or the previously predicted stable tetragonal $SnH_{8-x}$ [37].

**Table S5.** Summary table of the XRD experiments (without $SnH_{14}$ and Sn). Unit cell parameters of $Fd\bar{3}m$-$Sn_8H_{30}$ and $C2$-$Sn_{12}H_{18}$ compounds obtained in three experiments at different pressures are shown. Unit cell parameters of *fcc*-1 phase ($SnH_2$), *fcc*-2 ($SnH_6$ or $SnH_{8-x}$) and a gasket (e.g., *bcc*-Fe (?), one diffusive ring), obtained at PETRA-2020, are also shown. The experimental error in the pressure value is within 5 GPa.

| **Phase** | **$Fd\bar{3}m$-$Sn_8H_{30}$** | | | **$C2$-$Sn_{12}H_{18}$ ($\beta$ = 124.46°, fixed)** | | | | ***bcc*-Sn** | |
|---|---|---|---|---|---|---|---|---|---|
| **Experiment** | ***a*, Å (Z = 32)** | ***V*, Å$^3$ (Z = 32)** | ***V*, Å$^3$/Sn** | ***a*, Å** | ***b*, Å** | ***c*, Å** | ***V*, Å$^3$/Sn** | ***a*, Å** | ***V*, Å$^3$ (Z = 2)** |
| ESRF-2017, 180 GPa<br>PETRA-2020, 170 GPa | 8.567<br>8.619 | 628.7<br>640.5 | 19.64<br>20.0 | - | - | - | - | 2.996<br>3.012 | 26.9<br>27.32 |
| APS-2022, 192-196 GPa | 8.5075 | 615.7 | 19.24 | 7.007 | 4.095 | 4.020 | 15.85 | - | - |
| APS-2022, 165-171 GPa | 8.689 | 656.0 | 20.5 | 7.097 | 4.175 | 4.130 | 16.82 | - | - |
| APS-2022, 150-155 GPa | 8.719 | 662.8 | 21.2 | 7.087 | 4.205 | 4.100 | 16.85 | | - |
| **Phase** | ***fcc*-1 $SnH_2$** | | | ***fcc*-2, $SnH_6$ or pseudocubic *I*4/*mmm*-$SnH_{8-x}$** | | | | ***bcc* gasket (Fe ?)** | |
| PETRA-2020 | ***a*, Å (Z = 32)** | ***V*, Å$^3$ (Z = 32)** | ***V*, Å$^3$/Sn** | ***a*, Å** | ***c*, Å** | **$V_{tet}$, Å$^3$ (Z = 2)** | **$V_{cub}$, Å$^3$/Sn** | ***a*, Å** | ***V*, Å$^3$** |
| 185 GPa | 8.159<br>8.165 | 543.13<br>544.33 | 16.97<br>17.01 | 3.241 | 4.591 | 48.22 | 24.11 | 2.365 | 13.23 |
| 180-185 GPa | 8.180 | 547.34 | 17.10 | 3.229 | 4.614 | 48.11 | 24.11 | 2.365 | 13.23 |
| 190-196 GPa | 8.130<br>8.140<br>8.135 | 537.37<br>539.35<br>538.33 | 16.79<br>16.85<br>16.82 | 3.229<br>3.233 | 4.577<br>4.582 | 47.74<br>47.90 | 23.95 | 2.360 | 13.14 |
| 201-208 GPa | 8.1060 | 532.62 | 16.64 | 3.215 | 4.557 | 47.11 | 23.58 | 2.345<br>2.352 | 12.90<br>13.01 |
| 208-210 GPa | 8.090 | 529.47 | 16.54 | 3.195 | 4.561 | 46.57 | 23.47 | 2.342 | 12.87 |

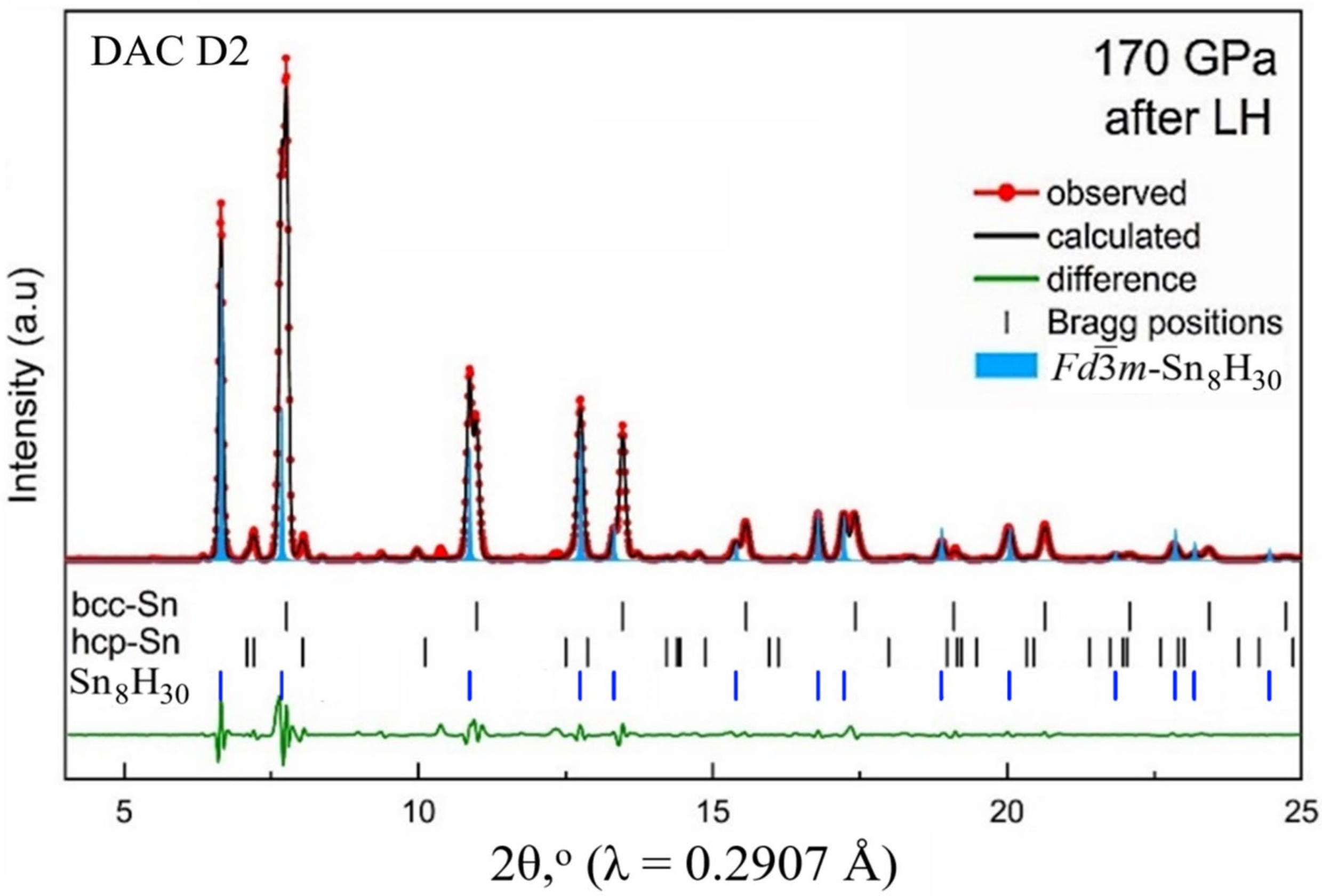

**Figure S3.** Experimental X-ray diffraction pattern of the results of laser heating of Sn/AB at a pressure of 170 GPa (DAC D2). XRD pattern was obtained at PETRA III in 2020. Cubic $SnH_4$ compound as the main product, along with unreacted Sn in hexagonal (*hcp*) and cubic (*bcc*) modifications are clearly seen. The Bragg peaks positions for the theoretical model, $Fd\bar{3}m$-$Sn_8H_{30}$, are also shown.

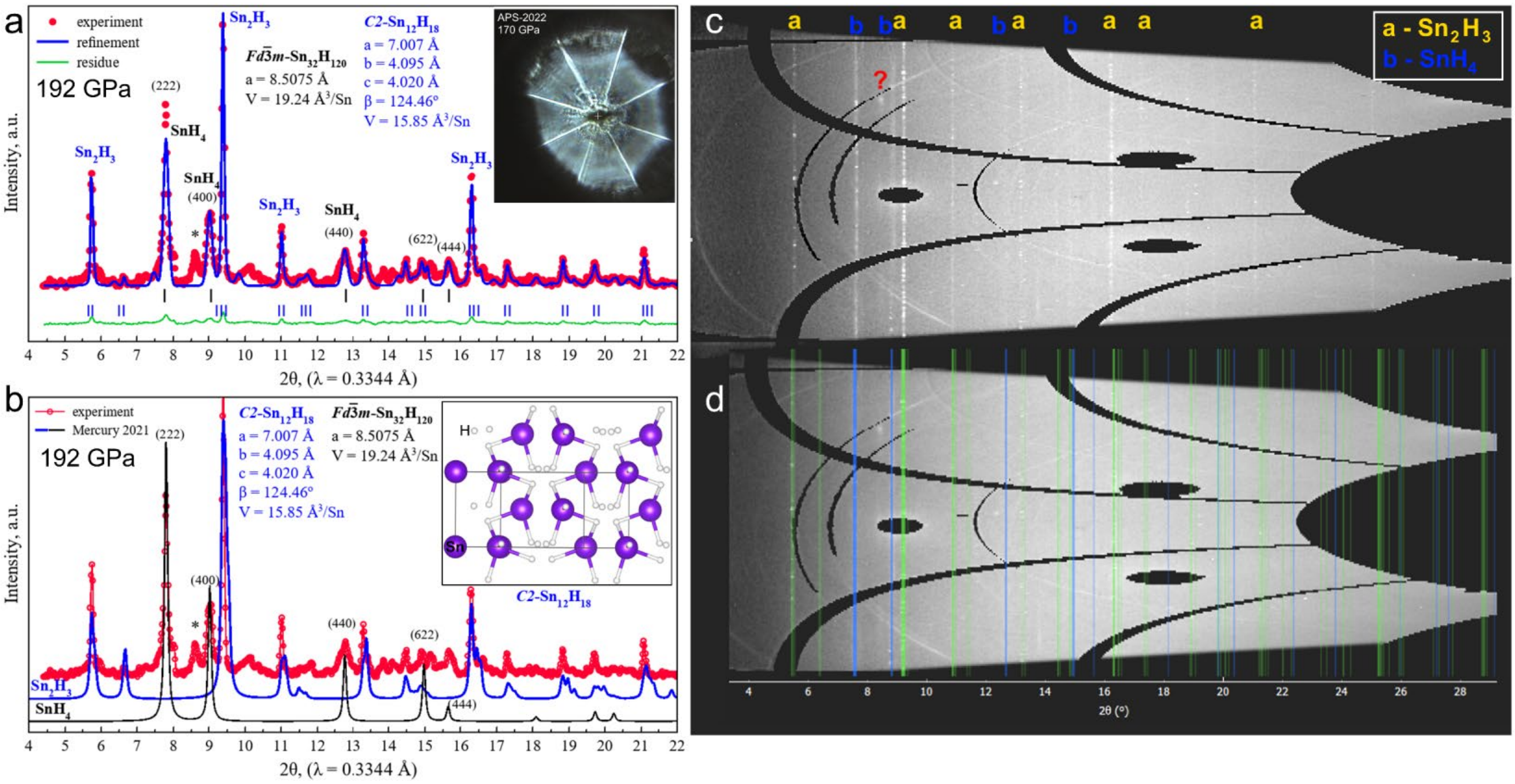

**Figure S4**. Main results of the XRD experiment performed at APS-2022. (a) The Le Bail refinement of the unit cell parameters of $C2$-$Sn_{12}H_{18}$ (= $Sn_2H_3$) and cubic $SnH_4$ (theoretical model is $Fd\bar{3}m$-$Sn_{32}H_{120}$ = $Sn_8H_{30}$) phases at a pressure of 192 GPa. Unidentified reflections are marked by asterisks. The experimental data, fitted line, and residues are shown in red, blue, and green, respectively. Inset: sample photograph at 171 GPa. (b) Comparison of experimental (red line) and calculated (blue and black lines) XRD patterns for $Sn_{12}H_{18}$ (= $Sn_2H_3$) and $SnH_4$ (theoretical model is $Fd\bar{3}m$-$Sn_{32}H_{120}$), respectively. Inset: crystal structure of $C2$-$Sn_{12}H_{18}$. (c, d) Diffraction images ("cake") where one can see two groups of diffraction rings: relatively diffuse rings (b) corresponding to the cubic $SnH_4$, which can be obtained in pure form, and the second fraction, "dotted" rings (a) corresponding to $Sn_{12}H_{18}$ (= $Sn_2H_3$).

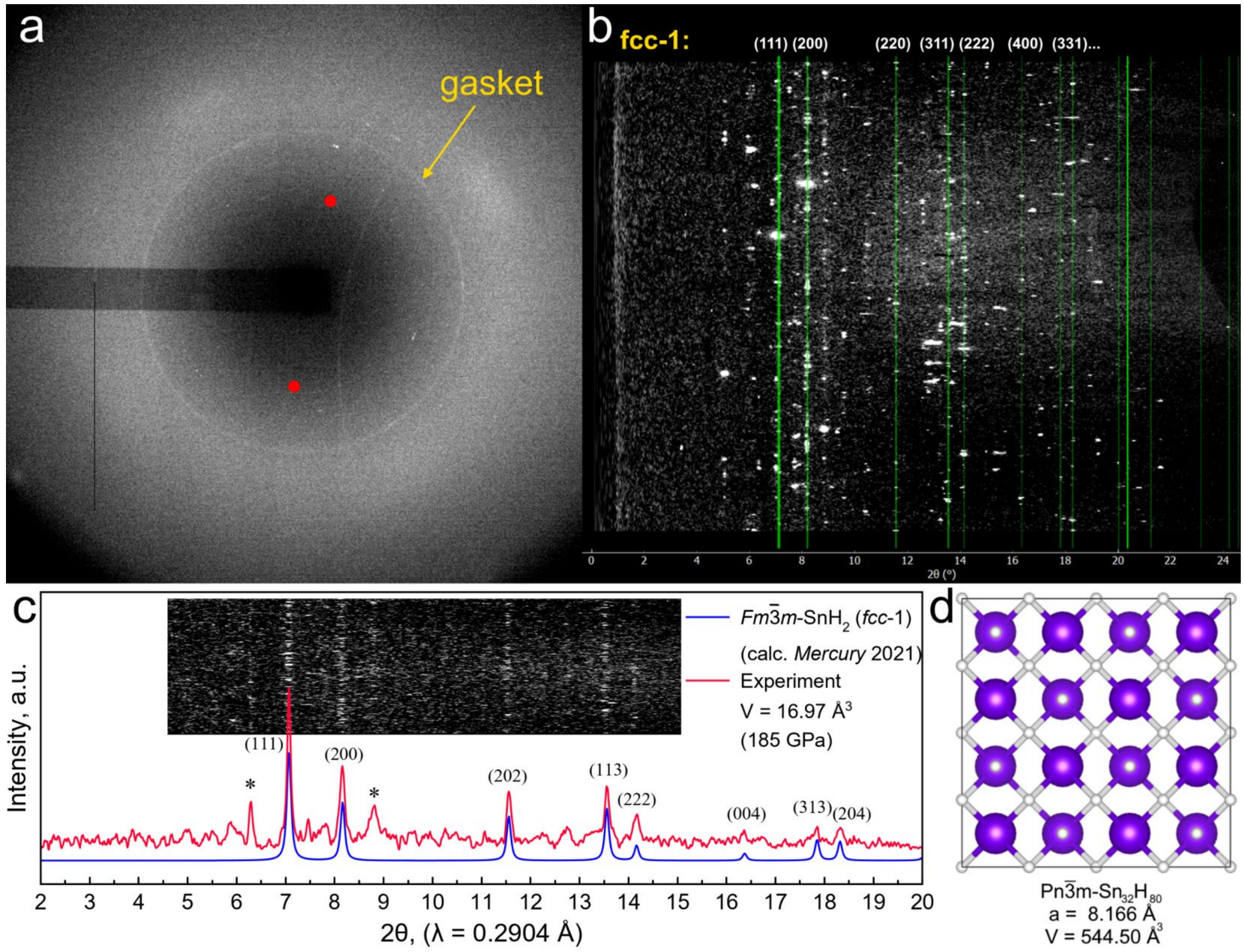

**Figure S5.** XRD patterns of hydrides synthesized in DAC M2 at 185 GPa. (a) A characteristic continuous diffraction ring from the gasket material. (b) A characteristic "dotted" pattern (coarse-crystalline) of tin hydrides. Green lines correspond to the *fcc*-1 phase ($SnH_2$). (c) X-ray diffraction pattern of the sample in DAC M2 at 185 GPa collected at a wavelength of λ = 0.2904 Å. Comparison of experimental and calculated XRD reflection intensities for the *fcc*-1 phase (refined using prototype $Fm\bar{3}m$-$YH_2$) is shown. Inset: diffraction image ("cake"). Unidentified reflections are marked by asterisks. (d) Crystal structure of one alternative candidate for the *fcc*-1 phase, $Sn_{32}H_{80}$ = $SnH_{2.5}$ which has a unit cell volume close to the experimental one.

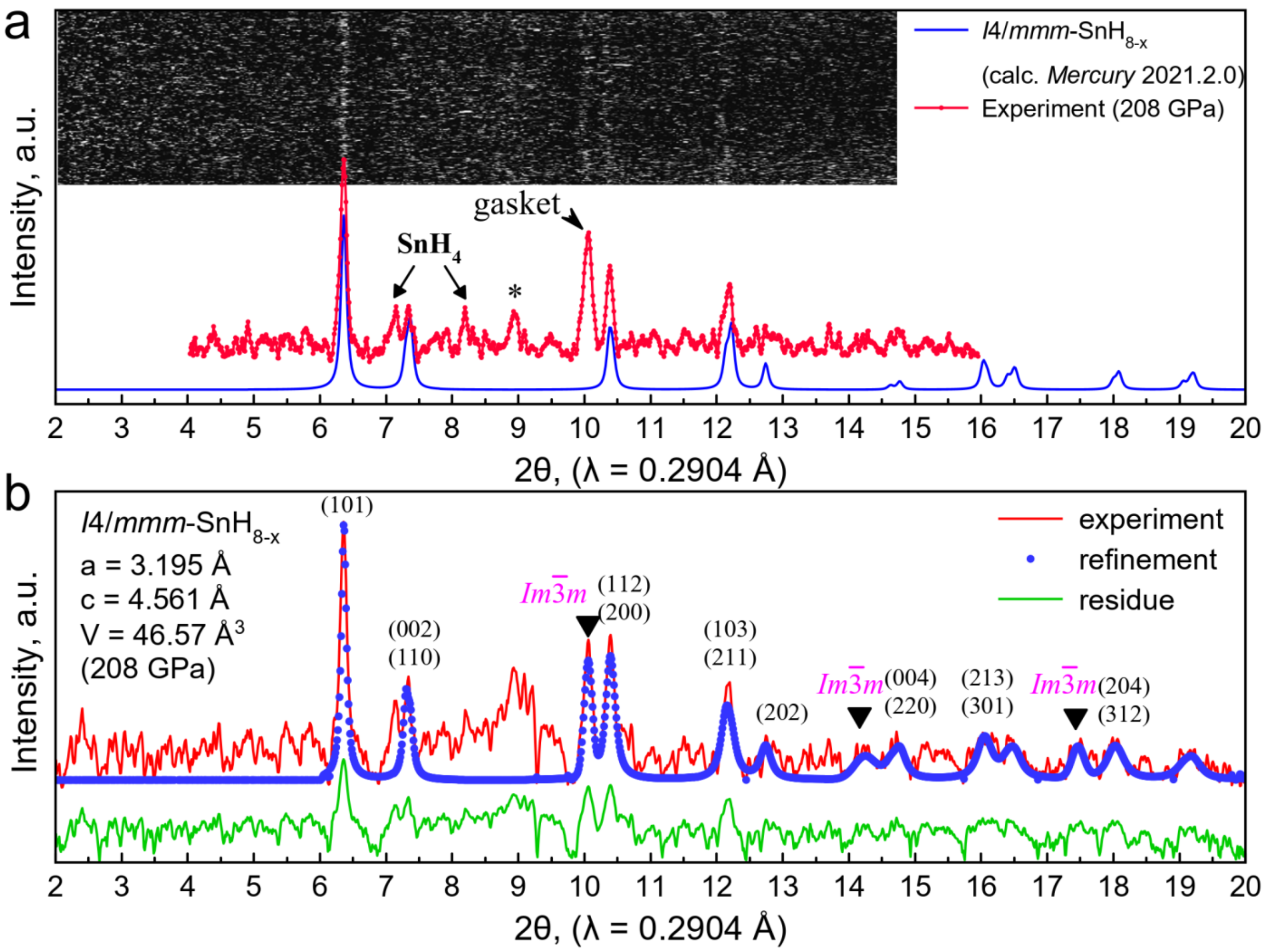

**Figure S6.** X-ray diffraction pattern of the edge of sample in DAC M2 at 208 GPa collected at a wavelength of λ = 0.2904 Å. (a) Comparison of experimental and calculated XRD intensities for the *fcc*-2 phase (refined as pseudocubic *I4/mmm*-$SnH_{8-x}$, Z = 2). Inset: the diffraction image ("cake"). (b) The Le Bail refinement of the unit cell parameters of the *I4/mmm*-$SnH_{8-x}$ (*fcc*-2) phase. Unidentified reflections are marked by asterisks. The experimental data, fitted line, and residues are shown in red, blue, and green, respectively. "*Im*$\bar{3}$*m*" denotes the unidentified reflection from the gasket or material of the DAC.

**Table S6**. Experimental and predicted structures (cif files) of various tin hydrides.

| ***fcc*** **$SnH_4$ at 171 GPa (experiment)** | ***C*2/*m*-$SnH_{14}$ at 180 GPa (calculations)** |
|---|---|
| data_cf_snh4<br>_audit_creation_date 2023-01-26<br>_audit_creation_method<br>;<br>Olex2 1.5<br>(compiled 2022.04.07 svn.rca3783a0 for OlexSys, GUI svn.r6498)<br>;<br>_shelx_SHELXL_version_number '2014/7'<br>_audit_contact_author_address ?<br>_audit_contact_author_email ?<br>_audit_contact_author_name ''<br>_audit_contact_author_phone ?<br>_publ_contact_author_id_orcid ?<br>_publ_section_references<br>;<br>Dolomanov, O.V., Bourhis, L.J., Gildea, R.J, Howard, J.A.K. & Puschmann, H.<br>(2009), J. Appl. Cryst. 42, 339-341.<br><br>Sheldrick, G.M. (2015). Acta Cryst. C71, 3-8.<br>;<br>_chemical_name_common ?<br>_chemical_name_systematic 'tin hydride'<br>_chemical_formula_moiety '0.021(Sn48)'<br>_chemical_formula_sum 'Sn'<br>_chemical_formula_weight 118.69 | data_SnH14_180GPa<br>_symmetry_cell_setting monoclinic<br>_symmetry_space_group_name_H-M 'C 2/m'<br>_symmetry_Int_Tables_number 12<br>_space_group_name_Hall '-C 2y'<br>loop_<br>_symmetry_equiv_pos_site_id<br>_symmetry_equiv_pos_as_xyz<br>1 x,y,z<br>2 -x,y,-z<br>3 -x,-y,-z<br>4 x,-y,z<br>5 1/2+x,1/2+y,z<br>6 1/2-x,1/2+y,-z<br>7 1/2-x,1/2-y,-z<br>8 1/2+x,1/2-y,z<br>_cell_length_a 7.756<br>_cell_length_b 2.871<br>_cell_length_c 3.797<br>_cell_angle_alpha 90<br>_cell_angle_beta 118.34<br>_cell_angle_gamma 90<br>_cell_volume 74.42 |

_chemical_melting_point ?
loop_
_atom_type_symbol
_atom_type_description
_atom_type_scat_dispersion_real
_atom_type_scat_dispersion_imag
_atom_type_scat_source
'Sn' 'Sn' -0.6925 2.4050
'International Tables Vol C Tables 4.2.6.8 and 6.1.1.4'

_shelx_space_group_comment
;
The symmetry employed for this shelxl refinement is uniquely defined by the following loop, which should always be used as a source of symmetry information in preference to the above space-group names. They are only intended as comments.
;
_space_group_crystal_system 'cubic'
_space_group_IT_number 225
_space_group_name_H-M_alt 'F m -3 m'
_space_group_name_Hall '-F 4 2 3'
loop_
_space_group_symop_operation_xyz
'x, y, z'
'-x, -y, z'
'-x, y, -z'
'x, -y, -z'
'z, x, y'
'z, -x, -y'
'-z, -x, y'
'-z, x, -y'
'y, z, x'
'-y, z, -x'
'y, -z, -x'
'-y, -z, x'
'y, x, -z'
'-y, -x, -z'
'y, -x, z'
'-y, x, z'
'x, z, -y'
'-x, z, y'
'-x, -z, -y'
'x, -z, y'
'z, y, -x'
'z, -y, x'
'-z, y, x'
'-z, -y, -x'
'x, y+1/2, z+1/2'
'-x, -y+1/2, z+1/2'
'-x, y+1/2, -z+1/2'
'x, -y+1/2, -z+1/2'
'z, x+1/2, y+1/2'
'z, -x+1/2, -y+1/2'
'-z, -x+1/2, y+1/2'
'-z, x+1/2, -y+1/2'
'y, z+1/2, x+1/2'
'-y, z+1/2, -x+1/2'
'y, -z+1/2, -x+1/2'
'-y, -z+1/2, x+1/2'
'y, x+1/2, -z+1/2'
'-y, -x+1/2, -z+1/2'
'y, -x+1/2, z+1/2'
'-y, x+1/2, z+1/2'
'x, z+1/2, -y+1/2'
'-x, z+1/2, y+1/2'
'-x, -z+1/2, -y+1/2'
'x, -z+1/2, y+1/2'
'z, y+1/2, -x+1/2'
'z, -y+1/2, x+1/2'
'-z, y+1/2, x+1/2'
'-z, -y+1/2, -x+1/2'
'x+1/2, y, z+1/2'
'-x+1/2, -y, z+1/2'

loop_
_atom_site_label
_atom_site_type_symbol
_atom_site_fract_x
_atom_site_fract_y
_atom_site_fract_z
_atom_site_U_iso_or_equiv
_atom_site_thermal_displace_type
Sn1 Sn 0 0.5 0 0.0000 Uiso
H1 H 0.26312 0.22116 0.40715 0.0000 Uiso
H3 H -0.18543 0 -0.01272 0.0000 Uiso
H4 H -0.06856 0 -0.3698 0.0000 Uiso
H6 H 0 0.5 0.5 0.0000 Uiso
H2 H 0.13137 0.5 -0.2995 0.0000 Uiso
H5 H -0.30277 0.5 -0.1822 0.0000 Uiso
H1 H -0.26312 0.22116 -0.40715 0.0000 Uiso
H3 H 0.18543 0 0.01272 0.0000 Uiso
H4 H 0.06856 0 0.3698 0.0000 Uiso
H2 H -0.13137 0.5 0.2995 0.0000 Uiso
H5 H 0.30277 0.5 0.1822 0.0000 Uiso
H1 H -0.26312 0.77884 -0.40715 0.0000 Uiso
H1 H 0.26312 0.77884 0.40715 0.0000 Uiso
Sn1 Sn 0 -0.5 0 0.0000 Uiso
Sn1 Sn 0 0.5 1 0.0000 Uiso
Sn1 Sn 0 1.5 0 0.0000 Uiso
H3 H -0.18543 1 -0.01272 0.0000 Uiso
H3 H 0.18543 1 0.01272 0.0000 Uiso
H4 H -0.06856 1 -0.3698 0.0000 Uiso
H4 H 0.06856 1 0.3698 0.0000 Uiso
H6 H 0 0.5 -0.5 0.0000 Uiso
H7 H 0 0 0.5 0.0000 Uiso

#END

**Symmetrized Sn sublattice (*R*-3*m*) in *C*2-$Sn_{12}H_{18}$ at 200 GPa (calculations)**

# CIF file
# This file was generated by FINDSYM
# Harold T. Stokes, Branton J. Campbell, Dorian M. Hatch
# Brigham Young University, Provo, Utah, USA

data_findsym-output

_symmetry_space_group_name_H-M 'R -3 2/m (hexagonal axes)'
_symmetry_Int_Tables_number 166

_cell_length_a 4.07646
_cell_length_b 4.07646
_cell_length_c 9.94541
_cell_angle_alpha 90.00000
_cell_angle_beta 90.00000
_cell_angle_gamma 120.00000
_cell_volume 143.1264

loop_
_space_group_symop_operation_xyz
x,y,z
-y,x-y,z
-x+y,-x,z
y,x,-z

'-x+1/2, y, -z+1/2'
'x+1/2, -y, -z+1/2'
'z+1/2, x, y+1/2'
'z+1/2, -x, -y+1/2'
'-z+1/2, -x, y+1/2'
'-z+1/2, x, -y+1/2'
'y+1/2, z, x+1/2'
'-y+1/2, z, -x+1/2'
'y+1/2, -z, -x+1/2'
'-y+1/2, -z, x+1/2'
'y+1/2, x, -z+1/2'
'-y+1/2, -x, -z+1/2'
'y+1/2, -x, z+1/2'
'-y+1/2, x, z+1/2'
'x+1/2, z, -y+1/2'
'-x+1/2, z, y+1/2'
'-x+1/2, -z, -y+1/2'
'x+1/2, -z, y+1/2'
'z+1/2, y, -x+1/2'
'z+1/2, -y, x+1/2'
'-z+1/2, y, x+1/2'
'-z+1/2, -y, -x+1/2'
'x+1/2, y+1/2, z'
'-x+1/2, -y+1/2, z'
'-x+1/2, y+1/2, -z'
'x+1/2, -y+1/2, -z'
'z+1/2, x+1/2, y'
'z+1/2, -x+1/2, -y'
'-z+1/2, -x+1/2, y'
'-z+1/2, x+1/2, -y'
'y+1/2, z+1/2, x'
'-y+1/2, z+1/2, -x'
'y+1/2, -z+1/2, -x'
'-y+1/2, -z+1/2, x'
'y+1/2, x+1/2, -z'
'-y+1/2, -x+1/2, -z'
'y+1/2, -x+1/2, z'
'-y+1/2, x+1/2, z'
'x+1/2, z+1/2, -y'
'-x+1/2, z+1/2, y'
'-x+1/2, -z+1/2, -y'
'x+1/2, -z+1/2, y'
'z+1/2, y+1/2, -x'
'z+1/2, -y+1/2, x'
'-z+1/2, y+1/2, x'
'-z+1/2, -y+1/2, -x'
'-x, -y, -z'
'x, y, -z'
'x, -y, z'
'-x, y, z'
'-z, -x, -y'
'-z, x, y'
'z, x, -y'
'z, -x, y'
'-y, -z, -x'
'y, -z, x'
'-y, z, x'
'y, z, -x'
'-y, -x, z'
'y, x, z'
'-y, x, -z'
'y, -x, -z'
'-x, -z, y'
'x, -z, -y'
'x, z, y'
'-x, z, -y'
'-z, -y, x'
'-z, y, -x'
'z, -y, -x'
'z, y, x'
'-x, -y+1/2, -z+1/2'
'x, y+1/2, -z+1/2'
'x, -y+1/2, z+1/2'
'-x, y+1/2, z+1/2'

-x,-x+y,-z
x-y,-y,-z
-x,-y,-z
y,-x+y,-z
x-y,x,-z
-y,-x,z
x,x-y,z
-x+y,y,z
x+1/3,y+2/3,z+2/3
-y+1/3,x-y+2/3,z+2/3
-x+y+1/3,-x+2/3,z+2/3
y+1/3,x+2/3,-z+2/3
-x+1/3,-x+y+2/3,-z+2/3
x-y+1/3,-y+2/3,-z+2/3
-x+1/3,-y+2/3,-z+2/3
y+1/3,-x+y+2/3,-z+2/3
x-y+1/3,x+2/3,-z+2/3
-y+1/3,-x+2/3,z+2/3
x+1/3,x-y+2/3,z+2/3
-x+y+1/3,y+2/3,z+2/3
x+2/3,y+1/3,z+1/3
-y+2/3,x-y+1/3,z+1/3
-x+y+2/3,-x+1/3,z+1/3
y+2/3,x+1/3,-z+1/3
-x+2/3,-x+y+1/3,-z+1/3
x-y+2/3,-y+1/3,-z+1/3
-x+2/3,-y+1/3,-z+1/3
y+2/3,-x+y+1/3,-z+1/3
x-y+2/3,x+1/3,-z+1/3
-y+2/3,-x+1/3,z+1/3
x+2/3,x-y+1/3,z+1/3
-x+y+2/3,y+1/3,z+1/3

loop_
_atom_site_label
_atom_site_type_symbol
_atom_site_fract_x
_atom_site_fract_y
_atom_site_fract_z
_atom_site_occupancy
Sn1 Sn 0.00000 0.00000 -0.24147 1.00000
Sn2 Sn 0.00000 0.00000 0.00000 1.00000

**$R\bar{3}m$-$Sn_{12}H_{45}$ at 180 GPa (calculations)**

# CIF file
# This file was generated by FINDSYM
# Harold T. Stokes, Branton J. Campbell, Dorian M. Hatch
# Brigham Young University, Provo, Utah, USA

data_findsym-output

_symmetry_space_group_name_H-M 'R -3 2/m (hexagonal axes)'
_symmetry_Int_Tables_number 166

_cell_length_a 6.05800

| | |
|---|---|
| '-z, -x+1/2, -y+1/2'<br>'-z, x+1/2, y+1/2'<br>'z, x+1/2, -y+1/2'<br>'z, -x+1/2, y+1/2'<br>'-y, -z+1/2, -x+1/2'<br>'y, -z+1/2, x+1/2'<br>'-y, z+1/2, x+1/2'<br>'y, z+1/2, -x+1/2'<br>'-y, -x+1/2, z+1/2'<br>'y, x+1/2, z+1/2'<br>'-y, x+1/2, -z+1/2'<br>'y, -x+1/2, -z+1/2'<br>'-x, -z+1/2, y+1/2'<br>'x, -z+1/2, -y+1/2'<br>'x, z+1/2, y+1/2'<br>'-x, z+1/2, -y+1/2'<br>'-z, -y+1/2, x+1/2'<br>'-z, y+1/2, -x+1/2'<br>'z, -y+1/2, -x+1/2'<br>'z, y+1/2, x+1/2'<br>'-x+1/2, -y, -z+1/2'<br>'x+1/2, y, -z+1/2'<br>'x+1/2, -y, z+1/2'<br>'-x+1/2, y, z+1/2'<br>'-z+1/2, -x, -y+1/2'<br>'-z+1/2, x, y+1/2'<br>'z+1/2, x, -y+1/2'<br>'z+1/2, -x, y+1/2'<br>'-y+1/2, -z, -x+1/2'<br>'y+1/2, -z, x+1/2'<br>'-y+1/2, z, x+1/2'<br>'y+1/2, z, -x+1/2'<br>'-y+1/2, -x, z+1/2'<br>'y+1/2, x, z+1/2'<br>'-y+1/2, x, -z+1/2'<br>'y+1/2, -x, -z+1/2'<br>'-x+1/2, -z, y+1/2'<br>'x+1/2, -z, -y+1/2'<br>'x+1/2, z, y+1/2'<br>'-x+1/2, z, -y+1/2'<br>'-z+1/2, -y, x+1/2'<br>'-z+1/2, y, -x+1/2'<br>'z+1/2, -y, -x+1/2'<br>'z+1/2, y, x+1/2'<br>'-x+1/2, -y+1/2, -z'<br>'x+1/2, y+1/2, -z'<br>'x+1/2, -y+1/2, z'<br>'-x+1/2, y+1/2, z'<br>'-z+1/2, -x+1/2, -y'<br>'-z+1/2, x+1/2, y'<br>'z+1/2, x+1/2, -y'<br>'z+1/2, -x+1/2, y'<br>'-y+1/2, -z+1/2, -x'<br>'y+1/2, -z+1/2, x'<br>'-y+1/2, z+1/2, x'<br>'y+1/2, z+1/2, -x'<br>'-y+1/2, -x+1/2, z'<br>'y+1/2, x+1/2, z'<br>'-y+1/2, x+1/2, -z'<br>'y+1/2, -x+1/2, -z'<br>'-x+1/2, -z+1/2, y'<br>'x+1/2, -z+1/2, -y'<br>'x+1/2, z+1/2, y'<br>'-x+1/2, z+1/2, -y'<br>'-z+1/2, -y+1/2, x'<br>'-z+1/2, y+1/2, -x'<br>'z+1/2, -y+1/2, -x'<br>'z+1/2, y+1/2, x'<br><br>_cell_length_a 4.244(3)<br>_cell_length_b 4.244(3)<br>_cell_length_c 4.244(3)<br>_cell_angle_alpha 90<br>_cell_angle_beta 90 | _cell_length_b 6.05800<br>_cell_length_c 7.37800<br>_cell_angle_alpha 90.00000<br>_cell_angle_beta 90.00000<br>_cell_angle_gamma 120.00000<br><br>loop_<br>_space_group_symop_operation_xyz<br>x,y,z<br>-y,x-y,z<br>-x+y,-x,z<br>y,x,-z<br>-x,-x+y,-z<br>x-y,-y,-z<br>-x,-y,-z<br>y,-x+y,-z<br>x-y,x,-z<br>-y,-x,z<br>x,x-y,z<br>-x+y,y,z<br>x+1/3,y+2/3,z+2/3<br>-y+1/3,x-y+2/3,z+2/3<br>-x+y+1/3,-x+2/3,z+2/3<br>y+1/3,x+2/3,-z+2/3<br>-x+1/3,-x+y+2/3,-z+2/3<br>x-y+1/3,-y+2/3,-z+2/3<br>-x+1/3,-y+2/3,-z+2/3<br>y+1/3,-x+y+2/3,-z+2/3<br>x-y+1/3,x+2/3,-z+2/3<br>-y+1/3,-x+2/3,z+2/3<br>x+1/3,x-y+2/3,z+2/3<br>-x+y+1/3,y+2/3,z+2/3<br>x+2/3,y+1/3,z+1/3<br>-y+2/3,x-y+1/3,z+1/3<br>-x+y+2/3,-x+1/3,z+1/3<br>y+2/3,x+1/3,-z+1/3<br>-x+2/3,-x+y+1/3,-z+1/3<br>x-y+2/3,-y+1/3,-z+1/3<br>-x+2/3,-y+1/3,-z+1/3<br>y+2/3,-x+y+1/3,-z+1/3<br>x-y+2/3,x+1/3,-z+1/3<br>-y+2/3,-x+1/3,z+1/3<br>x+2/3,x-y+1/3,z+1/3<br>-x+y+2/3,y+1/3,z+1/3<br><br>loop_<br>_atom_site_label<br>_atom_site_type_symbol<br>_atom_site_fract_x<br>_atom_site_fract_y<br>_atom_site_fract_z<br>_atom_site_occupancy<br>Sn1 Sn 0.00000 0.00000 0.50000 1.00000<br>Sn2 Sn 0.50000 0.00000 0.50000 1.00000<br>H1 H -0.46473 0.46473 0.01893 1.00000<br>H2 H 0.16805 -0.16805 0.08631 1.00000<br>H3 H 0.00000 0.00000 -0.24152 1.00000<br>H4 H 0.00000 0.00000 0.00000 1.00000 |

```
_cell_angle_gamma                90
_cell_volume                   76.44(14)
_cell_formula_units_Z             4
_cell_measurement_reflns_used      25
_cell_measurement_temperature      293(2)
_cell_measurement_theta_max        14.4550
_cell_measurement_theta_min        4.5250
_shelx_estimated_absorpt_T_max     0.988
_shelx_estimated_absorpt_T_min     0.988
_exptl_absorpt_coefficient_mu      24.888
_exptl_absorpt_correction_T_max    ?
_exptl_absorpt_correction_T_min    ?
_exptl_absorpt_correction_type     none
_exptl_absorpt_process_details     ?
_exptl_absorpt_special_details     ?
_exptl_crystal_colour              metallic
_exptl_crystal_colour_lustre       metallic
_exptl_crystal_density_diffrn      10.313
_exptl_crystal_density_meas        ?
_exptl_crystal_density_method      ?
_exptl_crystal_description         irregular
_exptl_crystal_F_000               200
_exptl_crystal_preparation         ?
_exptl_crystal_size_max            0.0005
_exptl_crystal_size_mid            0.0005
_exptl_crystal_size_min            0.0005
_exptl_transmission_factor_max     ?
_exptl_transmission_factor_min     ?
_diffrn_reflns_av_R_equivalents    0.1356
_diffrn_reflns_av_unetI/netI       0.0827
_diffrn_reflns_Laue_measured_fraction_full  0.778
_diffrn_reflns_Laue_measured_fraction_max  0.625
_diffrn_reflns_limit_h_max         5
_diffrn_reflns_limit_h_min         -6
_diffrn_reflns_limit_k_max         5
_diffrn_reflns_limit_k_min         -5
_diffrn_reflns_limit_l_max         2
_diffrn_reflns_limit_l_min         -1
_diffrn_reflns_number              29
_diffrn_reflns_point_group_measured_fraction_full  0.778
_diffrn_reflns_point_group_measured_fraction_max  0.625
_diffrn_reflns_theta_full          11.574
_diffrn_reflns_theta_max           14.428
_diffrn_reflns_theta_min           4.519
_diffrn_ambient_temperature        293(2)
_diffrn_detector                   'Pixel detector'
_diffrn_detector_area_resol_mean   5.8140
_diffrn_detector_type              'Pilatus 1M CdTe'
_diffrn_measured_fraction_theta_full  0.778
_diffrn_measured_fraction_theta_max  0.625
_diffrn_measurement_details
;
List of Runs (angles in degrees, time in seconds):

  # Type    Start    End  Width  t~exp~   \w    \q    \k    \f   Frames
--------------------------------------------------------------------------
  1  \w   -32.00  32.00   0.50    1.00    --   0.00   0.00   0.00  128
;
_diffrn_measurement_device         '13IDD @ APS'
_diffrn_measurement_device_type    'LH Table'
_diffrn_measurement_method         'omega rotation'
_diffrn_orient_matrix_type
 'CrysAlisPro convention (1999,Acta A55,543-557)'
_diffrn_orient_matrix_UB_11        -0.0136159167
_diffrn_orient_matrix_UB_12        -0.0060303500
_diffrn_orient_matrix_UB_13        0.0776876667
_diffrn_orient_matrix_UB_21        -0.0288921993
_diffrn_orient_matrix_UB_22        0.0731562500
_diffrn_orient_matrix_UB_23        0.0012496493
_diffrn_orient_matrix_UB_31        -0.0721157857
_diffrn_orient_matrix_UB_32        -0.0276921000
_diffrn_orient_matrix_UB_33        -0.0150944143
_diffrn_radiation_monochromator    synchrotron
_diffrn_radiation_probe            x-ray
```

### ***$Fd\bar{3}m$*-$Sn_{32}H_{120}$ at 180 GPa (calculations)**

---

```
# CIF file
# This file was generated by FINDSYM
# Harold T. Stokes, Branton J. Campbell, Dorian M. Hatch
# Brigham Young University, Provo, Utah, USA

data_findsym-output

_symmetry_space_group_name_H-M 'F 41/d -3 2/m (origin choice 2)'
_symmetry_Int_Tables_number 227

_cell_length_a       8.57900
_cell_length_b       8.57900
_cell_length_c       8.57900
_cell_angle_alpha   90.00000
_cell_angle_beta    90.00000
_cell_angle_gamma   90.00000
_cell_volume 631.407

loop_
_space_group_symop_operation_xyz
x,y,z
x,-y+1/4,-z+1/4
-x+1/4,y,-z+1/4
-x+1/4,-y+1/4,z
y,z,x
y,-z+1/4,-x+1/4
-y+1/4,z,-x+1/4
-y+1/4,-z+1/4,x
z,x,y
z,-x+1/4,-y+1/4
-z+1/4,x,-y+1/4
-z+1/4,-x+1/4,y
-y,-x,-z
-y,x+1/4,z+1/4
y+1/4,-x,z+1/4
y+1/4,x+1/4,-z
-x,-z,-y
-x,z+1/4,y+1/4
x+1/4,-z,y+1/4
x+1/4,z+1/4,-y
-z,-y,-x
-z,y+1/4,x+1/4
z+1/4,-y,x+1/4
z+1/4,y+1/4,-x
-x,-y,-z
-x,y+1/4,z+1/4
x+1/4,-y,z+1/4
x+1/4,y+1/4,-z
-y,-z,-x
-y,z+1/4,x+1/4
y+1/4,-z,x+1/4
y+1/4,z+1/4,-x
-z,-x,-y
-z,x+1/4,y+1/4
z+1/4,-x,y+1/4
z+1/4,x+1/4,-y
y,x,z
y,-x+1/4,-z+1/4
-y+1/4,x,-z+1/4
-y+1/4,-x+1/4,z
x,z,y
x,-z+1/4,-y+1/4
-x+1/4,z,-y+1/4
-x+1/4,-z+1/4,y
z,y,x
z,-y+1/4,-x+1/4
-z+1/4,y,-x+1/4
-z+1/4,-y+1/4,x
x,y+1/2,z+1/2
x,-y+3/4,-z+3/4
-x+1/4,y+1/2,-z+3/4
```

```
_diffrn_radiation_type            synchrotron
_diffrn_radiation_wavelength      0.3344
_diffrn_source                    synchrotron
_reflns_Friedel_coverage          0.000
_reflns_Friedel_fraction_full     .
_reflns_Friedel_fraction_max      .
_reflns_number_gt                 10
_reflns_number_total              10
_reflns_special_details
;
 Reflections were merged by SHELXL according to the crystal
 class for the calculation of statistics and refinement.

 _reflns_Friedel_fraction is defined as the number of unique
 Friedel pairs measured divided by the number that would be
 possible theoretically, ignoring centric projections and
 systematic absences.
;
_reflns_threshold_expression      'I > 2\s(I)'
_computing_cell_refinement        'CrysAlisPro 1.171.40.69a (Rigaku OD, 2020)'
_computing_data_collection        'CrysAlisPro 1.171.40.69a (Rigaku OD, 2020)'
_computing_data_reduction         'CrysAlisPro 1.171.40.69a (Rigaku OD, 2020)'
_computing_molecular_graphics     'Olex2 1.5 (Dolomanov et al., 2009)'
_computing_publication_material   'Olex2 1.5 (Dolomanov et al., 2009)'
_computing_structure_refinement   'SHELXL 2014/7 (Sheldrick, 2015)'
_computing_structure_solution     ?
_refine_diff_density_max          1.134
_refine_diff_density_min          -1.995
_refine_diff_density_rms          0.464
_refine_ls_extinction_coef        0.5(6)
_refine_ls_extinction_expression
 'Fc^*^=kFc[1+0.001xFc^2^\l^3^/sin(2\q)]^-1/4^'
_refine_ls_extinction_method      'SHELXL-2014/7 (Sheldrick 2014'
_refine_ls_goodness_of_fit_ref    1.240
_refine_ls_hydrogen_treatment     undef
_refine_ls_matrix_type            full
_refine_ls_number_parameters      3
_refine_ls_number_reflns          10
_refine_ls_number_restraints      0
_refine_ls_R_factor_all           0.0475
_refine_ls_R_factor_gt            0.0475
_refine_ls_restrained_S_all       1.240
_refine_ls_shift/su_max           0.000
_refine_ls_shift/su_mean          0.000
_refine_ls_structure_factor_coef  Fsqd
_refine_ls_weighting_details
 'w=1/[\s^2^(Fo^2^)+22.1894P] where P=(Fo^2^+2Fc^2^)/3'
_refine_ls_weighting_scheme       calc
_refine_ls_wR_factor_gt           0.0960
_refine_ls_wR_factor_ref          0.0960
_refine_special_details           ?
_olex2_refinement_description
;
;
_atom_sites_solution_hydrogens    .
_atom_sites_solution_primary      ?
_atom_sites_solution_secondary    ?
loop_
  _atom_site_label
  _atom_site_type_symbol
  _atom_site_fract_x
  _atom_site_fract_y
  _atom_site_fract_z
  _atom_site_U_iso_or_equiv
  _atom_site_adp_type
  _atom_site_occupancy
  _atom_site_site_symmetry_order
  _atom_site_calc_flag
```

```
-x+1/4,-y+3/4,z+1/2
y,z+1/2,x+1/2
y,-z+3/4,-x+3/4
-y+1/4,z+1/2,-x+3/4
-y+1/4,-z+3/4,x+1/2
z,x+1/2,y+1/2
z,-x+3/4,-y+3/4
-z+1/4,x+1/2,-y+3/4
-z+1/4,-x+3/4,y+1/2
-y,-x+1/2,-z+1/2
-y,x+3/4,z+3/4
y+1/4,-x+1/2,z+3/4
y+1/4,x+3/4,-z+1/2
-x,-z+1/2,-y+1/2
-x,z+3/4,y+3/4
x+1/4,-z+1/2,y+3/4
x+1/4,z+3/4,-y+1/2
-z,-y+1/2,-x+1/2
-z,y+3/4,x+3/4
z+1/4,-y+1/2,x+3/4
z+1/4,y+3/4,-x+1/2
-x,-y+1/2,-z+1/2
-x,y+3/4,z+3/4
x+1/4,-y+1/2,z+3/4
x+1/4,y+3/4,-z+1/2
-y,-z+1/2,-x+1/2
-y,z+3/4,x+3/4
y+1/4,-z+1/2,x+3/4
y+1/4,z+3/4,-x+1/2
-z,-x+1/2,-y+1/2
-z,x+3/4,y+3/4
z+1/4,-x+1/2,y+3/4
z+1/4,x+3/4,-y+1/2
y,x+1/2,z+1/2
y,-x+3/4,-z+3/4
-y+1/4,x+1/2,-z+3/4
-y+1/4,-x+3/4,z+1/2
x,z+1/2,y+1/2
x,-z+3/4,-y+3/4
-x+1/4,z+1/2,-y+3/4
-x+1/4,-z+3/4,y+1/2
z,y+1/2,x+1/2
z,-y+3/4,-x+3/4
-z+1/4,y+1/2,-x+3/4
-z+1/4,-y+3/4,x+1/2
x+1/2,y,z+1/2
x+1/2,-y+1/4,-z+3/4
-x+3/4,y,-z+3/4
-x+3/4,-y+1/4,z+1/2
y+1/2,z,x+1/2
y+1/2,-z+1/4,-x+3/4
-y+3/4,z,-x+3/4
-y+3/4,-z+1/4,x+1/2
z+1/2,x,y+1/2
z+1/2,-x+1/4,-y+3/4
-z+3/4,x,-y+3/4
-z+3/4,-x+1/4,y+1/2
-y+1/2,-x,-z+1/2
-y+1/2,x+1/4,z+3/4
y+3/4,-x,z+3/4
y+3/4,x+1/4,-z+1/2
-x+1/2,-z,-y+1/2
-x+1/2,z+1/4,y+3/4
x+3/4,-z,y+3/4
x+3/4,z+1/4,-y+1/2
-z+1/2,-y,-x+1/2
-z+1/2,y+1/4,x+3/4
z+3/4,-y,x+3/4
z+3/4,y+1/4,-x+1/2
-x+1/2,-y,-z+1/2
-x+1/2,y+1/4,z+3/4
x+3/4,-y,z+3/4
x+3/4,y+1/4,-z+1/2
-y+1/2,-z,-x+1/2
```

_atom_site_refinement_flags_posn
_atom_site_refinement_flags_adp
_atom_site_refinement_flags_occupancy
_atom_site_disorder_assembly
_atom_site_disorder_group
Sn01 Sn 0.0000 0.0000 0.0000 0.024(6) Uani 1 48 d S T P . .

loop_
_atom_site_aniso_label
_atom_site_aniso_U_11
_atom_site_aniso_U_22
_atom_site_aniso_U_33
_atom_site_aniso_U_23
_atom_site_aniso_U_13
_atom_site_aniso_U_12
Sn01 0.024(6) 0.024(6) 0.024(6) 0.000 0.000 0.000

_geom_special_details
;
All esds (except the esd in the dihedral angle between two l.s. planes) are estimated using the full covariance matrix. The cell esds are taken into account individually in the estimation of esds in distances, angles and torsion angles; correlations between esds in cell parameters are only
used when they are defined by crystal symmetry. An approximate (isotropic)
treatment of cell esds is used for estimating esds involving l.s. planes.
;
loop_
_geom_bond_atom_site_label_1
_geom_bond_atom_site_label_2
_geom_bond_distance
_geom_bond_site_symmetry_2
_geom_bond_publ_flag
Sn01 Sn01 3.0010(18) 73 ?
Sn01 Sn01 3.0010(18) 49_554 ?
Sn01 Sn01 3.0010(18) 25_554 ?
Sn01 Sn01 3.0010(18) 25_544 ?
Sn01 Sn01 3.0010(18) 73_545 ?
Sn01 Sn01 3.0010(18) 73_445 ?
Sn01 Sn01 3.0010(18) 49_455 ?
Sn01 Sn01 3.0010(18) 49 ?
Sn01 Sn01 3.0010(18) 25_545 ?
Sn01 Sn01 3.0010(18) 25 ?
Sn01 Sn01 3.0010(18) 73_455 ?
Sn01 Sn01 3.0010(18) 49_454 ?

loop_
_geom_angle_atom_site_label_1
_geom_angle_atom_site_label_2
_geom_angle_atom_site_label_3
_geom_angle
_geom_angle_site_symmetry_1
_geom_angle_site_symmetry_3
_geom_angle_publ_flag
Sn01 Sn01 Sn01 120.0 73 25_544 ?
Sn01 Sn01 Sn01 120.0 25_545 73_455 ?
Sn01 Sn01 Sn01 60.0 49 73_545 ?
Sn01 Sn01 Sn01 180.0 73 73_445 ?
Sn01 Sn01 Sn01 90.0 25 25_545 ?
Sn01 Sn01 Sn01 60.0 25_544 73_445 ?
Sn01 Sn01 Sn01 120.0 25_544 73_455 ?
Sn01 Sn01 Sn01 60.0 73 49 ?
Sn01 Sn01 Sn01 120.0 73 25_545 ?
Sn01 Sn01 Sn01 120.0 25_544 49 ?
Sn01 Sn01 Sn01 120.0 49_554 25_545 ?
Sn01 Sn01 Sn01 120.0 73_445 49 ?
Sn01 Sn01 Sn01 120.0 49_455 73_545 ?
Sn01 Sn01 Sn01 60.0 73 25 ?
Sn01 Sn01 Sn01 60.0 49_454 73_455 ?
Sn01 Sn01 Sn01 180.0 25_544 25 ?
Sn01 Sn01 Sn01 60.0 25_554 49_554 ?
Sn01 Sn01 Sn01 120.0 73_445 25 ?
Sn01 Sn01 Sn01 60.0 73_445 25_545 ?

-y+1/2,z+1/4,x+3/4
y+3/4,-z,x+3/4
y+3/4,z+1/4,-x+1/2
-z+1/2,-x,-y+1/2
-z+1/2,x+1/4,y+3/4
z+3/4,-x,y+3/4
z+3/4,x+1/4,-y+1/2
y+1/2,x,z+1/2
y+1/2,-x+1/4,-z+3/4
-y+3/4,x,-z+3/4
-y+3/4,-x+1/4,z+1/2
x+1/2,z,y+1/2
x+1/2,-z+1/4,-y+3/4
-x+3/4,z,-y+3/4
-x+3/4,-z+1/4,y+1/2
z+1/2,y,x+1/2
z+1/2,-y+1/4,-x+3/4
-z+3/4,y,-x+3/4
-z+3/4,-y+1/4,x+1/2
x+1/2,y+1/2,z
x+1/2,-y+3/4,-z+1/4
-x+3/4,y+1/2,-z+1/4
-x+3/4,-y+3/4,z
y+1/2,z+1/2,x
y+1/2,-z+3/4,-x+1/4
-y+3/4,z+1/2,-x+1/4
-y+3/4,-z+3/4,x
z+1/2,x+1/2,y
z+1/2,-x+3/4,-y+1/4
-z+3/4,x+1/2,-y+1/4
-z+3/4,-x+3/4,y
-y+1/2,-x+1/2,-z
-y+1/2,x+3/4,z+1/4
y+3/4,-x+1/2,z+1/4
y+3/4,x+3/4,-z
-x+1/2,-z+1/2,-y
-x+1/2,z+3/4,y+1/4
x+3/4,-z+1/2,y+1/4
x+3/4,z+3/4,-y
-z+1/2,-y+1/2,-x
-z+1/2,y+3/4,x+1/4
z+3/4,-y+1/2,x+1/4
z+3/4,y+3/4,-x
-x+1/2,-y+1/2,-z
-x+1/2,y+3/4,z+1/4
x+3/4,-y+1/2,z+1/4
x+3/4,y+3/4,-z
-y+1/2,-z+1/2,-x
-y+1/2,z+3/4,x+1/4
y+3/4,-z+1/2,x+1/4
y+3/4,z+3/4,-x
-z+1/2,-x+1/2,-y
-z+1/2,x+3/4,y+1/4
z+3/4,-x+1/2,y+1/4
z+3/4,x+3/4,-y
y+1/2,x+1/2,z
y+1/2,-x+3/4,-z+1/4
-y+3/4,x+1/2,-z+1/4
-y+3/4,-x+3/4,z
x+1/2,z+1/2,y
x+1/2,-z+3/4,-y+1/4
-x+3/4,z+1/2,-y+1/4
-x+3/4,-z+3/4,y
z+1/2,y+1/2,x
z+1/2,-y+3/4,-x+1/4
-z+3/4,y+1/2,-x+1/4
-z+3/4,-y+3/4,x

loop_
_atom_site_label
_atom_site_type_symbol
_atom_site_fract_x
_atom_site_fract_y
_atom_site_fract_z

```
 Sn01 Sn01 Sn01 60.0 49 25 ?
 Sn01 Sn01 Sn01 180.0 25_554 25_545 ?
 Sn01 Sn01 Sn01 120.0 73 49_454 ?
 Sn01 Sn01 Sn01 60.0 25_544 73_545 ?
 Sn01 Sn01 Sn01 60.0 25_544 49_454 ?
 Sn01 Sn01 Sn01 120.0 49_454 73_545 ?
 Sn01 Sn01 Sn01 60.0 73_445 49_454 ?
 Sn01 Sn01 Sn01 60.0 25_545 73_545 ?
 Sn01 Sn01 Sn01 180.0 49 49_454 ?
 Sn01 Sn01 Sn01 120.0 49 73_455 ?
 Sn01 Sn01 Sn01 120.0 25 49_454 ?
 Sn01 Sn01 Sn01 60.0 49_455 73_455 ?
 Sn01 Sn01 Sn01 60.0 73 25_554 ?
 Sn01 Sn01 Sn01 90.0 49_454 49_554 ?
 Sn01 Sn01 Sn01 90.0 25_544 25_554 ?
 Sn01 Sn01 Sn01 180.0 49_455 49_554 ?
 Sn01 Sn01 Sn01 120.0 73_445 25_554 ?
 Sn01 Sn01 Sn01 90.0 25_544 25_545 ?
 Sn01 Sn01 Sn01 120.0 49 25_554 ?
 Sn01 Sn01 Sn01 60.0 49 25_545 ?
 Sn01 Sn01 Sn01 90.0 25 25_554 ?
 Sn01 Sn01 Sn01 120.0 49_454 25_545 ?
 Sn01 Sn01 Sn01 60.0 49_454 25_554 ?
 Sn01 Sn01 Sn01 60.0 49_455 25_545 ?
 Sn01 Sn01 Sn01 120.0 73 49_455 ?
 Sn01 Sn01 Sn01 90.0 73 73_545 ?
 Sn01 Sn01 Sn01 120.0 25_544 49_455 ?
 Sn01 Sn01 Sn01 90.0 73_445 73_545 ?
 Sn01 Sn01 Sn01 60.0 73_445 49_455 ?
 Sn01 Sn01 Sn01 120.0 25 73_545 ?
 Sn01 Sn01 Sn01 90.0 49 49_455 ?
 Sn01 Sn01 Sn01 120.0 25_554 73_545 ?
 Sn01 Sn01 Sn01 60.0 25 49_455 ?
 Sn01 Sn01 Sn01 60.0 49_554 73_545 ?
 Sn01 Sn01 Sn01 90.0 49_454 49_455 ?
 Sn01 Sn01 Sn01 90.0 73 73_455 ?
 Sn01 Sn01 Sn01 120.0 25_554 49_455 ?
 Sn01 Sn01 Sn01 90.0 73_445 73_455 ?
 Sn01 Sn01 Sn01 60.0 73 49_554 ?
 Sn01 Sn01 Sn01 60.0 25 73_455 ?
 Sn01 Sn01 Sn01 60.0 25_544 49_554 ?
 Sn01 Sn01 Sn01 60.0 25_554 73_455 ?
 Sn01 Sn01 Sn01 120.0 73_445 49_554 ?
 Sn01 Sn01 Sn01 120.0 49_554 73_455 ?
 Sn01 Sn01 Sn01 90.0 49 49_554 ?
 Sn01 Sn01 Sn01 180.0 73_545 73_455 ?
 Sn01 Sn01 Sn01 120.0 25 49_554 ?

_shelx_res_file
;
   cf_snh4.res created by SHELXL-2014/7

TITL snh_p02_q3_pos130
CELL 0.3344 4.244 4.244 4.244 90 90 90
ZERR 4 0.002599 0.002599 0.002599 0 0 0
LATT 4
SYMM -X,-Y,+Z
SYMM -X,+Y,-Z
SYMM +X,-Y,-Z
SYMM +Z,+X,+Y
SYMM +Z,-X,-Y
SYMM -Z,-X,+Y
SYMM -Z,+X,-Y
SYMM +Y,+Z,+X
SYMM -Y,+Z,-X
SYMM +Y,-Z,-X
SYMM -Y,-Z,+X
SYMM +Y,+X,-Z
SYMM -Y,-X,-Z
SYMM +Y,-X,+Z
SYMM -Y,+X,+Z
SYMM +X,+Z,-Y
SYMM -X,+Z,+Y
SYMM -X,-Z,-Y
```

```
 _atom_site_occupancy
Sn1 Sn   0.50000   0.50000   0.50000   1.00000
Sn2 Sn   0.00000   0.00000   0.00000   1.00000
H1  H   -0.12413   0.12500   0.12500   1.00000
H2  H    0.28368   0.28368   0.28368   1.00000
H3  H    0.22628   0.22628   0.22628   1.00000
H4  H    0.12500   0.12500   0.12500   1.00000
```

```
SYMM +X,-Z,+Y
SYMM +Z,+Y,-X
SYMM +Z,-Y,+X
SYMM -Z,+Y,+X
SYMM -Z,-Y,-X
SFAC Sn
DISP Sn -0.6925 2.405 4756.1985
UNIT 4

L.S. 10 0 0
PLAN  5
SIZE 0.0005 0.0005 0.0005
CONF
MORE -1
fmap 2 53
ACTA
REM <olex2.extras>
REM <HklSrc "%.\\cF_SnH4.hkl">
REM </olex2.extras>

WGHT    0.000000   22.189402
EXTI    0.541500
FVAR       1.01538
SN01  1     0.000000    0.000000    0.000000    10.02083    0.02443
0.02443 =
       0.02443    0.00000    0.00000    0.00000
HKLF 4

REM  snh_p02_q3_pos130
REM R1 =  0.0475 for       10 Fo > 4sig(Fo)  and  0.0475 for all       10
data
REM      3 parameters refined using      0 restraints

END

WGHT      0.0000     19.9852

REM Highest difference peak  1.134,  deepest hole -1.995,  1-sigma
level  0.464
Q1    1   0.2500  0.2500 -0.1231  10.25000  0.05    1.13
Q2    1   0.0000  0.3032  0.0000  10.12500  0.05    0.94
Q3    1   0.0000  0.0000  0.0773  10.12500  0.05    0.91
Q4    1   0.1739  0.0000  0.0000  10.12500  0.05    0.90
Q5    1   0.0000 -0.3415  0.1585  10.25000  0.05    0.72
;
_shelx_res_checksum              69794
_shelx_hkl_file
;
  -2   0   010079.50   48.90   1
   0  -2   012715.90   83.54   1
  -2  -2   0 7816.28   71.28   1
   1  -3   1 6288.93  151.82   1
  -1   3   1 5994.86  153.40   1
  -3   1  -1 7997.15  113.20   1
  -1   3  -1 7219.69  163.26   1
  -3   3   1 3471.32  189.81   1
   3  -3   1 3498.64  184.45   1
   3   3   1 2777.11   98.18   1
  -3   3  -1 4243.44  191.87   1
   0  -4   0 3709.52  173.03   1
   0   4   0 3934.57  176.98   1
   4  -2   2 2634.75  170.81   1
   4   0   2 3645.09  107.95   1
  -2   4   0 3888.53  210.43   1
   2  -4   0 3751.22  208.44   1
   2   4   2 2135.76  161.07   1
   4   2   2  751.01   68.82   1
   5  -1   1  970.27  147.15   1
  -1  -5  -1 1070.62  209.66   1
   1   5   1 2075.23  203.40   1
  -1  -5   1 1685.69  208.09   1
  -5   1  -1 2010.98  150.14   1
   5  -3   1  940.45  232.90   1
  -3  -5  -1  438.57  178.09   1
```

| | |
|---|---|
| 3 5 1 1245.79 187.15 1<br>-5 3 -1 1368.40 233.43 1<br>-6 2 0 672.27 208.70 1<br>0 0 0 0.00 0.00 0<br>;<br>_shelx_hkl_checksum 70650<br>_olex2_submission_special_instructions 'No special instructions were received'<br>_oxdiff_exptl_absorpt_empirical_details<br>;<br>Empirical correction (ABSPACK) includes:<br>;<br>_oxdiff_exptl_absorpt_empirical_full_max 1.000<br>_oxdiff_exptl_absorpt_empirical_full_min 1.000 | |

**Table S7.** Experimental unit cell parameters of proposed distorted $Im\bar{3}m$ (Z = 2) phase and $P6_3/mmc$-Sn (Z = 2) obtained from the exteriments at PETRA-III in 2020 at different pressures (DAC M2).

| Experiment and pressure | ***a*, Å** | ***b*, Å** | ***c*, Å** | ***V*, Å$^3$** | ***a*, Å** | ***c*, Å** | ***V*, Å$^3$** |
|---|---|---|---|---|---|---|---|
| P02 (h1), 180 GPa | 2.895 | 2.895 | 2.895 | 24.26 | 2.633 | 4.581 | 27.51 |
| P02 (h3), 185 GPa | 2.885 | 2.935 | 2.845 | 24.09 | 2.634 | 4.595 | 27.61 |
| P03, 190-196 GPa | 2.885<br>2.885 | 2.865<br>2.925 | 2.895<br>2.835 | 23.93<br>23.92 | 2.716 | 4.093 | 26.15 |
| P04, 201-208 GPa | 2.863<br>2.870 | 2.863<br>2.870 | 2.863<br>2.870 | 23.47<br>23.64 | 2.634 | 4.435 | 26.65 |
| P05, 208-210 GPa | 2.855 | 2.905 | 2.825 | 23.43 | 2.621 | 4.415 | 26.32 |

**Table S8.** Calculated unit cell parameters of $Im\bar{3}m$-Sn (Z=2) and $Fd\bar{3}m$-$Sn_8H_{30}$ (Z = 8 or 32).

| Pressure, GPa | a, Å (Sn) | V, Å$^3$ (Sn) | a, Å (Z=8) | V, Å$^3$ (Z= 8) | V, Å$^3$ (Z= 32) | V, Å$^3$/Sn |
|---|---|---|---|---|---|---|
| 160 | 3.0503 | 28.38 | 5.4737 | 164.0 | 656.0 | 20.50 |
| 180 | 3.0195 | 27.53 | 5.4115 | 158.47 | 633.88 | 19.81 |
| 200 | 2.9922 | 26.79 | 5.3556 | 153.61 | 614.44 | 19.20 |
| 210 | 2.9799 | 26.46 | 5.3294 | 151.37 | 605.48 | 18.92 |
| 220 | 2.9674 | 26.13 | 5.3044 | 149.25 | 597.0 | 18.65 |
| 230 | 2.955 | 25.82 | - | - | - | - |
| 240 | 2.944 | 25.52 | - | - | - | - |

**Table S9.** Experimental unit cell parameters of $C2/m$-$SnH_{14}$ (Z = 2) at different pressures, the experiment was done at PETRA III in 2020. "h1" denotes the first laser heating, "h3" – the third laser heating.

| Experiment and pressure | a, Å | b, Å | c, Å | β,° | V, Å$^3$ | V, Å$^3$/Sn |
|---|---|---|---|---|---|---|
| DAC D2, 193 GPa | 7.450 | 2.870 | 3.860 | 120.0 | 71.6 | 35.80 |
| DAC M2,<br>P02 (h1), 180 GPa | 7.756 | 2.871 | 3.797 | 118.34 | 74.42 | 37.21 |
| DAC M2,<br>P02 (h3), 185 GPa | 7.529 | 2.888 | 3.806 | 118.34 | 72.87 | 36.43 |
| DAC M2,<br>P03, 190-196 GPa | 7.542 | 2.876 | 3.815 | 118.74 | 72.57 | 36.28 |
| DAC M2,<br>P04, 201-208 GPa | 7.495 | 2.863 | 3.813 | 117.21 | 72.79 | 36.39 |
| DAC M2,<br>P05, 208-210 GPa | 7.482 | 2.854 | 3.790 | 117.74 | 71.66 | 35.83 |

**Table S10.** Calculated unit cell parameters of $C2/m$-$SnH_{14}$ at different pressures.

| Pressure, GPa | a, Å | b, Å | c, Å | (180 – β), ° | V, Å$^3$ | V, Å$^3$/Sn |
|---|---|---|---|---|---|---|
| 160 | 7.628 | 2.940 | 3.907 | 61.56 | 77.07 | 38.53 |
| 180 | 7.534 | 2.901 | 3.866 | 61.38 | 74.19 | 37.09 |
| 200 | 7.445 | 2.866 | 3.828 | 61.27 | 71.66 | 35.83 |
| 210 | 7.406 | 2.850 | 3.812 | 61.17 | 70.51 | 35.25 |
| 220 | 7.368 | 2.834 | 3.796 | 61.11 | 69.43 | 34.71 |
| 230 | 7.334 | 2.820 | 3.779 | 61.05 | 68.41 | 34.21 |
| 240 | 7.297 | 2.805 | 3.766 | 61.00 | 67.44 | 33.72 |

**Table S11.** Calculated unit cell parameters of cubic $SnH_2$ (prototype is $Fm\bar{3}m$-$YH_2$) and $SnH_3$ (prototype is $Fm\bar{3}m$-$YH_3$).

| Pressure, GPa | *Fm*-3*m*-$Sn_4H_{12}$ ($SnH_3$) | | | *Fm*-3*m*-$Sn_4H_8$ ($SnH_2$) | | |
|---|---|---|---|---|---|---|
| | a, Å (Z=4) | V, Å$^3$ (Z=4) | V, Å$^3$/Sn | a, Å (Z=4) | V, Å$^3$ (Z=4) | V, Å$^3$/Sn |
| 160 | 4.232 | 75.79 | 18.95 | 4.188 | 73.43 | 18.36 |
| 180 | 4.186 | 73.34 | 18.33 | 4.139 | 70.95 | 17.74 |
| 200 | 4.146 | 71.26 | 17.81 | 4.098 | 68.82 | 17.20 |
| 210 | 4.126 | 70.26 | 17.56 | 4.078 | 67.81 | 16.95 |
| 220 | 4.107 | 69.26 | 17.32 | 4.058 | 66.85 | 16.71 |

**Table S12.** Calculated unit cell parameters of $Fm\bar{3}m$-$SnH_5$, $Fm\bar{3}m$-$SnH_6$, $Fm\bar{3}m$-$SnH_7$ (prototype is $YMg_7$), $Fm\bar{3}m$-$SnH_8$ (prototype is $UH_8$), $F\bar{4}3m$-$SnH_9$ (prototype is $PrH_9$) and $Fm\bar{3}m$-$SnH_{10}$ (prototype is $LaH_{10}$).

| Pressure, GPa | V, Å$^3$ /Sn ($SnH_5$) | V, Å$^3$ /Sn ($SnH_6$) | V, Å$^3$ /Sn ($SnH_7$) | V, Å$^3$ /Sn ($SnH_8$) | V, Å$^3$ /Sn ($SnH_9$) | V, Å$^3$ /Sn ($SnH_{10}$) |
|---|---|---|---|---|---|---|
| 180 | 21.50 | 26.54 | 27.00 | 26.32 | - | - |
| 200 | 20.86 | 25.68 | 26.17 | 25.56 | 26.93 | 28.52 |
| 210 | 20.56 | 25.38 | 25.82 | 25.20 | 26.54 | 28.10 |
| 220 | 20.30 | 25.01 | 25.45 | 24.84 | 26.17 | 27.70 |
| 230 | - | - | - | - | 25.80 | 27.32 |
| 240 | - | - | - | - | 25.47 | 26.94 |

## 4. Raman spectra

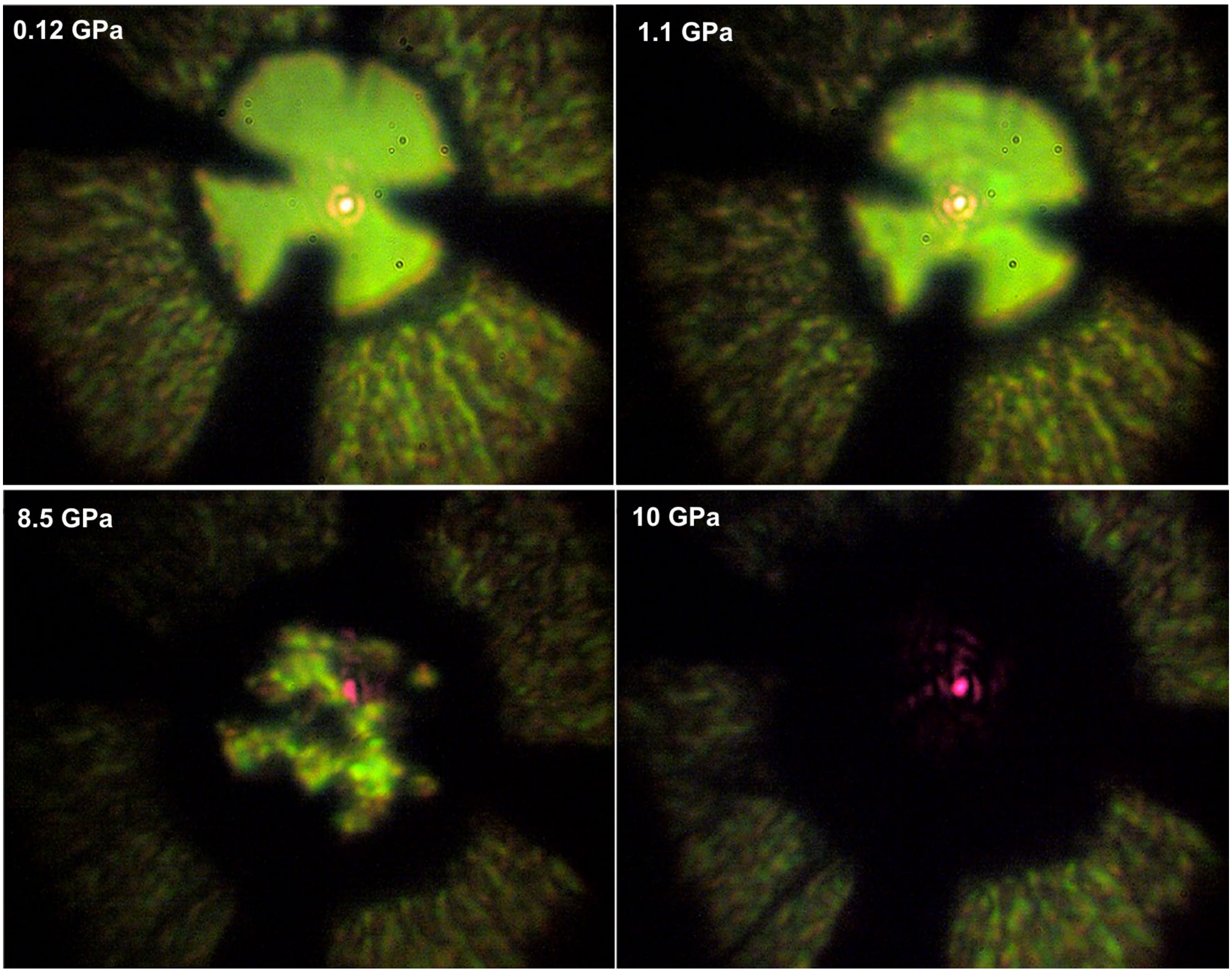

**Figure S7**. Metallization of molecular stannane compressed in a diamond anvil cell to 10 GPa. A cryogenic loading of the DAC was used.

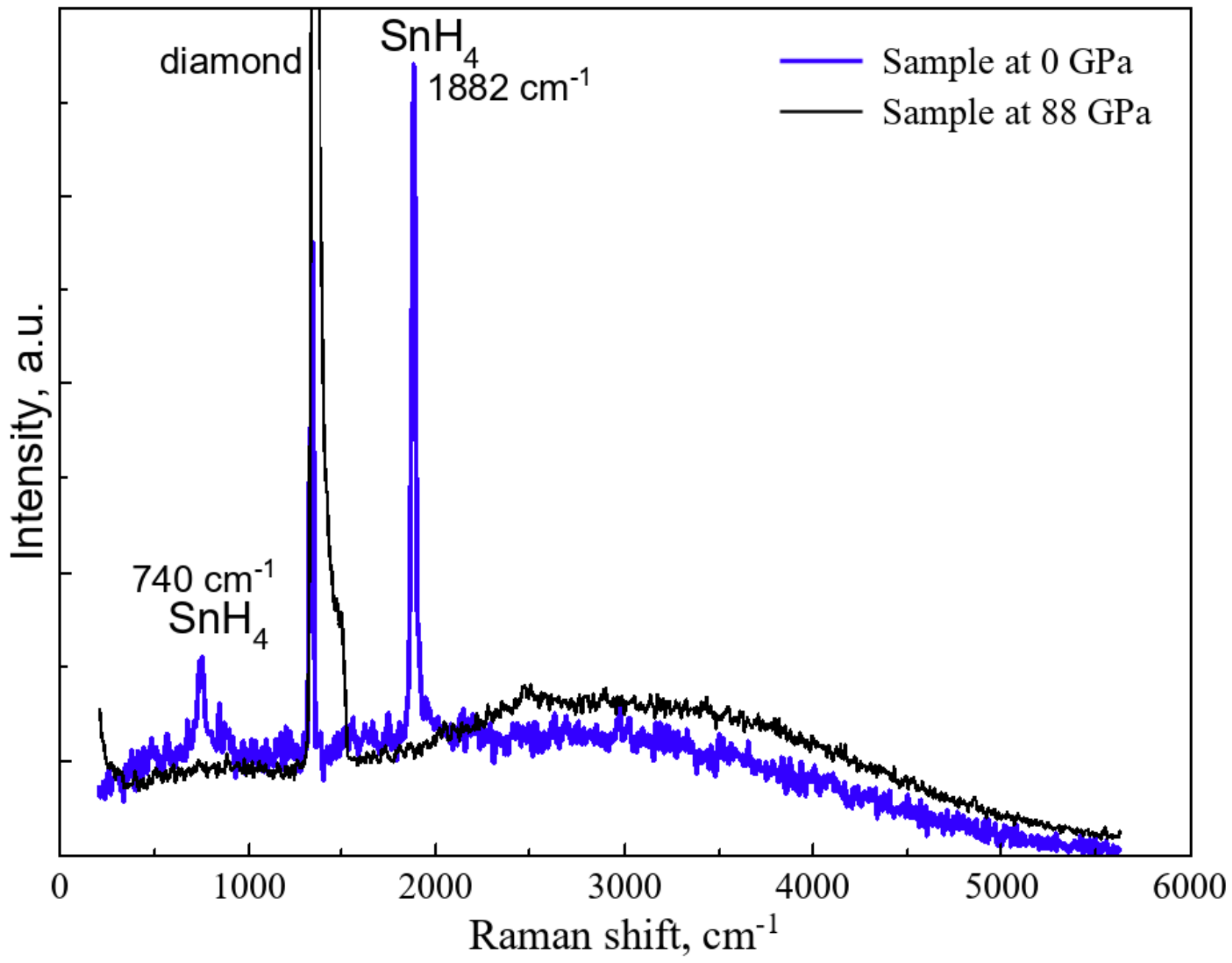

**Figure S8.** Raman spectrum of stannane ($SnH_4$) immediately after loading into the diamond cell (0 GPa, blue line) and after metallization (at 88 GPa, black line). The wavelength of the exciting laser is 633 nm.

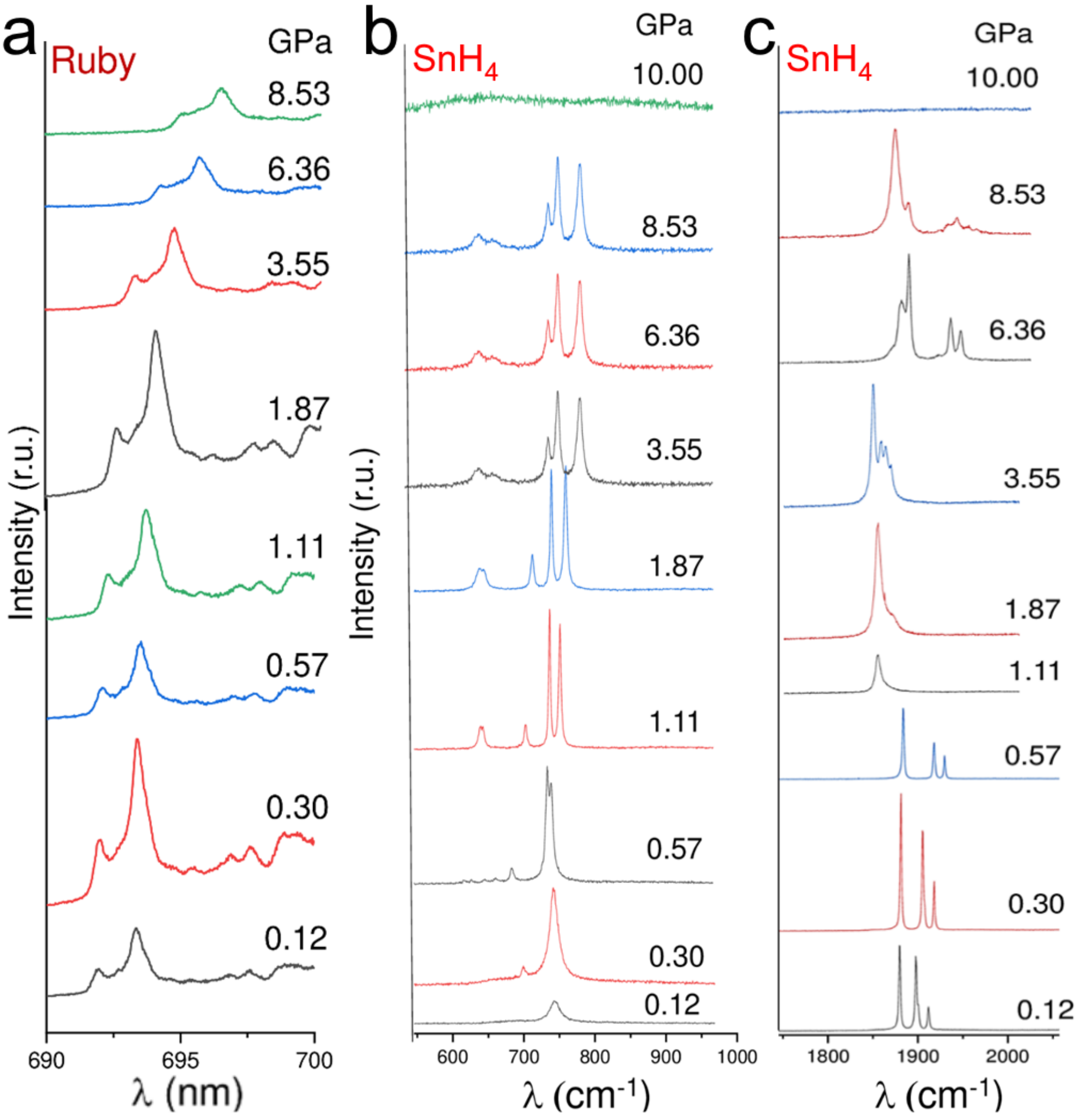

**Figure S9**. Raman spectra of stannane below 10 GPa measured at 157-164 K. (a) The ruby fluorescence peak and its changes with pressure. (b) Raman spectra of $SnH_4$ in 600-1000 $cm^{-1}$ range at different pressures. There are signs of phase transitions in $SnH_4$ as the pressure increases. (c) Raman spectra of $SnH_4$ in 1700-2100 $cm^{-1}$ range at different pressures. Above 10 GPa, the Raman peaks of stannane disappear. The wavelength of the exciting laser is 633 nm.

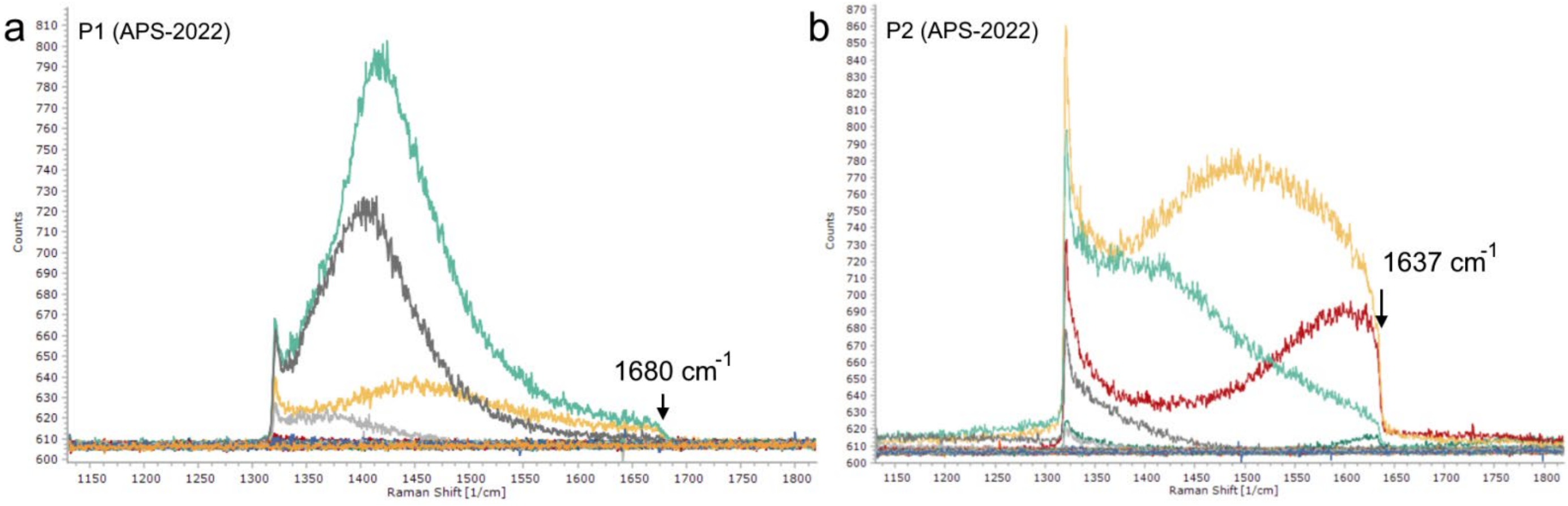

**Figure S10.** Raman spectra of the diamond anvils in the region of C-C vibrations, obtained at APS-2022. Different curves correspond to different points on the diamond culet.

## 5. Electron and phonon band structures of tin hydrides

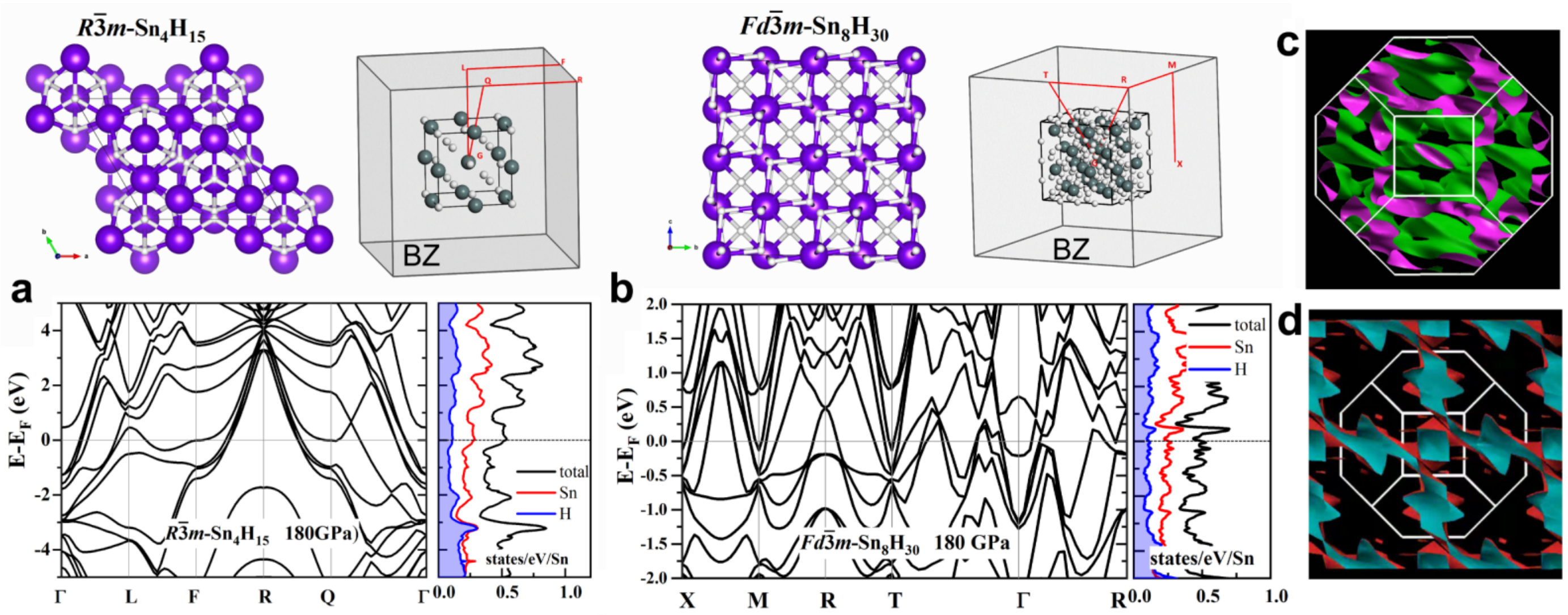

**Figure S11.** Crystal and electron band structures of (a) $R\bar{3}m$-$Sn_{12}H_{45}$ ($Sn_4H_{15}$) and (b) $Fd\bar{3}m$-$Sn_8H_{30}$ calculated at 180 GPa (PAW PBE). (c, d) Sections of the Fermi surface of $Fd\bar{3}m$-$Sn_8H_{30}$ showing its open character (XCrySDen). Bands correspond to $E_{min}$ = 15.39 eV, $E_{max}$ = 18.07 eV (c) and $E_{min}$ = 16.36 eV, $E_{max}$ = 18.11 eV (d) intervals.

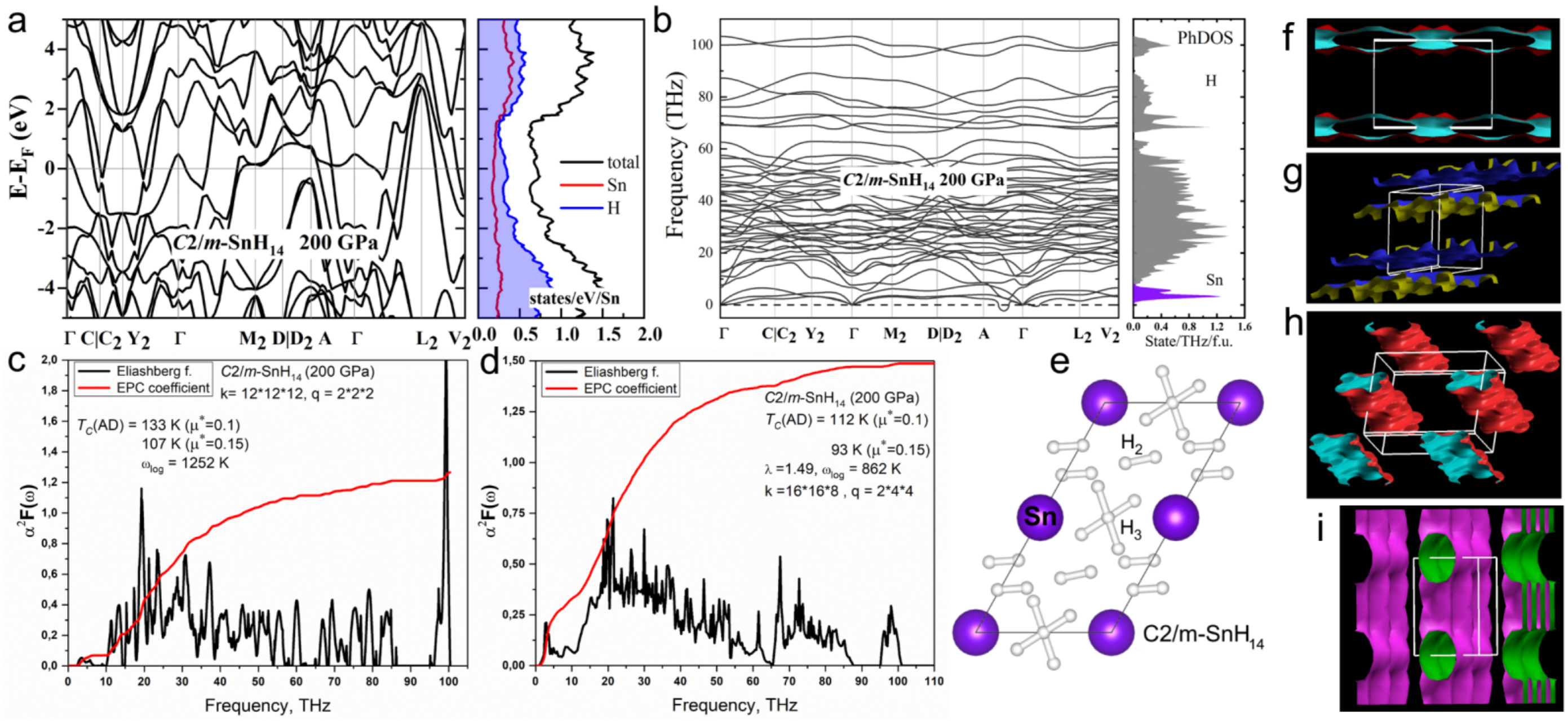

**Figure S12.** Electron (a), phonon (b) band structures and crystal structure (e), as well as Eliashberg functions (c, d) of $C2/m$-$SnH_{14}$ at 200 GPa. Red lines in (c, d) correspond the dependence of the electron-phonon coupling coefficient on the frequency. (f-i) Sections of the Fermi surface of $C2/m$-$SnH_{14}$ at 200 GPa showing its open character (XCrySDen).

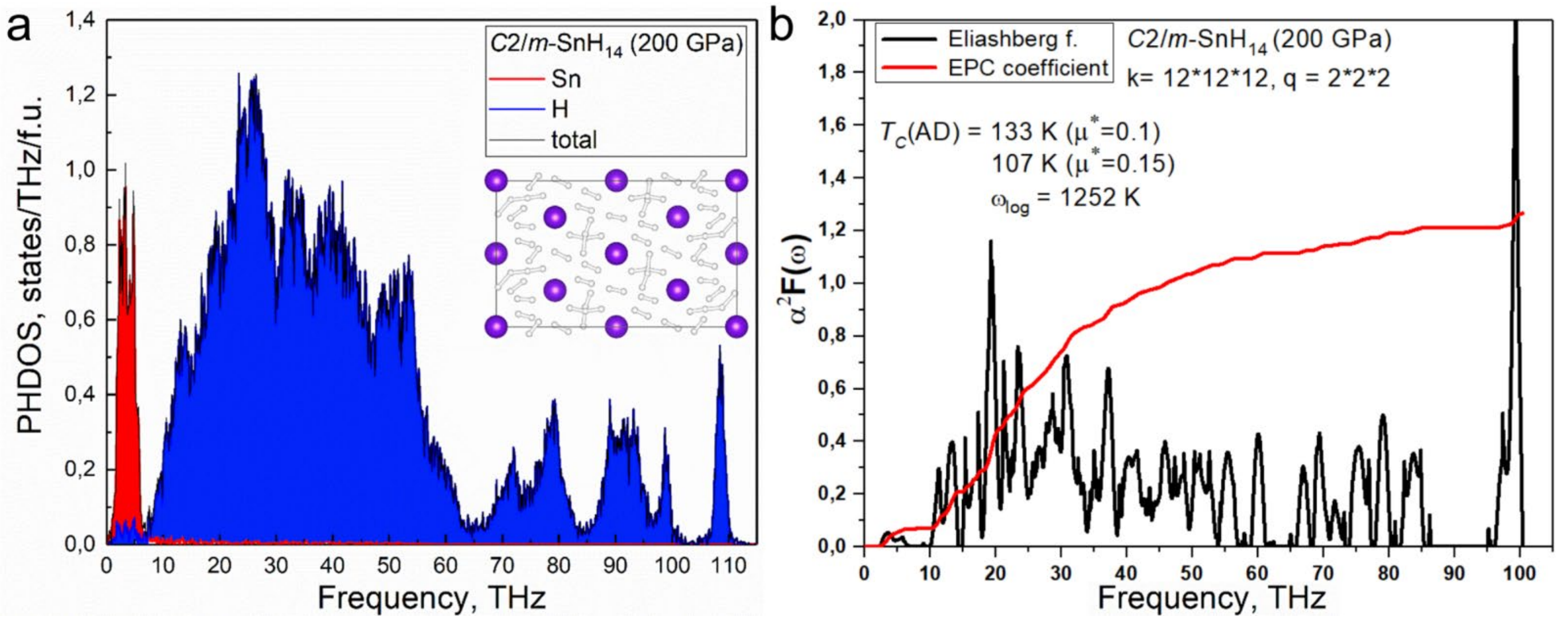

**Figure S13**. (a) Anharmonic density of phonon states at 200 GPa and 300 K obtained using molecular dynamics with MTP potentials, and (b) harmonic Eliashberg function with parameters of superconductivity of $C2/m$-$SnH_{14}$.

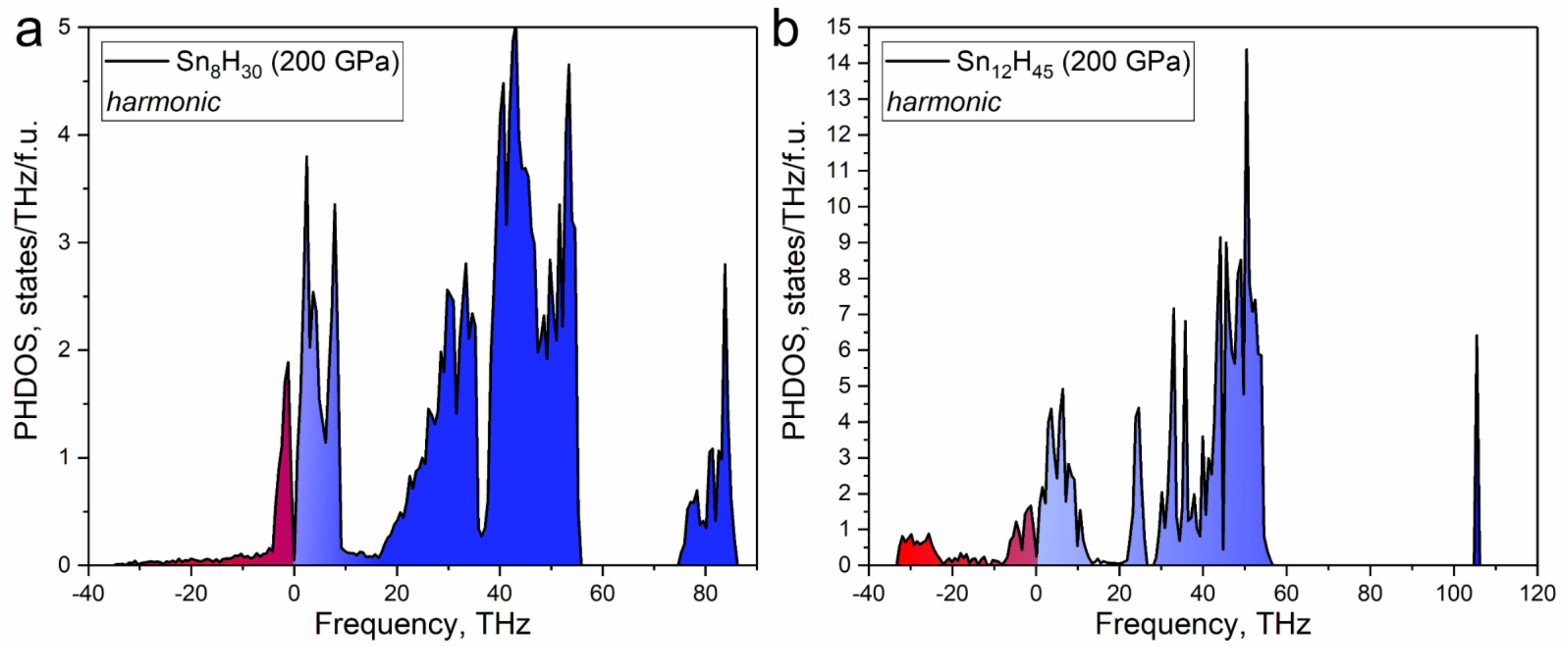

**Figure S14**. Harmonic density of phonon states of (a) $Fd\bar{3}m$-$Sn_8H_{30}$ (f.u. = $Sn_8H_{30}$) and (b) $R\bar{3}m$-$Sn_{12}H_{45}$ (f.u. = $Sn_{12}H_{45}$) at 200 GPa and 0 K obtained using VASP [10-12] and Phonopy [13,14] codes. The imaginary part is marked in red.

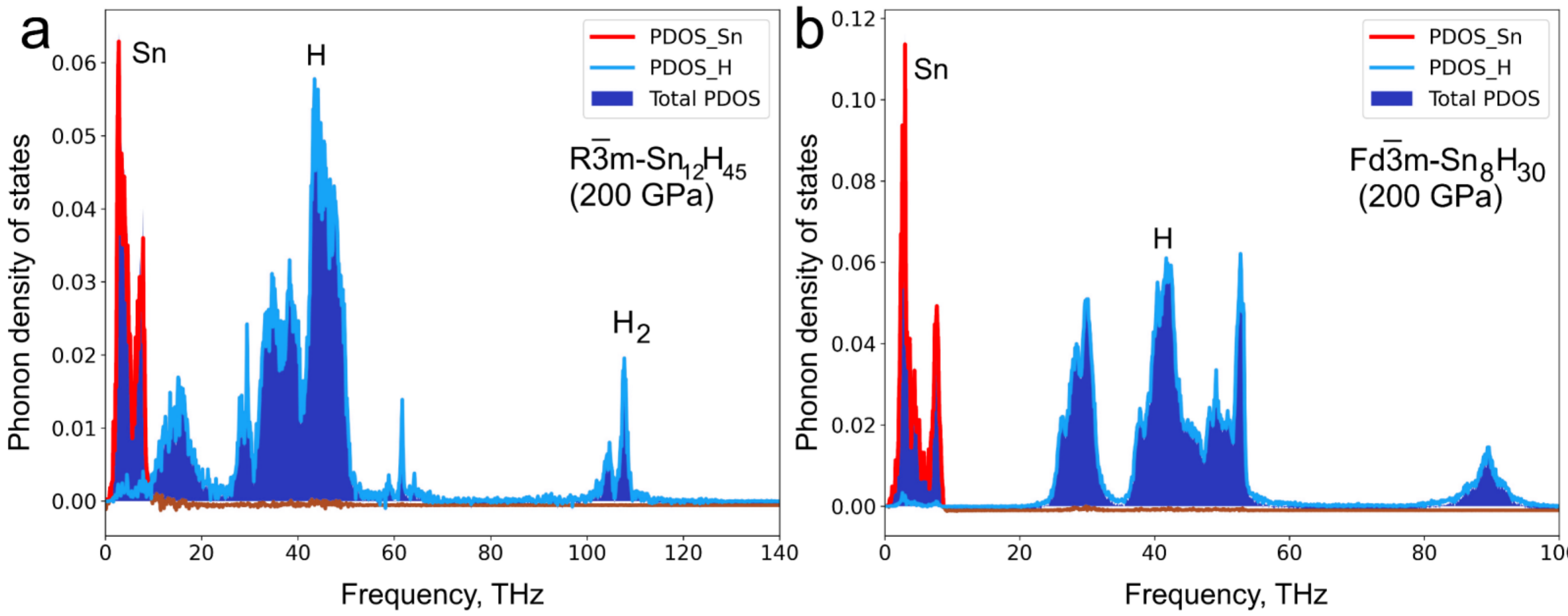

**Figure S15**. Anharmonic density of phonon states (integral of PDOS is normalized to 1, $THz^{-1}$) of (a) $R\bar{3}m$-$Sn_{12}H_{45}$ and (b) $Fd\bar{3}m$-$Sn_8H_{30}$ at 200 GPa and 300 K obtained using molecular dynamics with MTP potentials.

## 6. Transport and superconducting properties

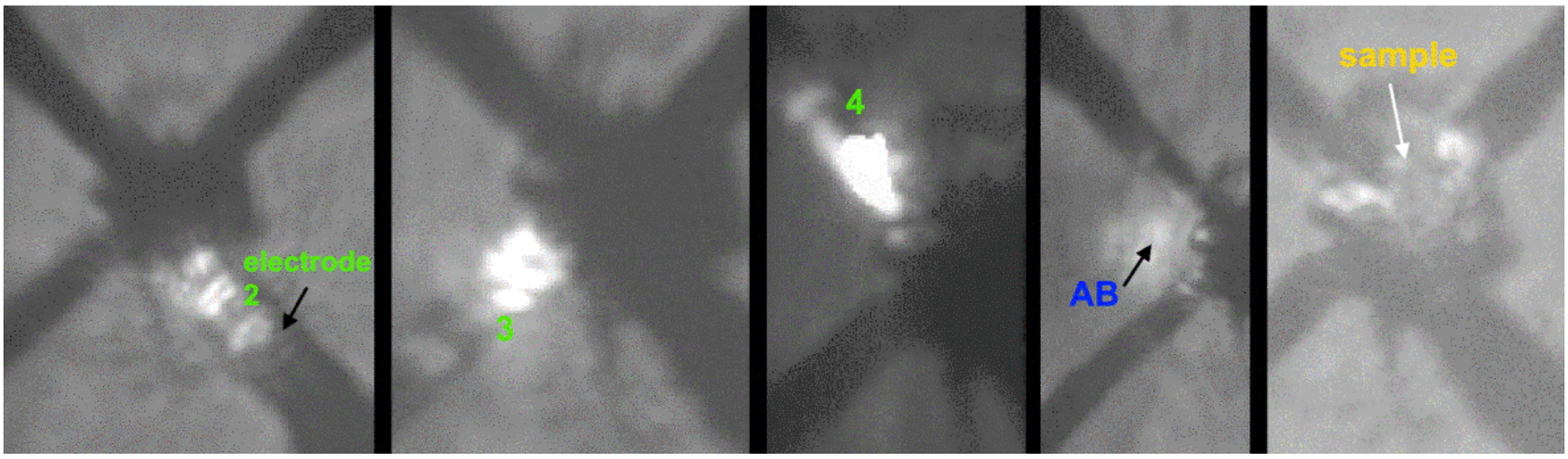

**Figure S16**. A series of optical photographs of a Sn hydride sample on a 50-µm diameter diamond anvil with four sputtered electrical contacts. Arrows and numbers show electrodes, ammonia borane (AB), and the sample.

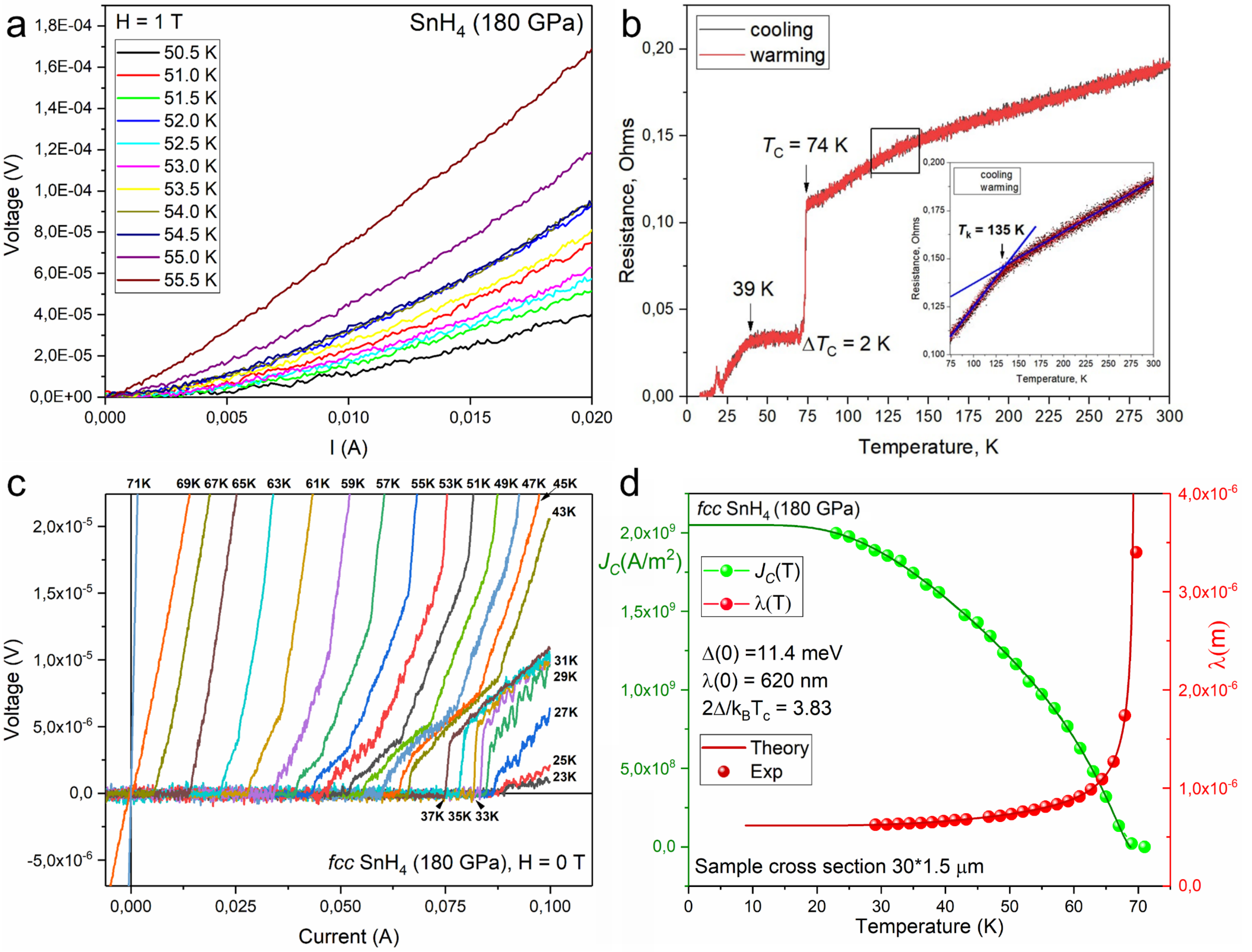

**Figure S17**. (a) Voltage-current (V-I) characteristic of the cubic $SnH_4$ sample at 180 GPa in a 1T magnetic field (H) in the temperature range of 50.5 - 55.5 K. The moment of transition to the superconducting state is weakly expressed probably due to heating of the sample (there is a quadratic ~$T^2$ component in V(I)). (b) Temperature dependence of the electrical resistance in a multiphase Sn hydride sample at 190 GPa (3rd laser heating). The inset shows a previously discovered anomaly (kink) at 135 K. Signs of a second, probable, superconducting transition are observed at ~39 K. (c) Voltage-current (V-I) characteristic of the cubic $SnH_4$ sample at 180 GPa in the absence of magnetic field in the temperature range of 23 - 71 K. (d) Temperature dependence of critical current density (green balls) and penetration depth (red balls) and fitting curves based on the results of Talantsev et al. [26,27].

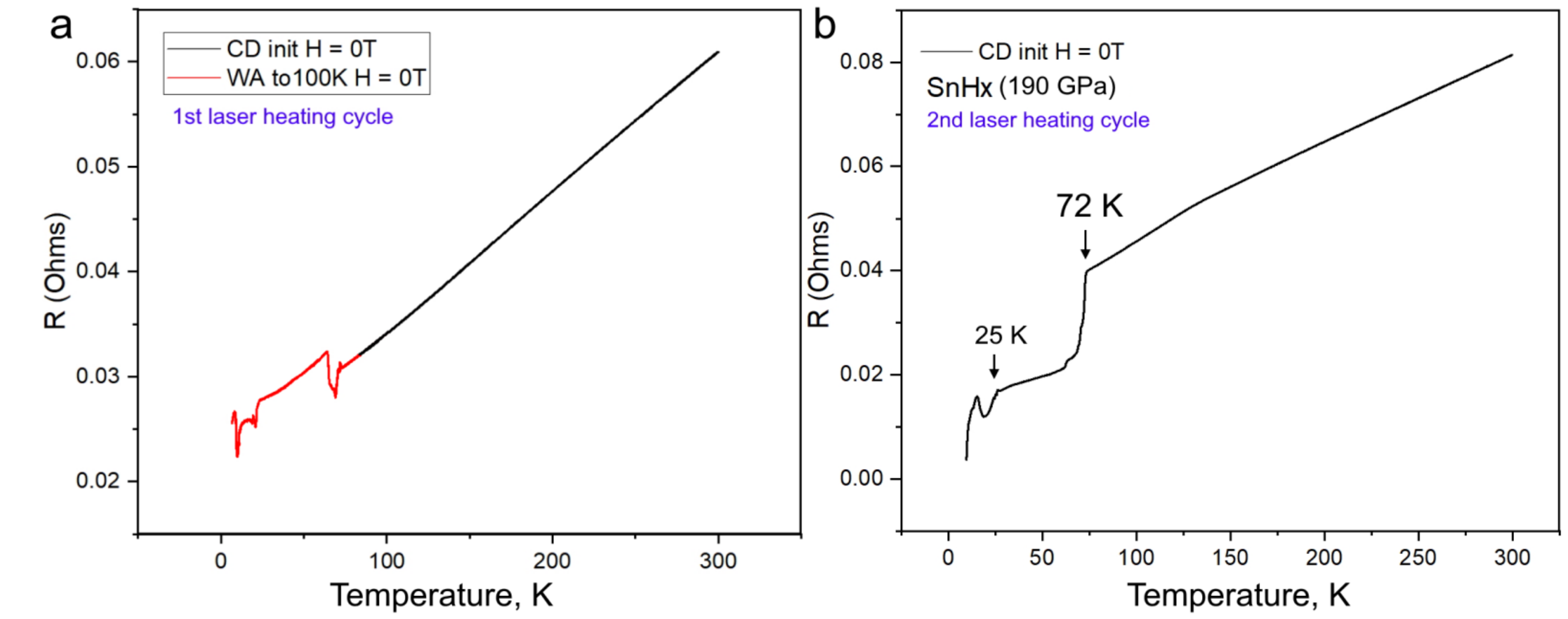

**Figure S18.** Temperature dependence of the electrical resistance for insufficiently heated Sn samples at 190 GPa. (a) The first cycle of laser heating. Only a local resistance anomaly is observed in the region of the expected superconducting transition (70 – 75 K). (b) The second cycle of laser heating. Due to an increase in the concentration of superconducting phase in the sample, a partial resistive transition at 72 K was observed. The third laser heating cycle results in the curve shown in Figure 17b.

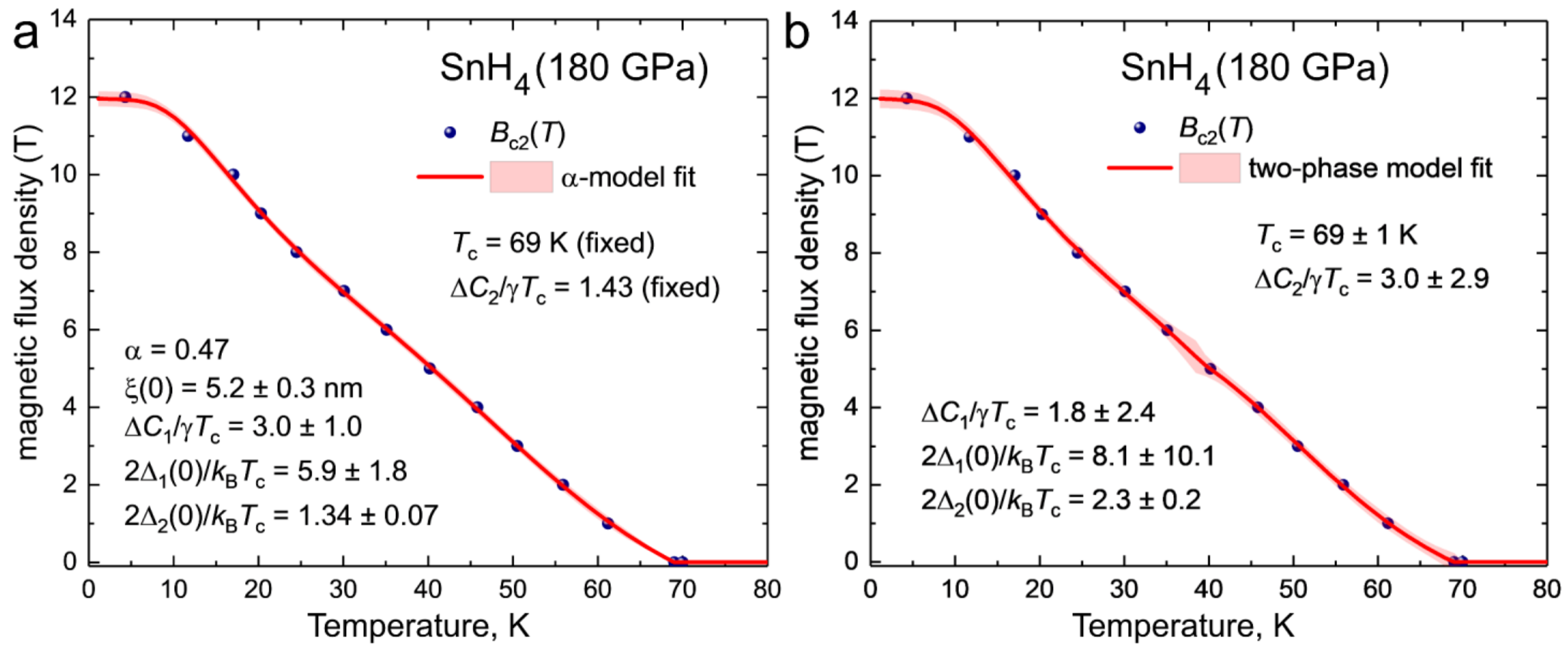

**Figure S19**. Possible explanation of the quasilinear dependence of the upper magnetic field on temperature $B_{C2}$(T) for cubic $SnH_4$ using two-gap models: (a) α-model, (b) two-phase model. The critical temperature $T_C$ = 69 K is taken at the first point where *R(T)* < 0. The analysis was carried out by Dr. Evgueni Talantsev (M.N. Mikheev Institute of Metal Physics, Ural Branch of RAS).

Estimation of the upper critical magnetic field of $Fd\bar{3}m$-$Sn_8H_{30}$ in the clean limit by a simplified model for hydrides, suggested in Ref. [41], gives:

$$B_{C2}^{cl}(0) = 2.45 \times 10^{-4} T_C^2 (1+\lambda)^2, \qquad \text{(S7)}$$

where $T_C$ = 72 K, and λ is the electron-phonon interaction parameter which varies from 1.24 (harmonic approximation) to 2.98 (anharmonic approximation). Depending on the λ value, Eq. (4) gives $B_{C2}(0)$ = 6.4 – 20.3 T, which is in acceptable agreement with the experimental data (14-16 T). From eq. (S7), considering the anomalously linear dependence *R(T),* it follows that the value of the electron-phonon interaction parameter in Sn tetrahydride is quite large (λ ~ 2.5).

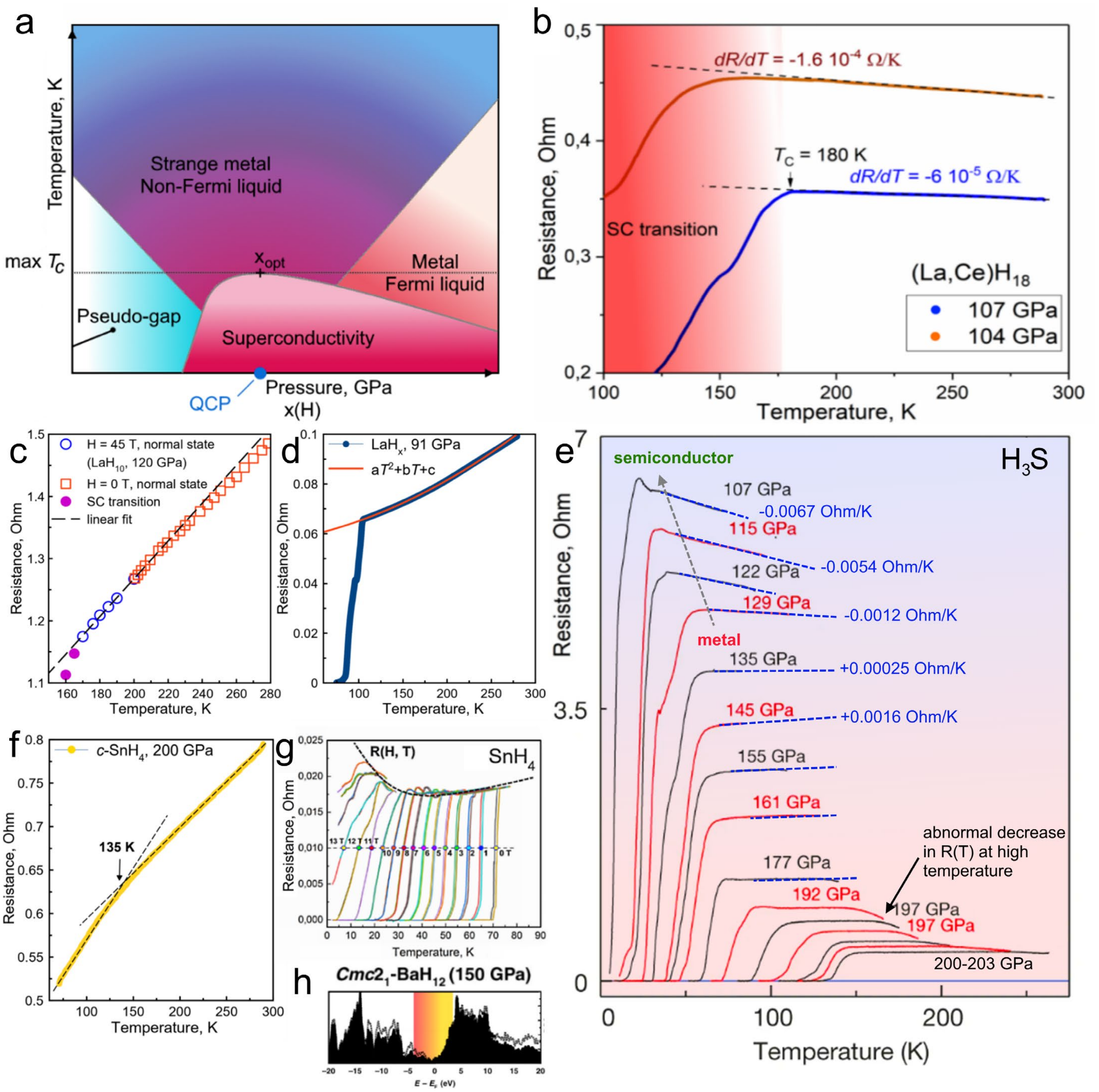

**Figure S20**. Examples of anomalous behavior of superhydrides in the normal resistivity state and their hypothetical phase diagram. (a) Qualitative *P-T* phase diagram of compressed hydrides, constructed by analogy with cuprates. Similar to doping level in cuprates, pressure (P) plays the same role in hydrides. (b) Anomalous linear decrease in electrical resistance ($d\rho/dT < 0$) with increasing temperature in (La, Ce)$H_9$ [42]. (c) Anomalous temperature dependence (close to T-linear) of electrical resistance in $LaH_{10}$, observed at 120 GPa after suppression of the superconducting state by a magnetic field of 45 T [43]. (d) The Fermi liquid behavior of *R(T)* in $LaH_x$ at 91 GPa. The electrical resistance can be approximated by a quadratic function [42]. (e) Gradual transition of sulfur hydride ($H_3S$) from a metallic (non-Fermi liquid) phase to a semiconducting pseudo-gap phase (below 135 GPa) with the inversion of sign of the temperature coefficient of resistance and the disappearance of superconductivity below 107 GPa [44]. (f) Anomalous reproducible kink of *R(T)* in *fcc* $SnH_4$ which cannot be explained in terms of the Fermi liquid model. (g) Growth of electrical resistance of cubic $SnH_4$ at low temperatures after suppression of superconductivity by magnetic field. This is a probable manifestation of pseudo-gap phase. (h) Pronounced pseudo-gap in the electronic structure of a superconducting molecular $Cmc2_1$-$BaH_{12}$ at 150 GPa [45].

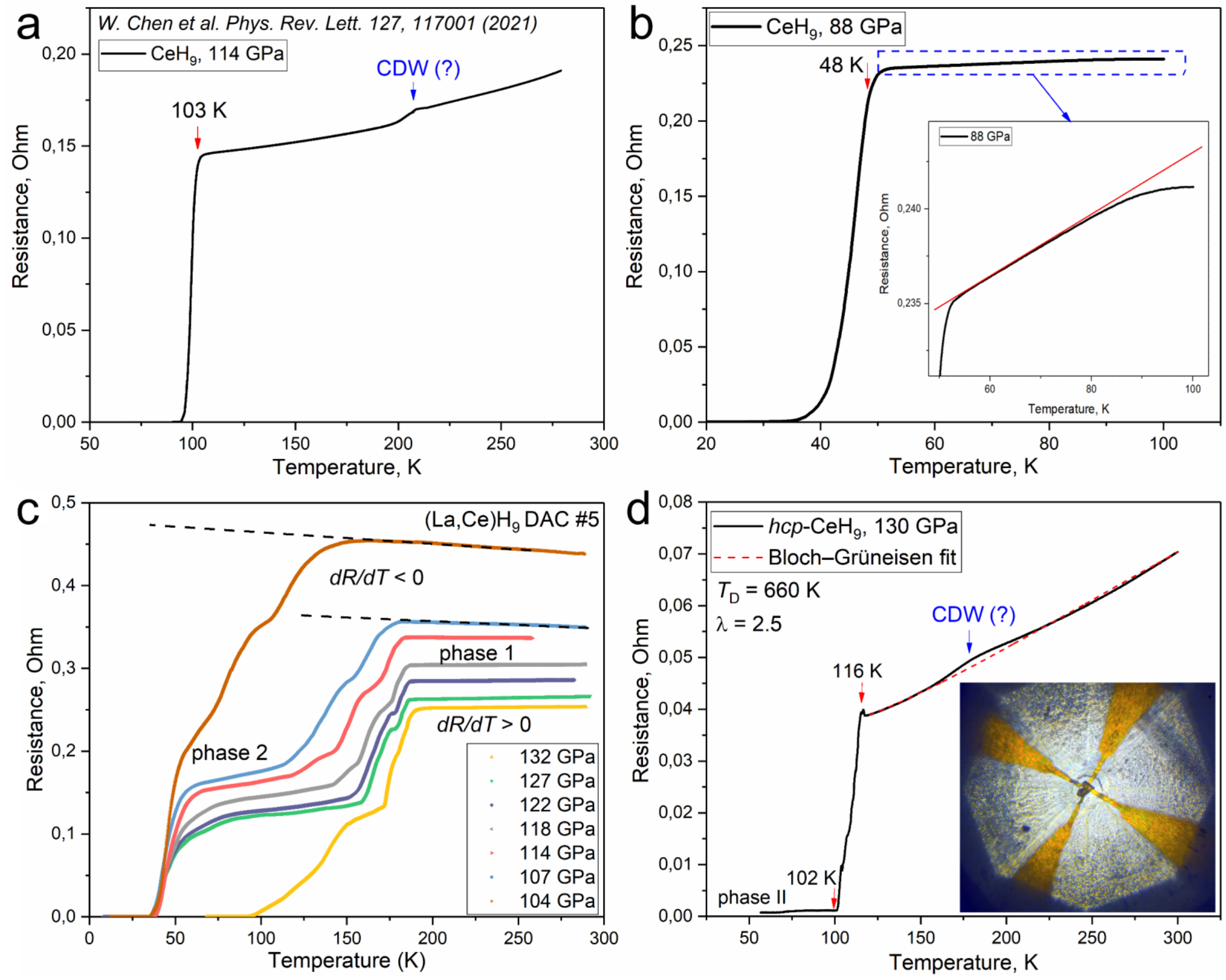

**Figure S21.** Examples of anomalous behavior of electrical resistance in samples of cerium and lanthanum-cerium hydrides. (a) Temperature dependence *R(T)* in $CeH_9$ at 114 GPa, (b) the same dependence at 88 GPa [46]. In both cases, *R(T)* cannot be interpolated using the Bloch–Grüneisen formula [47,48]. (c) Reversal of the sign of the temperature coefficient of electrical resistance during decompression of $(La,Ce)H_9$ [42]. A similar effect is observed in *bcc* $H_3S$ with decreasing pressure [44]. (d) Reproducible anomaly of the *R(T)* dependence in $P6_3/mmc$-$CeH_9$. This experiment was carried out in Moscow (IC RAS) in 2022. CDW means the charge density wave, which is one of the possible reasons of such anomalies in $CeH_9$ at high pressure.

**Table S13.** Parameters of the normal resistivity state of the cubic $SnH_4$ found in the study of the magnetoresistance. The mobility of electrons and their relaxation time were calculated from the quadratic part of the ρ(*B*) dependence and, in order of magnitude, they correspond to the parameters of metals (e.g., Sr, Ba, Pb…) [49].

| **Temperature, K** | **Crit. magnetic field ($B_{cr}$), T** | **Cyclotron time ($\tau_c$), $s^{-1}$** | **Electron mobility ($\rho=\mu^2B^2$), $m^2/s\times V$** | **Electron relaxation time, ($\tau = \mu\times m_e/e$), $s^{-1}$** |
|---|---|---|---|---|
| 75* | 5 | 1.1e-12 | 0.033 | 1.9e-13 |
| 80 | 6 | 9.5e-13 | 0.035 | 2e-13 |
| 100* | 7.5 | 7.6e-13 | 0.022 | 1.25e-13 |
| 145.5 | 12 | 4.7e-13 | 0.017 | 9.45e-14 |
| 216 | 14 | 4e-13 | 0.015 | 8.53e-14 |

*PPMS measurements

It should also be noted that the linear behavior of the magnetoresistance (MR) is also typical for polycrystalline metals with an open Fermi surface, for example, for Li, Cu, Ag, Au, etc. [50]. This behavior was explained in 1959 by Lifshits and Peschanskii [51], who took into account the shape of the Fermi surface and the anisotropy of MR. When averaged over all directions ($\Delta\theta \sim H_0/H$) and open sections of the Fermi surface (where $\rho \sim H^2$) for polycrystals, the MR acquires a linear dependence on the magnetic field [51]. The open topology of the Fermi surface is confirmed by DFT calculations (Supporting Figures S11-S12).

## 7. Eliashberg functions

Calculations of the anharmonic correction to the critical temperature of superconductivity in this work are based on the constant DOS approximation [52], which is applicable if the density of electronic states $N(E)$ in the vicinity of the Fermi energy has no pronounced features (e.g., van Hove singularities [53]) and can be approximately represented as a constant. In this case, the main contribution to the shape of Eliashberg function is made by the phonon density of states (denoted here for convenience as $F(\omega)$), and an auxiliary function of electron contribution $\alpha^2(\omega) = \alpha^2F(\omega)/F(\omega)$ is close to a constant. The exceptions are those frequency zones where both functions $\alpha^2F(\omega)$ and $F(\omega) \sim 0$, and where the calculation of their ratio is not sufficiently stable and accurate.

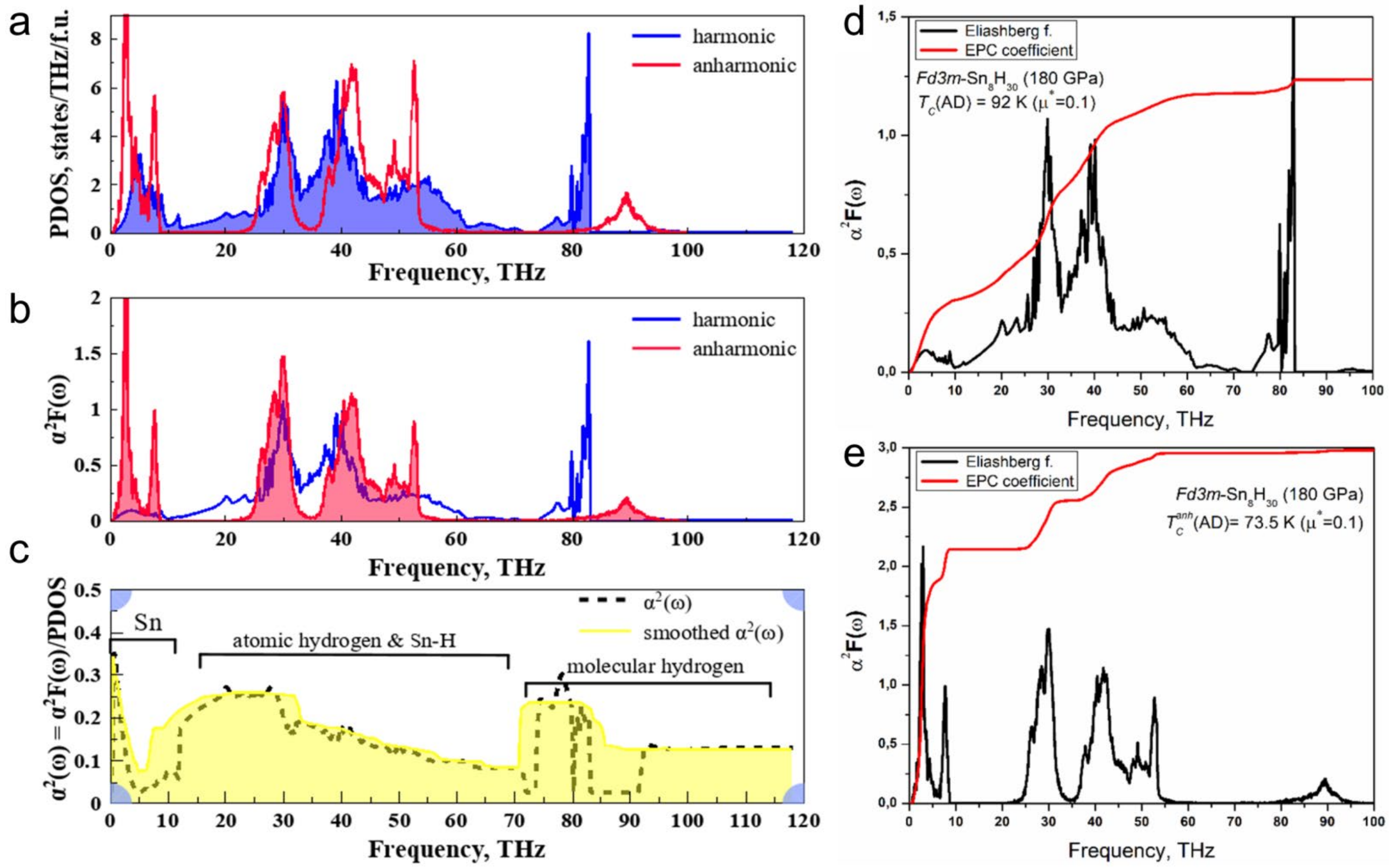

**Figure S22.** Harmonic and anharmonic spectral functions of $Fd\bar{3}m$-$Sn_8H_{30}$ at 180 GPa, f.u. = $Sn_8H_{30}$. (a) Harmonic (blue) and anharmonic (red) density of phonon states at 100 K. (b) Harmonic (blue) and anharmonic (red) Eliashberg functions. (c) Smoothed (yellow) and original (black dotted line) function of contribution of the density of electronic states (formally, $\alpha^2(\omega)$). (d) Superconducting state parameters of $Fd\bar{3}m$-$Sn_8H_{30}$ at 180 GPa in harmonic (d) and anharmonic (e) approximations. Attention is drawn to the presence of high-intensity soft modes in molecular dynamics calculations of $Sn_8H_{30}$ which can lead to the emergence of the charge density waves (CDW).

In this work, the procedure for calculating the Eliashberg anharmonic functions of Sn hydrides consisted of several steps:

1. Calculation of the harmonic Eliashberg function and phonon density of states in Quantum ESPRESSO [17,18] using the tetrahedral method [54].

2. Calculation of the anharmonic phonon spectrum using the molecular dynamics [55] at a temperature near the superconducting transition ($T_C$). Here it is a temperature of 100 K.

3. Equalization of unit cell (N - is number of atoms in the formula unit) by calibration of integrals

$$\int F_{harm}(\omega)d\omega = \int F_{anh}(\omega)d\omega = 3N \qquad \text{(S8)}$$

4. Calculation of the contribution of the electron density of states to the Eliashberg function for the harmonic case: $\alpha^2(\omega) = \alpha^2F(\omega)/F(\omega)$.

5. Reduction of the function $\alpha^2(\omega) \geq 0$ to the same mesh on the energy scale that was used to calculate the anharmonic phonon spectrum by linear interpolation. All obtained negative values $\alpha^2(\omega) < 0$ were set to zero.

6. Smoothing the obtained $\alpha^2(\omega)$ to eliminate the influence of "empty" frequency zones where both $\alpha^2F(\omega)$ and $F(\omega) \sim 0$. In this work, we used the percentile filter (points of window is 140, percentile is 70) in the OriginLab 2017 [56] program.

7. Calculation of the Eliashberg anharmonic function as the product of the electronic contribution and the anharmonic density of phonon states $\alpha^2(\omega)F_{anh}(\omega)$.

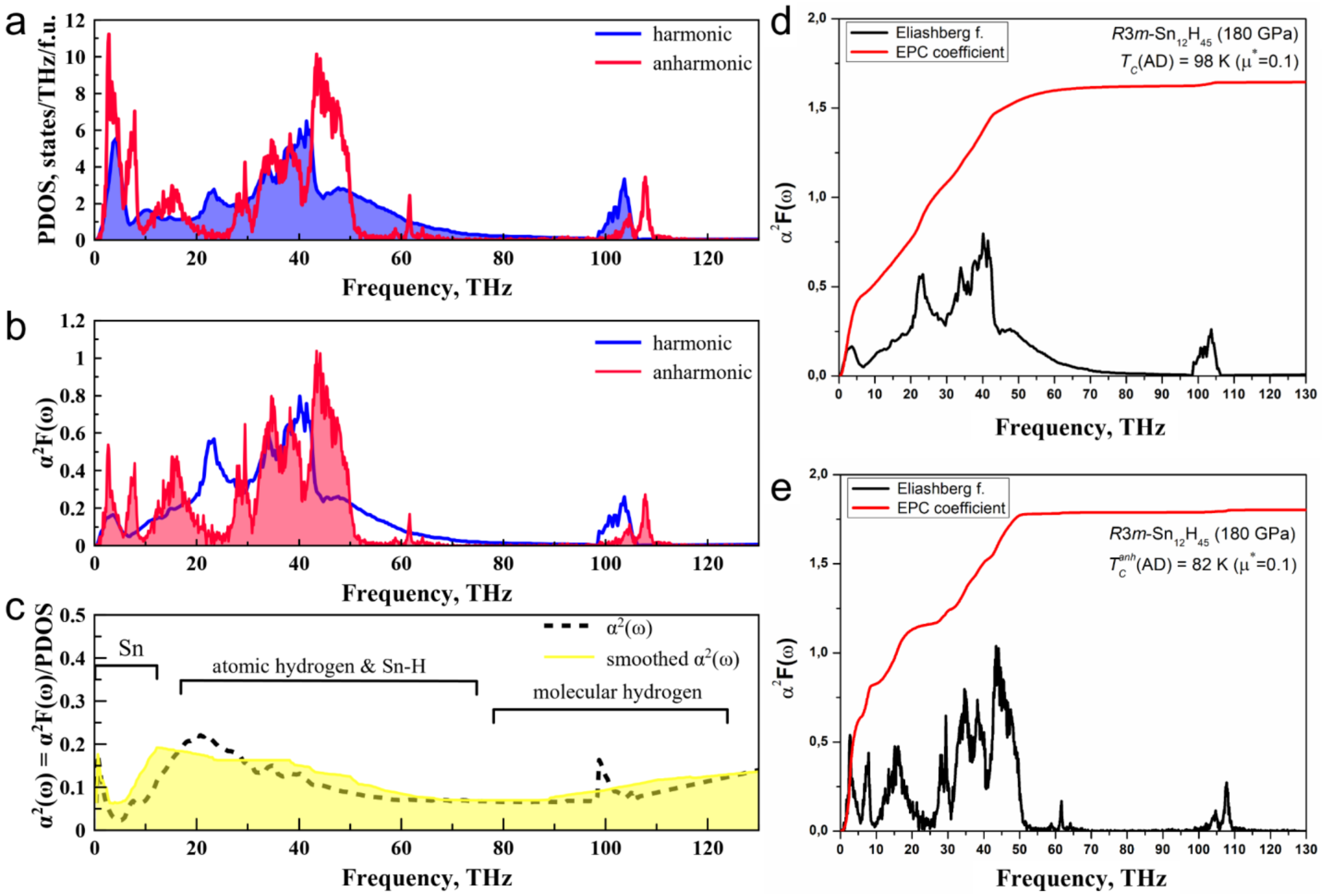

**Figure S23.** Harmonic and anharmonic spectral functions of $R\bar{3}m$-$Sn_{12}H_{45}$ at 180 GPa, f.u. = $Sn_{12}H_{45}$. (a) Harmonic (blue) and anharmonic (red) density of phonon states at 100 K. (b) Harmonic (blue) and anharmonic (red) Eliashberg functions. (c) Smoothed (yellow) and original (black dotted line) function of contribution of the density of electronic states (formally, $\alpha^2(\omega)$). (d) Superconducting state parameters of $R\bar{3}m$-$Sn_{12}H_{45}$ at 180 GPa in harmonic (d) and anharmonic (e) approximations.

Some details of the obtained results require discussion. First, the significant growth of the electron-phonon coupling (EPC) coefficient $\lambda$ from 1.24 to $\approx 3$ in the anharmonic approximation attracts attention. The opposite effect is usually observed, which is associated with acoustic phonons hardening [57]. However, in our case, within the harmonic approximation, $Sn_8H_{30}$ is not dynamically stable at 180-200 GPa, and has a series of imaginary acoustic modes that we did not take into account in our calculations of the superconducting state (see Figure S14). Considering anharmonic corrections, these imaginary modes, roughly speaking, move to the real acoustic part of the phonon spectrum, and cause a sharp maximum in $F(\omega)$ around 0-7 THz (Figure S15, S22), as well as lead to an increase in the EPC coefficient.

The intermediate range of frequencies, that corresponds to the atomic hydrogen vibrations (10-60 THz), is subjected to the aforementioned phonons hardening. As a result, the 10-25 THz frequency zone becomes almost empty, and the modes, that were previously in this region, cause a significant increase in the intensity of the $\alpha^2F(\omega)$ and $F(\omega)$ functions in the range 25-55 THz (Figures S22a, b).

In the range of molecular hydrogen vibrations, anharmonic interactions of $H_2$ molecules with the framework of the atomic H-sublattice and with Sn atoms lead to the splitting and smoothing of the phonon density of states in the range of 75-95 THz (Figure S22a, b). There is also a strong high-frequency shift of these modes, which is characteristic of the anharmonic approximation (for example, $H_3S$ [58], high-frequency region).

Thus, despite the pronounced phonons hardening for hydrogen modes, the effect of soft modes of Sn atoms (compare with *hcp*-$NdH_9$ [59]) leads to an unexpected increase in the electron-phonon interaction strength by more than two times and a sharp decrease in the logarithmically averaged frequency ($\omega_{log}$) of $Sn_8H_{30}$ at 180 GPa. As a result, the critical temperature ($T_C$) is reduced by about 20 K compared to harmonic calculations, which is in qualitative agreement with the results obtained for other hydrides (e.g. $AlH_3$ [60], $YH_6$ [16,61], $LaH_{10}$ [62,63], $H_3S$ [44]).

Summing up, we can say that the properties of cubic $SnH_4$ are unusual: this compound exhibits much higher electron-phonon interaction strength ($\lambda$ = 2.5 – 3) than one would expect from a hydride with such a low hydrogen content. $SnH_4$ can rather be described as a covalent superconductor, unlike lanthanum ($LaH_{10}$) and yttrium ($YH_6$) superhydrides.

**Table S14.** Superconducting properties of $Fd\bar{3}m$-$Sn_8H_{30}$ at 180 GPa. AD – corresponds to calculations made by Allen-Dynes formula [20]. Calculations of elastic properties give the following estimates: $T_\theta$ = 380 K, $\omega_{log}$ ≈ 315 K (see below).

| **Parameter** | **λ** | **$\omega_{log}$, K** | **$\omega_2$, K** | **$T_C$(AD), K (μ*=0.1)** |
|---|---|---|---|---|
| Harmonic* | 1.24 | 892 | 1664 | 92.3 |
| Anharmonic | 2.98 | 282 | 1016 | 73.5 |

*Coherence length in this case is about 5 nm, the Ginzburg-Landau parameter is κ = 43, and the expected lower critical field is $H_{C1}$ = 9 mT.

**Table S15.** Superconducting properties of $C2/m$-$SnH_{14}$ at 200 GPa. AD – corresponds to calculations made by Allen-Dynes formula [20]. Calculations of elastic properties give the following estimates: $T_\theta$ = 1404 K, $\omega_{log}$ ≈ 1160 K (see below).

| **Parameter** | **k-mesh** | **q-mesh** | **λ** | **$\omega_{log}$, K** | **$\omega_2$, K** | **$T_C$(AD), K (μ*=0.15-0.1)** |
|---|---|---|---|---|---|---|
| Harmonic | 12×12×12 | 2×2×2 | 1.27 | 1252 | 1930 | 107-133 |
| Anharmonic | 16×16×8 | 2×4×4 | 1.49 | 862 | 1585 | 93-112 |

**Table S16.** Critical temperatures of superconductivity for various tin polyhydrides predicted in previous years. It can be concluded that, despite the incorrect structures, the theoretical estimates of $T_C$ are not very far from the experimental $T_C$($SnH_4$).

| **Phase** | **Pressure, GPa** | **$T_C$, K** | **Consistency (±20 %) with the experimental $T_C$** | **Reference** |
|---|---|---|---|---|
| *Ama2*-$SnH_4$ | 200 | 15-22 | No | G. Gao et al. [38] |
| $P6_3/mmc$-$SnH_4$ | 200 | 52-62 | No | G. Gao et al [38] |
| *I*4/*mmm*-$SnH_4$ | 220 | 80-91 | Yes | M. Davari et al.[37] |
| *I*4*m*2-$SnH_8$ | 220 | 72-81 | Yes | M. Davari et al.[37] |

## 8. Elastic properties

The elastic constants of tin hydrides were evaluated using VASP code (with IBRION=6, ISIF=4) [10-12]. The kinetic energy cutoff for plane waves was 600 eV. Tetrahedron method with Blöchl corrections was used [64]. The purpose of the performed calculations is to get an estimate for the Debye temperature, and to confirm the conclusions about the "softness" of $Sn_8H_{30}$ obtained earlier on the basis of experimental data.

**Table S17.** Elastic properties of tin hydrides in harmonic approximation: $Fd\bar{3}m$-$Sn_8H_{30}$ and $C2/m$-$SnH_{14}$.

| Compound | $C_{11}$-$C_{12}$, GPa | $C_{11}$+2$C_{12}$, GPa | $C_{44}$, GPa | B, GPa | G, GPa | E, GPa | Poisson's ratio |
|---|---|---|---|---|---|---|---|
| $Sn_8H_{30}$ (180 GPa) | 179.1 | 1257.3 | 20.8 | 467.9 | 33.3* | 97.5* | 0.46 |
| $SnH_{14}$ (200 GPa) | 409.5 | 1067.1 | 190.6 | 408.9 | 196.4 | 507.6 | 0.29 |

* Rather low values for a pressure of 200 GPa, characterizing unusual "softness" of this hydride.

**Table S18.** Elastic tensor $C_{ij}$ (in kbar) of $Fd\bar{3}m$-$Sn_8H_{30}$ at 180 GPa in harmonic approximation. The Debye temperature is 380 K [65].

| | $C_{i1}$ | $C_{i2}$ | $C_{i3}$ | $C_{i4}$ | $C_{i5}$ | $C_{i6}$ |
|---|---|---|---|---|---|---|
| $C_{1j}$ | 5385.0973 | 3593.9683 | 4998.1048 | 19.781 | 53.2054 | -67.0337 |
| $C_{2j}$ | 3593.9683 | 6194.2358 | 4768.1404 | 19.276 | -218.8843 | 4.2065 |
| $C_{3j}$ | 4998.1048 | 4768.1404 | 3706.2409 | 91.0908 | 169.6859 | 328.4853 |
| $C_{4j}$ | 19.781 | 19.276 | 91.0908 | 207.8296 | 59.2744 | 134.4097 |
| $C_{5j}$ | 53.2054 | -218.8843 | 169.6859 | 59.2744 | 496.8327 | -14.6745 |
| $C_{6j}$ | -67.0337 | 4.2065 | 328.4853 | 134.4097 | -14.6745 | 222.0507 |

**Table S19.** Elastic tensor $C_{ij}$ (in kbar) of $C2/m$-$SnH_{14}$ at 200 GPa in harmonic approximation. The Debye temperature is 1404 K [65].

| | $C_{i1}$ | $C_{i2}$ | $C_{i3}$ | $C_{i4}$ | $C_{i5}$ | $C_{i6}$ |
|---|---|---|---|---|---|---|
| $C_{1j}$ | 6286.7031 | 2192.0071 | 3813.1104 | 0.0 | 0.0 | -71.3417 |
| $C_{2j}$ | 2192.0071 | 8677.1942 | 1284.712 | 0.0 | 0.0 | 384.5683 |
| $C_{3j}$ | 3813.1104 | 1284.712 | 7304.6125 | 0.0 | 0.0 | 117.3508 |
| $C_{4j}$ | 0.0 | 0.0 | 0.0 | 1906.2537 | -357.1259 | 0.0 |
| $C_{5j}$ | 0.0 | 0.0 | 0.0 | -357.1259 | 1359.3217 | 0.0 |
| $C_{6j}$ | -71.3417 | 384.5683 | 117.3508 | 0.0 | 0.0 | 2313.0626 |